\documentclass[defaultstyle,11pt]{thesis}
\pdfoutput=1

\usepackage{enumerate}
\usepackage{graphicx}
\usepackage{latexsym}
\usepackage{rotating}
\usepackage{amsfonts}
\usepackage{bm}
\usepackage{amssymb}

\title{Paired fermionic superfluids with $s$- and $p$-wave interactions}

\author{Jesper~F.}{Levinsen}

\otherdegrees{B.A. Physics, University of Copenhagen, 2001 \\ 
M.S. Physics, University of Copenhagen, 2003} 

\degree{Doctor of Philosophy}{Ph.D., Physics}

\dept{Department of}{Physics}

\degreeyear{2007}

\advisor{Assistant Professor}{Victor Gurarie}

\reader{Eric Cornell}
\readerThree{Thomas DeGrand}
\readerFour{Andrew Hamilton}
\readerFive{James Shepard}

\abstract{\OnePageChapter

  This thesis presents theoretical work in $s$- and $p$-wave
  resonantly paired Fermi gases at zero temperature.  In the BEC
  regime of the wide-resonance $s$-wave BCS-BEC crossover, the
  chemical potential, speed of sound, condensate depletion, and ground
  state energy are computed. The quantities are calculated in a low
  density approximation. The bosonic scattering length is computed
  diagrammatically to be approximately $0.60a$ with $a$ the fermionic
  scattering length, confirming a well-known result. The perturbative
  approach breaks down in the intermediate BCS-BEC crossover regime
  which is not investigated. Quantum corrections are computed in the
  low density expansion, and the mass-imbalanced two-species Fermi gas
  is investigated. For identical fermions, it is discussed how
  $p$-wave Feshbach resonances naturally fall into two classes of
  ``weak'' and ``strong'', depending on the short-distance physics. It
  is shown how bound fermionic trimers appear in the strongly-resonant
  superfluid.  The appearance of the trimer under the strengthening of
  the resonance is investigated, and the lifetime of $p$-wave diatomic
  molecules due to collisional relaxation into trimers is estimated.

}

\acknowledgements{

  My journey towards becoming a physicist began at a very early age
  when I accompanied my dad to work at the Niels Bohr Institute. I was
  immediately captivated by all his different experiments and the
  magical feel of the place. The magic is still intact all these years
  later, although somehow I ended up with pen and paper as my primary
  tools. I guess it is difficult to foresee your future when you are
  only 7 years old? Having a pastor and a physicist as parents
  certainly proved an inspiration in my life and gave me a strong
  desire to discover and understand the world around me. I wish to
  thank my family for giving me a wonderful home to grow up in.

  I enjoyed a very dynamic and exciting physics education at the Niels
  Bohr Institute with inspirational teachers such as Mogens Dam, Poul
  Henrik Damgaard, Henrik Smith, Jens Martin Knudsen, Finn Berg, Jan
  Ambj{\o}rn, Andrew Jackson, and many others. In particular, I should
  thank Poul Henrik Damgaard, not only for his careful guidance of my
  master's thesis, but also for suggesting that I apply to CU when I
  considered pursuing my graduate studies in the US. He told me I would
  love it here, and he was absolutely right!

  From the day I arrived in Boulder, Victor has been a great advisor,
  always full of ideas for new and exciting projects. I have very much
  appreciated the independence he granted me in working on those
  projects which I found interesting. Over the past four years I have
  learned an immense amount from Victor, not only from the many times
  I would knock on his door to discuss different subjects, but also
  from his clear way of approaching physics problems. Thanks a lot for
  always taking the time to explain difficult subjects to me!

  Thanks also go to Jim Shepard who helped me take those important
  first steps towards solving difficult few-body problems. I should
  also mention his hugely useful class, through which I learned not to
  underestimate the power of Fetter and Walecka.

  Over the past few years, I have been happy to share F425 with a host
  of officemates. I thoroughly appreciate our many stimulating
  discussions. In particular, I enjoyed the company of James
  McCollough and Quan Zhang, and our many discussions of physics and
  whatever else came to mind. Many more people have helped make my
  time in Boulder rich. I would especially like to thank Rune
  Niclasen, Mike Wilking, Jim Hirschauer, and Sungsoo Choi.

  Last but not least I want to mention my girlfriend Klara. Klara
  continues to make me happy every day. She has taught me a lot about
  life and through her I have come to realize that life is not only
  about physics and football. I have immensely enjoyed our walks in
  the foothills, our trips, and our nights on the sofa watching a good
  movie. I look forward to every day with her.

  Finally, I am grateful for the generous support I have enjoyed from
  the NSF, the Fulbright Program (through the Danish-American
  Fulbright Commission), and the Danish Research Agency.

}

\emptyLoF
\emptyLoT

\begin{document}

\def\kx{\left( {p k_x \over 2} \right)}
\def\ky{\left( {p k_y \over 2} \right)}
\def\bk{{\bf k}}
\def\bx{{\bf x}}
\def\by{{\bf y}}
\newcommand{\boldgreek}[1]{\mbox{\boldmath$\mathbf{#1}$\unboldmath}}
\def\bpartial{\boldgreek{\partial}}
\def\8{\infty}
\def\ohm{\frac{1}{2m}}
\def\oh{\frac{1}{2}}
\def\ot{\frac{1}{3}}
\def\oq{\frac{1}{4}}
\def\tt{\frac{2}{3}}
\def\ft{\frac{4}{3}}
\def\tq{\frac{3}{4}}
\def\d{\partial}
\def\i{\imath\,}
\def\ih{\frac{\imath}{2}\,}
\def\undertext#1{\vtop{\hbox{#1}\kern 1pt \hrule}}
\def\ra{\rightarrow}
\def\lfa{\leftarrow}
\def\Ra{\Rightarrow}
\def\lra{\longrightarrow}
\def\ler{\leftrightarrow}
\def\lrb#1{\left(#1\right)}
\def\O#1{O\left(#1\right)}
\def\VEV#1{\left\langle\,#1\,\right\rangle}
\def\tr{\hbox{tr}\,}
\def\trb#1{\tr\lrb{#1}}
\def\dd#1{\frac{d}{d#1}}
\def\dbyd#1#2{\frac{d#1}{d#2}}
\def\pp#1{\frac{\partial}{\partial#1}}
\def\pbyp#1#2{\frac{\partial#1}{\partial#2}}
\def\ff#1{\frac{\delta}{\delta#1}}
\def\fbyf#1#2{\frac{\delta#1}{\delta#2}}
\def\br{\\ \nonumber & &}
\def\brr{\right. \\ \nonumber & &\left.}
\def\inv#1{\frac{1}{#1}}
\def\be{\begin{equation}}
\def\ee{\end{equation}}
\def\bea{\begin{eqnarray} & &}
\def\eea{\end{eqnarray}}
\def\ct#1{\cite{#1}}
\def\rf#1{(\ref{#1})}
\def\EXP#1{\exp\left(#1\right)}
\def\TEXP#1{\hat{T}\exp\left(#1\right)}
\def\INT#1#2{\int_{#1}^{#2}}
\def\MAT{{\it Mathematica }}
\def\LHS{left-hand side }
\def\RHS{right-hand side }
\def\COM#1#2{\left\lbrack #1\,,\,#2\right\rbrack}
\def\AC#1#2{\left\lbrace #1\,,\,#2\right\rbrace}

\def\grsq{{\bf \nabla}^2}

\def\cH{{\cal H}}
\def\rf#1{(\ref{#1})}
\def\t{\tilde}
\def\cE{{\cal E}}
\def\cD{{\cal D}}
\def\sign{{\rm sign}}

\def\rfs#1{Eq.~\rf{#1}}

\def\nn{\nonumber}

\def\del{\partial}
\def\up{\uparrow}
\def\down{\downarrow}

\chapter{Introduction \label{chap:intro}}

The history of superfluidity dates back to the discovery of
superconductivity (charged superfluidity) in mercury by Kamerlingh
Onnes \cite{Onnes1911}. Onnes observed that the electrical resistance
of mercury disappears completely as the system is cooled below a
critical temperature.
A detailed microscopic explanation of the phenomenon of
superconductivity proved elusive until the breakthrough in the 1950's
by Bardeen, Cooper, and Schrieffer
\cite{Bardeen1957a,Bardeen1957b}. The work of these authors (BCS
theory) described how an arbitrarily weak attractive interaction in a
degenerate Fermi gas leads to the formation of Cooper pairs
\cite{Cooper1956}, composed of weakly correlated pairs of
fermions. The BCS theory explained the resulting appearance of a
minimum excitation energy, or energy gap, and proved hugely succesful
in understanding properties of superconductors. In particular, the BCS
theory was able to correctly describe superfluid $^3$He (fermionic
atoms) but not superfluidity in liquid $^4$He (bosonic atoms).

The development of the theory of Bose-Einstein condensation (BEC)
occurred in parallel. In 1925 Einstein predicted \cite{Einstein1925}
the macroscopic occupation of the single-particle ground state below a
critical temperature in systems of particles obeying Bose
statistics. This followed ideas by Bose on the quantum statistics of
phonons \cite{Bose1924}. In 1938 it was proposed by London
\cite{London1938b,London1938a} that superfluidity in $^4$He appears as
a consequence of Bose-Einstein condensation. However, liquid $^4$He is
a strongly interacting system, very different from the weakly
interacting limit studied by Einstein. In particular, the number of
particles occupying the single-particle ground state is strongly
suppressed even at zero temperature, making the verification of
Bose-Einstein condensation in liquid $^4$He extremely difficult.

Considerable experimental effort was spent on attempting to obtain a
BEC in a weakly interacting system. This was first achieved in 1995 in
dilute gases of alkali atoms consisting of rubidium
\cite{Cornell1995}, sodium \cite{Ketterle1995}, and lithium
\cite{Hulet1995}. In these experiments, the gas was cooled to
temperatures for which the thermal de Broglie wavelength was of the
order of the interparticle spacing and the BEC of weakly interacting
atoms could be observed very clearly from the bimodal profile of the
momentum distribution of the gas below the critical temperature. For a
review of the subject of Bose-Einstein condensation in dilute gases
see e.g. Refs. \cite{Dalfovo1999,Leggett2001,Pethickbook}.

Soon after the observation of BEC in dilute gases of bosonic atoms,
considerable attention was shifted to the study of Fermi
gases. Applying the same experimental methods as in the bosonic gases
proved challenging, as the $s$-wave interactions necessary for
evaporative cooling are absent in spin-polarized Fermi gases. Thus the
cooling of Fermi gases requires the presence of two distinguishable
atoms, either in the form of two hyperfine spin components of the same
atom or in mixtures of two atomic species.  The first realization of a
quantum degenerate two-component Fermi gas was obtained in a system of
$^{40}$K \cite{Demarco1999} where a temperature of $T\approx0.2T_F$
was achieved ($T_F$ is the Fermi temperature). Soon experiments
followed in which quantum degeneracy was observed in $^6$Li
\cite{Truscott2001,Schreck2001} using sympathetic cooling with the
$^7$Li isotope.

In two-component ultracold Fermi gases, it should be possible to
observe superfluidity according to the BCS theory. However, in these
first experiments which reached quantum degeneracy, the critical
temperature required to enter the superfluid phase was very small due
to the diluteness and weak interactions in these gases. It was soon
recognized \cite{Holland2001,Timmermans2001} that it is possible to
enhance the weak interactions by the use of a scattering resonance
known as a Feshbach resonance \cite{Feshbach1958}, thereby increasing
the critical temperature.

The use of Feshbach resonances opened up the possibility of studying
both BCS superfluidity and BEC in the same system. These resonances
are a characteristic of the two-body interactions and they allow for
the precise control of the strength of interparticle interactions
through the tuning of an applied magnetic field. For a weak and
attractive interaction in a two-component Fermi gas, BCS theory is
valid and the system is expected to form a BCS superfluid below a
critical temperature. If the two-body interaction is made
progressively stronger by the tuning of a magnetic field, at some
critical strength of the interaction a two-body bound state becomes
possible. The bound states are then bosons and form a BEC at
sufficiently low temperatures. A BEC in a two-component Fermi gas was
first observed in 2003 through the typical bimodal distributions
\cite{Jin2003,Jochim2003,Zwierlein2003}. The observation of the BCS
superfluid phase is more subtle as the normal and superfluid phases
share many properties. Convincing proof of superfluidity was obtained
in 2005 through the observation of quantized vortices on both sides of
a Feshbach resonance \cite{Zwierlein2005}.

The possibility of a BCS-BEC crossover was originally proposed by
Eagles in 1969 \cite{Eagles1969}, and later by Leggett
\cite{Leggett1980}, and by Nozi{\` e}res and Schmitt-Rink
\cite{Nozieres1985}. These authors extended the mean field BCS model,
valid in the weakly interacting BCS regime, to an arbitrary
interaction strength, i.e. to well outside the regime of the original
model's validity. It was discovered that under the strengthening of
the interaction, the BCS gap equation, the mean field theory which
describes the BCS state, evolves into the Schr{\" o}dinger equation of
a bound pair of fermionic atoms, correctly predicting the binding
energy of the diatomic molecule. Furthermore, the results of these
authors suggested that no phase transition occurs in the crossover,
and thus the BCS and BEC phases are qualitatively the same phase.
These results connected the two types of paired superfluidity, within
mean field theory, and demonstrated how BCS and BEC are two extreme
limits of the same system.

Despite the success of the gap equation in describing the BCS
superfluid and the bound pairs in the BEC regime, it remained an open
question whether it could correctly predict properties of the
many-body system on the BEC side of a Feshbach resonance. The short
answer is that it cannot; the gap equation, when applied to the BEC
side of the crossover, implicitly assumes that the perturbative
parameter is the interaction strength between the Feshbach molecules,
i.e. that interactions are weak and scattering proceeds in the Born
approximation. However, the Born approximation is known to fail to
describe interactions between molecules properly
\cite{Petrov2005a}. It follows that while the BCS-BEC gap equation may
describe the physics of the BEC regime at a qualitative level due to
the lack of a phase transition, it is not to be trusted
quantitatively. Thus there is a need for a reliable technique in the
BEC regime, beyond mean field theory. This is one of the main subjects
of this thesis (see also Ref. \cite{Levinsen2006}).

In this thesis, the small parameter used in the BEC regime is the gas
parameter $na^3$. Here, $n$ is the density of fermions and $a$ the
$s$-wave scattering length between the two components of the Fermi
gas. In the gas parameter expansion the chemical potential, ground
state energy, speed of sound, and condensate depletion are
computed. The results are seen to match the results of the standard
dilute Bose gas \cite{AbrikosovBook,FetterBook}, however with the
boson-boson scattering length related to the underlying
fermion-fermion scattering length by
\begin{equation}
a_{\rm b} \approx 0.60 a.
\label{eq:abh}
\end{equation}
This result was first computed in Ref. \cite{Petrov2005a} using a
coordinate space formalism and the result is confirmed in this thesis
using a diagrammatic technique. While the work reported in
Ref. \cite{Levinsen2006} and also in this thesis was in progress, the
result (\ref{eq:abh}) was computed independently by Brodsky {\it et
  al} \cite{Brodsky2005} using essentially the same diagrammatic
technique.

At the crossover, the scattering length $a$ diverges. Thus, the
parameter $na^3$ is no longer small, and properties of the gas close
to the Feshbach resonance will not be computed.

In the 1950's, perturbative corrections in powers of $na^3$ to the
chemical potential and ground state energy were computed in the
standard dilute Bose gas
\cite{Lee1957a,Lee1957b,Wu1959,Sawada1959,Hugenholtz1959}. It is
explicitly demonstrated in this thesis how the fact that the bosons
are composed of fermions does not change the first two such
corrections, apart from replacing $a_{\rm b}$ by $0.60a$. Even higher
order contributions will depend explicitly on the presence of fermions
in the problem but will not be considered here.

An interesting system is a two-component Fermi gas where the two
components have different masses. For small mass imbalances, this
system still displays the usual BCS-BEC crossover physics, with the
few-body couplings slightly changed. However, as the mass imbalance is
increased, the system becomes less stable until finally above a
mass-ratio of 13.6 between the fermion components, the three-body
system has been found \cite{Petrov2004a} to contain bound states
similar to the Efimov states \cite{Efimov1971} in systems of
bosons. Properties of the mass imbalanced system such as the
atom-molecule and molecule-molecule scattering lengths and the loss
rate due to collisional relaxation have been calculated in
Refs. \cite{Petrov2003,Petrov2004a,Petrov2005a} by use of a real-space
technique, and the results will be confirmed in this thesis using a
diagrammatic technique.

To date, experiments in ultracold gases of fermionic atoms have
primarily studied gases consisting of two hyperfine states of the same
atom. However, recently there has been a considerable interest in
gases of identical fermionic atoms interacting close to a $p$-wave
Feshbach resonance.  The observation of $p$-wave Feshbach resonances
has been reported in $^{40}$K \cite{Ticknor2004} and $^6$Li
\cite{Schunck2005}, and theoretical work on the subject includes
Refs. \cite{Ho2005,Botelho2005,Gurarie2005,Cheng2005,Gurarie2007a,Levinsen2007}.
In the $p$-wave superfluid, the BCS and BEC phases are no longer
extreme limits of the same phase, rather they are separated by a
genuine phase transition \cite{VolovikBook,Botelho2005}, or in some
cases a sequence of phase transitions
\cite{Gurarie2005,Cheng2005,Gurarie2007a}. The phase transition may be
topological in some cases \cite{Read2000,Gurarie2005,Volovik2004}.

$p$-wave Feshbach resonances naturally fall into two categories of
weak and strong. Mean field theory in the crossover, as developed in
Refs. \cite{Gurarie2005,Cheng2005,Gurarie2007a}, is only strictly
valid for the weak resonances. On the other hand, the $p$-wave
resonance studied in the experiment of
Refs. \cite{Ticknor2004,Gaebler2007} is strong \cite{Gurarie2007a}. It
is therefore of interest to study the strong resonances. It was first
noticed by Y. Castin and collaborators \cite{CastinPrivate} that in
this regime, three fermions will form a bound state with angular
momentum 1. These trimers are very strongly bound, with a size of the
order of the short range physics which led to the formation of the
Feshbach molecule. In this thesis, the appearance of the trimer under
the strengthening of the resonance will be studied, and the decay rate
due to inelastic collisions in which two molecules turn into a trimer
and an atom will be estimated.

In this thesis the effects of a trapping potential will not be
considered. Of course, the trapping potential is an essential
ingredient in the experimental study of ultracold gases. However, the
local density approximation has proven quite succesful in describing
most properties of the system as long as the trapping potential is
sufficiently smooth. For a review of trapped gases see
e.g. Ref. \cite{Dalfovo1999}.

The thesis is organized as follows. Chapter \ref{chap:ultracold}
presents some simple properties of the BCS-BEC crossover, and a model
of the Feshbach resonance physics is introduced. Chapter
\ref{chap:bcsbec} contains calculations of properties of the BEC-BCS
crossover in the BEC regime, and in chapter \ref{chap:wu} higher order
corrections to the chemical potential and ground state energy are
computed in the gas parameter expansion. The two-component
mass-imbalanced Fermi gas in the BEC regime is studied in chapter
\ref{chap:mass}, with emphasis on few body scattering lengths and
collisional losses. The focus then shifts to $p$-wave resonantly
coupled superfluids. In chapter \ref{chap:pwavesetup} the system is
described and the model introduced. Chapter \ref{chap:pwave}
investigates the behavior of strongly-resonant $p$-wave superfluids
and the appearance of bound trimer states under the strengthening of
the resonance.

For simplicity of notation, Planck's constant $\hbar$ and Boltzmann's
constant $k_B$ have been set to 1.

\chapter{Physics of the BCS-BEC crossover \label{chap:ultracold}}

Soon after the achievement of Bose-Einstein condensation in dilute
gases of bosonic alkali atoms, considerable attention was instead
focused on gases comprised by fermionic atoms. The difference in
quantum statistics makes these systems quite distinct at ultracold
temperatures. As temperature is lowered, a bosonic system goes through
a phase-transition to a BEC and this happens even in the absence of
interactions. The Pauli principle, on the other hand, prevents two
identical fermions from occupying the same single-particle state.
Thus, in a non-interacting Fermi gas at zero temperature, the fermions
will instead occupy the lowest energy states with exactly one fermion
in each state. Hence, superfluidity in a gas of fermions can only stem
from the presence of interactions. As the temperature of a Fermi gas
is decreased, the crossover from a classical to a quantum behavior is
smooth.

Consider a gas of spin $1/2$ fermions at low temperature. If the
interactions in the spin singlet channel are arbitrarily weak and
attractive, the Fermi sea will be destabilized towards the formation
of Cooper pairs \cite{Cooper1956}.  The pairing occurs between
fermions with equal and opposite momenta, the pairs being weakly
correlated in momentum space. The pairs are highly overlapping in real
space and should not be thought of as molecules. The ground state of
the system is then no longer a Fermi sea, rather it is a superposition
of Cooper pairs, the primary contribution arising from fermions close
to the Fermi surface. This is explained in the famous work of Bardeen,
Cooper and Schrieffer \cite{Bardeen1957a,Bardeen1957b}. The system
studied by these authors consisted of pairs of spin-up and spin-down
electrons and the resulting many-body system was a superconductor (a
charged superfluid). In the context of ultracold two-species Fermi
gases, the regime described is known as the BCS regime.

\begin{figure}[tb]
\begin{center}
\includegraphics[height=2 in]{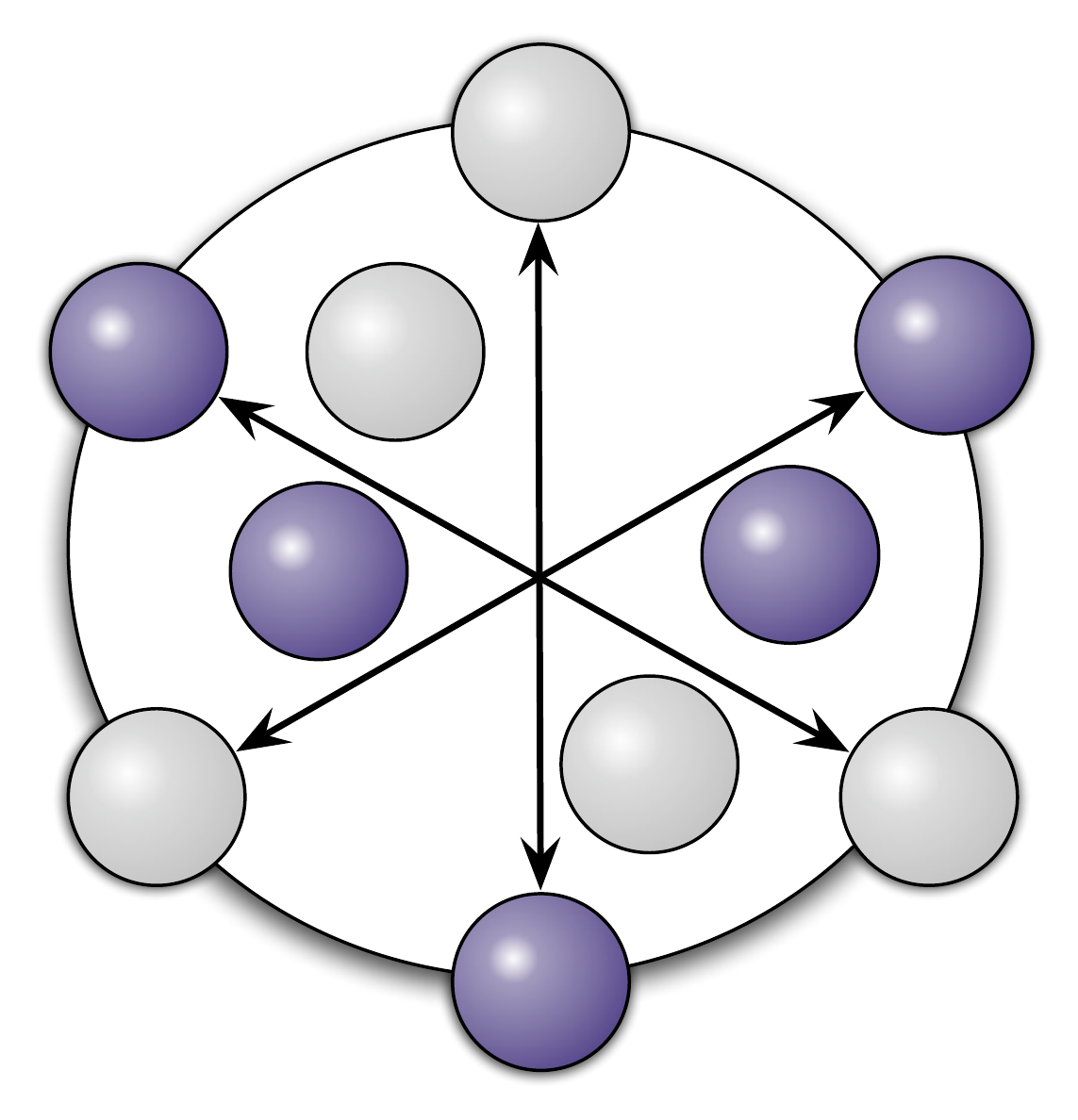}
\hspace{2cm}
\includegraphics[height=2 in]{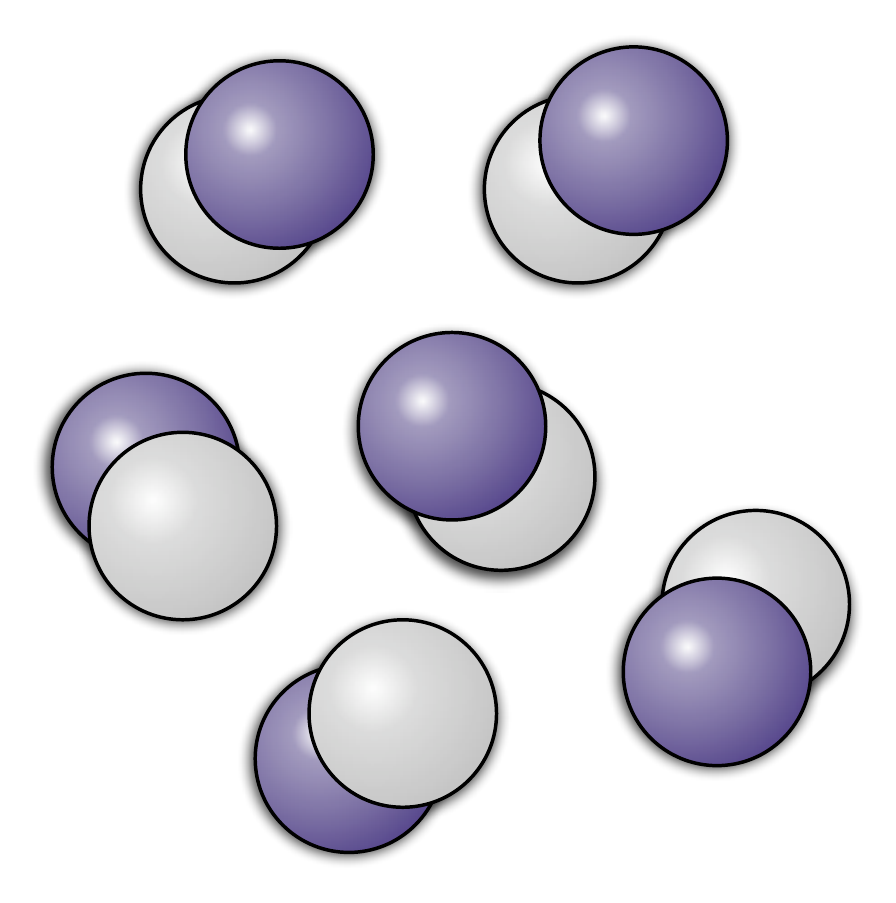}
\caption{An illustration of the BCS-BEC crossover. The BCS regime on
  the left is illustrated as the pairing in momentum space of
  fermions, while the BEC regime on the right is illustrated as the
  real-space pairing of atoms into molecules. Paired fermions in the
  central regime between the two limits share features with the paired
  fermions in both limits.}
\label{fig:bcsbec}
\end{center}
\end{figure}

Imagine then that by some means the attractive interaction is
increased. As the strength of the interaction is increased, a bound
molecular state between two fermions becomes possible. The resulting
bound state is then a boson, and a gas of these diatomic molecules may
form a BEC. This is the essence of the BCS-BEC crossover; as the
interaction between fermions is increased, there will be a continuous
crossover from a BCS superfluid to a BEC of diatomic molecules. Thus
the BCS and BEC regimes are merely limiting cases of the same
system. The crossover is illustrated in Fig. \ref{fig:bcsbec}. In the
central region between the two limiting regimes, the bosonic pairs retain some
properties of both Cooper pairs and diatomic molecules.

With the advent of BCS theory in the 1950's, the limiting regimes in
the BCS-BEC crossover were well understood. However, the connection
between these types of superfluids was not appreciated until the work
of Eagles in 1969 \cite{Eagles1969}, and the later works of Leggett
\cite{Leggett1980} as well as Nozi{\` e}res and Schmitt-Rink
\cite{Nozieres1985} in the 1980's. These authors started from a BCS
mean-field theory which is valid only for weak attractive
interactions. They extended the original model to arbitrarily strong
attractive interactions, well outside the range of validity of the
model. The mean field equation which describes properties of the BCS
superfluid is known as the gap equation and it was discovered
\cite{Leggett1980,Nozieres1985} that the gap equation evolves into the
Schr{\" o}dinger equation of a bound pair of fermions under the
strengthening of the interaction. Thus the results of these authors
connected the BCS superfluid with the BEC of condensed pairs of
fermionic atoms, within a mean field approach. Finite temperature
studies \cite{Nozieres1985,melo1993} further showed that the
transition temperature evolves smoothly between the two regimes. The
results suggested a smooth crossover without any phase transition. It
should be emphasized that the results obtained by this method, while
qualitatively correct due to the lack of a phase transition, are not
quantitatively correct in the BEC regime, since the mean field
approach is not strictly valid. This point will be discussed further
in chapter \ref{chap:bcsbec}.

\begin{figure}[tb]
\begin{center}
\includegraphics[height=2 in]{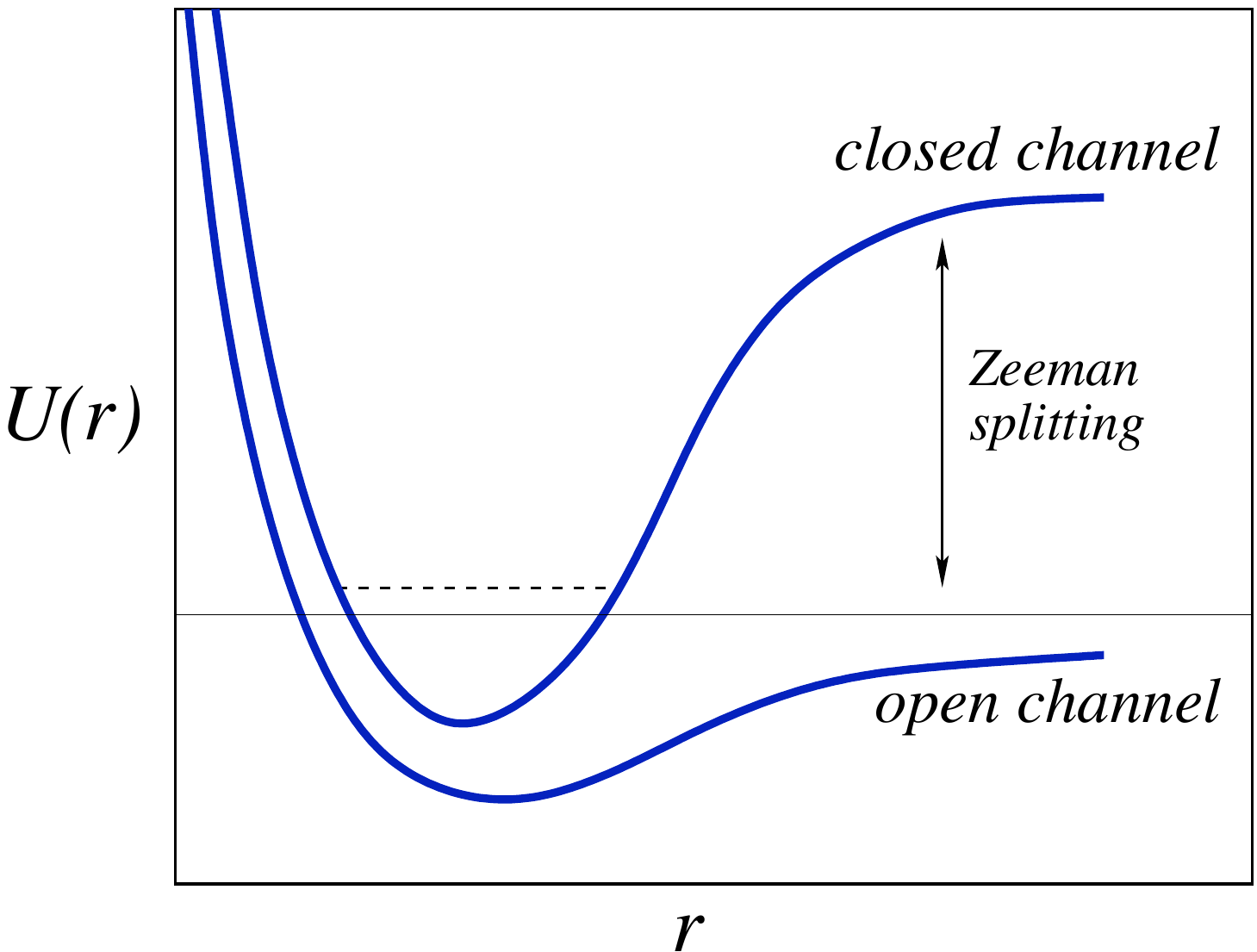}
  \caption{A Feshbach resonance occurs as the energy of the bound
    state in the closed channel crosses the energy of free particles
    scattering in the open channel.}
\label{fig:feshbach}
\end{center}
\end{figure}

The crucial tool which allows the experimental control over the
interaction strength in a two-component Fermi gas is a {\it Feshbach
  resonance} \cite{Feshbach1958,Feshbach1962,Timmermans1999}. A
Feshbach resonance is an intrinsic two-body phenomenon. Consider the
interaction between two fermionic atoms; at short distances the
interaction is repulsive, while at large distances it is dominated by
the van der Waals attraction which goes as $-C_6/r^6$. In general, the
interatomic potentials contain several vibrational bound states. A
Feshbach resonance occurs when the energy of a bound state in one
scattering channel (usually called the closed channel) coincides with
the energy of free particles scattering in another channel (the open
channel). The two spin channels will in general have different
magnetic moments. Typically, the closed channel is an approximate spin
singlet, the open channel an approximate spin triplet, and the
(large-distance) splitting between the energy levels is provided by
the Zeeman effect. The Zeeman splitting may be tuned by adjusting the
magnitude of an external magnetic field.
The situation is illustrated in Fig. \ref{fig:feshbach}.
The use of Feshbach resonances allows a
very precise tuning of the strength of the interaction, which may even
be changed in real time.

\begin{figure}[tb]
\begin{center}
\includegraphics[height=2.7 in]{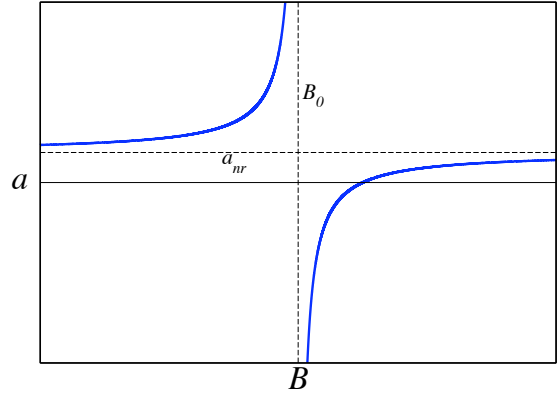}
  \caption{The scattering length $a$ across a Feshbach resonance. On
    the left side of the resonance a real bound state exists.}
\label{fig:avsh}
\end{center}
\end{figure}

The presence of the (real or virtual) bound state in the closed
channel can significantly alter scattering in the open channel. Across
a Feshbach resonance, the (open channel) scattering length behaves as
\begin{equation}
  a(B) = a_{\rm nr}\left[1+\frac{\Delta B}{B-B_0}\right].
\end{equation}
Here, $a_{\rm nr}$ is the non-resonant scattering length, the value of
the scattering length $a$ far off-resonance, $B_0$ is the magnetic
field position of the resonance, and $\Delta B$ is known as the
magnetic resonance width (defined as the difference in magnetic field,
$B_0-B$, at which the scattering length becomes zero). The behavior of
the scattering length is illustrated in Fig. \ref{fig:avsh}.

Below, low-energy scattering properties relevant for the present work
will be reviewed. A model of the resonant interactions, the so-called
two-channel model, will be considered and its relevance to the
Feshbach resonance physics described above will be discussed.

\section{Low energy scattering}
This section contains a short discussion of the scattering theory
relevant for this thesis. For elastic scattering of two particles via
a centrally symmetric interaction potential, the scattering amplitude
may be written as \cite{LL}
\begin{equation}
  f(k,\theta) = \sum_{\ell = 0}^{\infty}(2\ell+1)  P_\ell(\cos\theta)
  f_\ell(k).
\end{equation}
Here, $|\vec k|=|\vec k'|\equiv k$ is the magnitude of the incoming,
$\pm \vec k$, and outgoing, $\pm \vec k'$, momenta, while $\theta$ is
the scattering angle, with $\vec k\cdot \vec k'=k^2\cos\theta$. The
probability of scattering through the solid angle $\Omega$ is related
to the scattering amplitude by $d\sigma/d\Omega=|f|^2$. A centrally
symmetric interaction $U(r)$ conserves the angular momentum $\ell$ and
in this thesis, focus will be on $s$- and $p$-wave scattering, with
\begin{eqnarray}
f_s(k) \equiv f_{\ell = 0}(k), \nn \\
f_p(k) \equiv f_{\ell = 1}(k).
\end{eqnarray}
The partial wave amplitudes may also be written in the form \cite{LL}
\begin{equation}
f_\ell(k) = \frac1{k^{-2\ell}F_\ell(k^2)-ik}.
\label{eq:partialwav}
\end{equation}
The functions $F_\ell(k^2)$ are real and may be Taylor expanded in
their arguments. Assuming that all the expansion coefficients take
values of the order of some ``natural'' scale, it is seen that for
low-energy scattering (e.g. in ultracold gases) the functions
$F_\ell(k^2)$ are well approximated by the first few terms of their
Taylor expansions. In particular it is important to note that the
scattering amplitude at low energy only depends very weakly on the
precise details of the complicated inter-atomic interaction potential
$U(r)$.

For $s$-wave scattering, keeping only one term in the Taylor expansion
of $F_\ell(k^2)$ amounts to
\begin{equation}
  f_s(k) = \frac1{-a^{-1}-ik},
  \label{eq:lowestscat}
\end{equation}
where $a$ is the $s$-wave scattering length. The scattering amplitude
has one pole, located at $k_{\rm pole}=+ia^{-1}$. This pole only
corresponds to a real bound state if $a>0$, otherwise it corresponds
to a virtual bound state \cite{LL}. For $a>0$ the real bound state has
the energy
\begin{equation}
E = -\frac1{2m_{\rm r}a^2},
\end{equation}
where $m_{\rm r}$ is the reduced mass in the two body system. The
result already nicely matches the Feshbach resonance picture described
above. However, the approximation Eq. (\ref{eq:lowestscat}) is
deficient in one respect, it lacks a true resonance, i.e. a
quasistationary state with a finite lifetime \cite{Gurarie2007a}. In
order to correctly capture the possible existence of such states, the
next order expansion
\begin{equation}
F_s(k^2) = -a^{-1}+\frac12r_0k^2
\end{equation}
is needed. The parameter $r_0$ is known as the effective range of the
interaction potential $U(r)$.

An important energy scale is set by the effective range
\begin{equation}
  \Gamma_0 = \frac{4}{mr_0^2}.
\end{equation}
This energy scale is related to the magnetic field width by the
approximate relationship $\Gamma_0\approx 4m\mu_B^2a_{\rm nr}^2\Delta
B^2$ \cite{Gurarie2007a}. The gas contains a finite atom density $n$,
and consequently an additional length scale is set by the average
particle spacing $n^{-1/3}$. The corresponding energy scale is the
Fermi energy\footnote{In the absence of two-body bound states this
  would have been the conventional Fermi energy.}  $\epsilon_F$ given
by
\begin{equation}
\epsilon_F = \frac{k_F^2}{2m}.
\end{equation}
$k_F$ is the Fermi momentum. The two energy scales allow the
definition of a dimensionless parameter \cite{Gurarie2007a}
\begin{equation}
\gamma_s = \frac{\sqrt8}\pi\sqrt{\frac{\Gamma_0}{\epsilon_F}},
\end{equation}
which measures the {\it width} of the resonance. Resonances for which
$\gamma_s\ll1$ are termed {\it narrow} while those with $\gamma_s\gg1$
are called {\it wide}. The precise criterion for the resonances to be
wide was first derived in Ref. \cite{Bruun2001}.

The distinction between wide and narrow resonances is quite
important. Since the effective range only varies weakly across a
Feshbach resonance, for narrow resonances $\gamma_s$ provides a small
parameter in which a quantitatively correct (in the zero width limit)
perturbation theory may be formulated across the whole crossover
\cite{Gurarie2007a}. In contrast, no such simplification occurs for
wide resonances; the other dimensionless parameter which may be
constructed from the three length scales $a,$ $r_0$, and $n^{-1/3}$ is
known as the {\it gas parameter}, $n^{1/3}a$. Since the scattering
length diverges across the resonance, no small parameter exists close
to a wide Feshbach resonance.

Most current experiments in ultracold two-component Fermi gases are
carried out close to wide resonances. As these are the most relevant,
the focus in the parts of this thesis concerning $s$-wave Feshbach
resonances (chapters \ref{chap:bcsbec}, \ref{chap:wu}, and
\ref{chap:mass}) will be on wide resonances.

It is precisely for the narrow resonances that the two-body problem
may display a true resonance (a positive energy state with a finite
lifetime) \cite{Gurarie2007a}. Since the $s$-wave Feshbach resonances
studied in this thesis are wide, these quasistationary states will be
ignored. It is, however, interesting to note that although most
experiments are performed close to wide resonances, since
$\gamma_s\sim n^{-1/3}$ resonances which are wide in current
experiments may become narrow if higher densities are achieved.

\section{The two-channel model \label{sec:2channel}}
A model which reflects the physics of the interaction in a
two-component Fermi gas interacting close to a Feshbach resonance is
provided by the two-channel model. It was first studied by Rumer in
1959 \cite{Rumer1959} and in the context of atomic Bose-Einstein
condensates it was studied by Timmermans {\it et al} in 1999
\cite{Timmermans1999}. Without interactions, the model describes the
open-channel atoms and closed-channel molecules
separately. Interaction between atoms and molecules is provided by an
interconversion term.

The model is conveniently defined in terms of a functional integral
\begin{equation} 
Z=\int \cD \bar \psi_\uparrow \cD \psi_\uparrow \cD \bar
\psi_\downarrow \cD \psi_\downarrow \cD b \cD \bar b~e^{iS_{\rm tc}}.
\label{eq:tc1}
\end{equation}
$\psi_\up$ and $\psi_\down$ are the fermionic atoms in two different
hyperfine states (for simplicity denoted $\up$ and $\down$) and $b$ is
the closed-channel molecule. The action $S_{\rm tc}$ consists of three
parts, $S_{\rm tc}=S_0+S_{0\rm b}+S_{\rm fb}$. Of these, $S_0$ is the
free action of fermions
\begin{equation}
S_0 = \sum_{\sigma=\up,\down}\int d^3x\,dt\,
\bar\psi_\sigma\left(i\frac\partial{\partial t}+\frac1{2m}
\frac{\partial^2}{\partial {\bf x}^2}
+\mu\right)\psi_\sigma,
\end{equation}
where $\mu$ is the chemical potential of the fermions. The free action
of the closed-channel molecules is
\begin{equation}
  S_{0\rm b}= \int d^3x\, dt\, \bar b \left( i
    \frac\partial{\partial t}+
    \frac{1}{4m}\frac{\partial^2}{\partial {\bf x}^2}+2\mu
    -\epsilon_0\right)b.
\label{eq:freebosonaction}
\end{equation}
$\epsilon_0$ is called the detuning and is the parameter
which is typically controlled by a magnetic field in the experimental
setup. That is, the detuning controls the proximity to the Feshbach
resonance. The interconversion term between open-channel atoms and the
closed-channel molecule is given by
\begin{equation}
S_{\rm fb}= -g \int d^3x\,dt\, \left[ b \, \bar \psi_{\uparrow} \bar
\psi_{\downarrow} + \bar b \, \psi_{\downarrow} \psi_{\uparrow} \right].
\label{eq:interconversion}
\end{equation}
The coupling $g$ is the probability of converting a $b$ particle to a
pair of atoms. In itself, the interaction occurs at a point in space
and is unphysical; the action needs to be supplemented by a cut-off,
$\Lambda\sim\frac1{R_e}$ where $R_e$ denotes the range of the
(short-ranged) forces.

It is important to note that the molecule $b$ is only the true bound
state in the case of a vanishing coupling between atoms and
molecules. Otherwise, the true bound state is the ``bare'' molecule
$b$ surrounded by a cloud of atoms. Thus, while the bare molecule's
size is of the range of the forces which led to its creation, $R_e$,
the size of the physical molecule is of the order of the atom-atom
scattering length $a\gg R_e$. In fact, the contribution of the bare
molecule to the physical molecule effectively disappears for
infinitely wide Feshbach resonances, where the two-channel model may
be replaced by a one-channel model. This will be discussed in more
detail in chapter \ref{chap:bcsbec}.

The two parameters which characterize the scattering between the atoms
are the scattering length and the effective range. These may be
calculated in the model and are given by \cite{Andreev2004}
\begin{equation}
  a=-\frac{m g^2}{4 \pi \left(\epsilon_0-\frac{g^2 m \Lambda}{2 \pi^2}
    \right)},
  \hspace{10mm}  r_0=-\frac{8\pi}{m^2 g^2}.
\label{eq:ar0tc}
\end{equation}
It is apparent that the position of the resonance is shifted from the
``bare'' value $\epsilon_0$ to the {\it physical detuning},
\begin{equation}
\omega_0 = \epsilon_0-\frac{g^2 m \Lambda}{2 \pi^2}.
\label{eq:physdetuning}
\end{equation}
The shift again occurs because the bare molecule does not coincide
with the physical molecule.

In terms of the parameters of the model,
\begin{equation}
\gamma_s = \frac 8\pi\frac1{k_F|r_0|}.
\end{equation}
In the two-channel model, it is again seen that $\gamma_s$ is
independent of detuning $\omega_0$ and scattering length $a$. Thus, if
$\gamma_s\ll1$ away from resonance then this is the case across the
whole crossover. On the other hand, if $r_0$ is small then the
resonance will be wide. In the next chapter the effective range will
be taken to zero in order to model the wide resonances.

\chapter{The BCS-BEC crossover in the BEC regime
  \label{chap:bcsbec}} In the previous chapter some basic properties
of the BCS-BEC crossover were described. Consider a two-component
Fermi gas with short-ranged interactions. If the interactions between
the two components are weak, the fermions will pair into the usual
Cooper pairs and the system becomes a BCS superconductor at low
temperatures. If the strength of the interaction is then increased by
some means, at some point it becomes possible for two fermions to form
a bound molecular state. As the strength is tuned further away from
resonance, the bosonic molecules will form a regular BEC, and it is
properties of this BEC which will be described in the present chapter.

The $s$-wave scattering length of two different fermionic atoms
diverges across the Feshbach resonance. However, the BCS-BEC gap
equation which describes the weakly attractive Fermi two-component
Fermi gas implicitly assumes that the scattering is weak. Thus, even
though the gap equation correctly describes the physics on the BCS
side of the resonance there is no reason why the gap equation should
be trusted on the BEC side of a Feshbach resonance. Yet, in
Refs. \cite{Leggett1980,Nozieres1985} it was discovered that under the
strengthening of the interaction, the gap equation develops into the
Schr{\"o}dinger equation for the bound pair of fermionic atoms and
correctly predicts the binding energy of this bound pair. Studies by
these authors furthermore demonstrated that the BCS superfluid evolves
into the BEC without undergoing a phase transition. Thus, even though
the gap equation is not quantitatively correct on the BEC side of a
resonance, it still yields qualitatively correct results.

In order to systematically investigate the BEC regime of the weakly
bound bosonic molecules, it is necessary to identify a small
parameter. This is provided by the so-called gas parameter
$n^{1/3}a$. $n$ is the density of fermions while $n_{\rm b}=n/2$ is
the density of bosonic molecules. The size of a molecule is roughly of
order $a$, as discussed in chapter \ref{chap:ultracold}, while the
average intermolecular distance is $n_{\rm b}^{-1/3}$. Thus, even
though the molecules are rather large compared to the short-distance
physics which led to their formation, for small values of the gas
parameter the gas is a dilute Bose gas.

Using the low-density (or small gas parameter) expansion allows the
computation of the molecular chemical potential $\mu_{\rm b}$, the speed of
sound wave propagation $u$, the condensate depletion, and the ground
state energy $E_0$. In the lowest order these are
\begin{eqnarray}
\mu_{\rm b} & = & \frac{4\pi n_{\rm b} a_{\rm b}}{m_{\rm b}}, \label{eq:res1}\\
u & = & \sqrt{\frac{4\pi n_{\rm b} a_{\rm b}}{m_{\rm b}^2}}, \label{eq:res2}\\
n_{0,{\rm b}} & = & n_{\rm b}\left[1-\frac83\sqrt{\frac{n_{\rm b}a_{\rm b}^3}\pi}\right], \label{eq:res3}\\
E_0/V & = & \frac{2\pi n_{\rm b}^2a_{\rm b}}{m_{\rm b}}. \label{eq:res4}
\end{eqnarray}
$m_{\rm b}=2m$ is the mass of a molecule and $V$ is the volume of the
system. The molecule-molecule scattering length $a_{\rm b}$ appearing
in these equations is proportional to the atom-atom scattering
length. The coefficient of proportionality was first calculated by
Petrov {\it et al} \cite{Petrov2005a} who found
\begin{equation}
a_{\rm b} \approx 0.60a.
\end{equation}
This result was later confirmed in
Refs. \cite{Brodsky2005,Levinsen2006} with the calculations of
Ref. \cite{Levinsen2006} presented in section \ref{sec:4body} below.
It should be noted that the gap equation predics $a_{\rm b}=2a$. Hence
it is seen how the gap equation does not take interactions between the
molecules properly into account.

At the crossover between BEC and BCS physics, the scattering length $a$
diverges. This means that the gas parameter is no longer small,
instead the scattering length will be much larger than the
inter-molecular distance. The point at which $a$ diverges is known as
the unitary point. The physical behavior of the superfluid at zero
temperature can only depend on one parameter, the density. The
physical properties of the gas in this regime may be obtained by
methods such as a large $N$ expansion \cite{Veillette2007}, Monte
Carlo techniques (see e.g. the review \cite{Giorgini2007} for
references), or other methods. The perturbative approximation employed
in this thesis breaks down close to the resonance and here only
systems with $n^{1/3}a\ll1$ will be considered.

In this chapter, wide resonances are investigated. As discussed in
chapter {\ref{chap:ultracold}} this requires $a\gg r_0$. It is
important to note that while $a$ diverges across a Feshbach resonance
the effective range stays roughly constant. Thus the validity of the
low-density expansion is restricted to scattering lengths such that
$|r_0|\ll a \ll n^{-1/3}$. This limitation of the model's validity is
not too restrictive; typical experimental densities of two-component
Fermi gases are of order $10^{13}$ cm$^{-3}$ corresponding to an
interparticle distance of order $10^4a_0$ ($a_0$ is the Bohr radius)
while $r_0$ is typically of order $50a_0$.

The fact that the investigated Feshbach resonances are wide implies
that the dynamics of the closed-channel molecule is suppressed. In the
limit of a vanishing effective range it is then simpler to describe
the gas by a one-channel model rather than the two-channel model
described in chapter \ref{chap:ultracold}. The equivalence of these
two approaches will be demonstrated below.

It should be reiterated that all calculations presented in this thesis
proceed at zero temperature. A finite temperature extension of these
results will not be discussed here. For further references see
Ref. \cite{Andersen2004} and references therein.

The systematic perturbative expansion in the gas parameter may be
continued below the lowest order. The first two corrections to the
results (\ref{eq:res1}) and (\ref{eq:res4}) depend only on the gas
parameter after which further details of the interaction become
important. The discussion of these higher order contributions to the
chemical potential and ground state energy will be studied in chapter
\ref{chap:wu}.

Another natural extension of the work presented here is to consider a
two-component gas with a mass-imbalance. This is treated in chapter
\ref{chap:mass}.

This chapter is organized as follows. First, the one-channel model is
introduced and the diagrammatic approach to the BEC regime of the
weakly bound molecules is discussed. Next, in section
\ref{sec:bcsbec}, physical properties of the BEC are computed in the
lowest order gas parameter expansion. Finally, sections
\ref{sec:3body} and \ref{sec:4body} contain the few-body diagrammatic
calculations relevant for the BEC regime.

\section{The one-channel model}
In the previous chapter, the two-channel model was introduced to
describe a system where a bound state of two fermions appeared as the
magnetic field was tuned across a resonance. The Feshbach resonances
described in this thesis (and indeed relevant to most BCS-BEC
experiments) are wide, and for wide resonances it is more convenient
to treat the gas of weakly bound molecules using what is known as the
one-channel model. Below, it will be shown that these two approaches
are equivalent in the limit of infinitely wide resonances.

The one-channel model may be expressed by a functional integral
\begin{equation}
Z=\int \cD \bar \psi_\uparrow \cD \psi_\uparrow \cD \bar
\psi_\downarrow \cD \psi_\downarrow ~e^{iS_{\rm f}}.
\label{eq:sf}
\end{equation}
The action $S_{\rm f}=S_0+S_{\rm int}$ consists of a free part and
an interacting part. The free part is
\begin{equation}
S_0=\sum_{\sigma=\uparrow,\downarrow} \int d^3x~ dt~ \bar
\psi_{\sigma} \left( i \frac{\del}{\del t}+\frac{1}{2m}\frac{\del^2}{\del
{\bf x}^2}+\mu \right) \psi_\sigma,
\label{eq:S}
\end{equation}
while the interacting part is
\begin{equation}
S_{\rm int} = \lambda \int d^3x~dt~ \bar \psi_{\uparrow} \bar \psi_{
\downarrow} \psi_{\downarrow} \psi_{\uparrow}.
\label{eq:Sint}
\end{equation}
Here $\lambda$ is chosen positive because the interaction potential is
attractive. The interactions described by $S_{\rm int}$ take
place at one point in space which by itself is unphysical. The results
should be independent of the precise short distance physics and thus
$S_{\rm int}$ needs to be supplemented by a cut-off
$\Lambda$. Note that the one-channel model described by
Eqs. (\ref{eq:sf}), (\ref{eq:S}), and (\ref{eq:Sint}) does not
explicitly contain the closed channel molecule associated with a
Feshbach resonance.

\subsection{Diagrammatic approach}
Consider now the one-channel model of a two-component Fermi gas with
$m_\uparrow=m_\downarrow\equiv m$ and $\mu_\uparrow=\mu_\downarrow\equiv \mu$. The
propagator of fermions in the model described by
Eqs. (\ref{eq:sf}), (\ref{eq:S}), and (\ref{eq:Sint}) is
\begin{equation}
G_0(p,\omega) = \frac1{\omega-p^2/2m+\mu+i0},
\end{equation}
where $m$ is the mass of either one of the fermionic atoms.

\begin{figure}[bt]
\begin{center}
\includegraphics[height=.65 in]{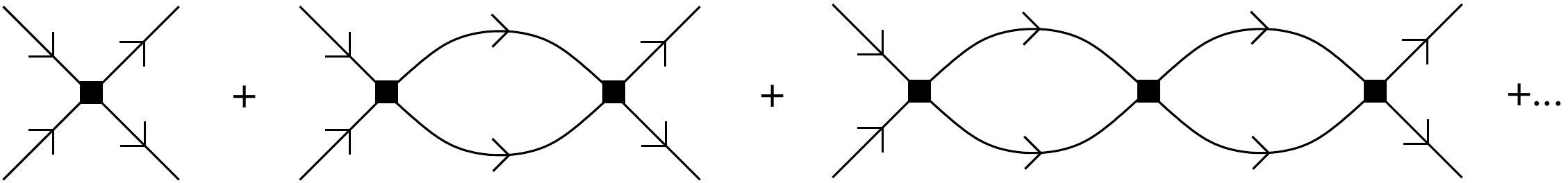}
\caption{The diagrams which sum to give the atom-atom scattering
  matrix $T$. The straight line is a propagator of fermions while the
  square at each vertex is the coupling $\lambda$.}
\label{fig:bubble}
\end{center}
\end{figure}

The scattering length of fermions in the vacuum may be computed by
summing the diagrams shown in Fig. \ref{fig:bubble}. The result of
this summation is the $T$-matrix and the relationship between the
scattering length and the $T$-matrix is\footnote{The $T$-matrix is a
  quantity defined for any kind of scattering. In this thesis,
  scattering between two fermions, a fermion and a bosonic molecule,
  and between two molecules will be considered. These result in
  $T$-matrices $T^{\rm ff}$, $T^{\rm bf}$, and $T^{\rm bb}$. In
  general the superscripts will be suppressed to ease notation except
  where any confusion might occur.}
\begin{equation}
a = \frac{m_r}{2\pi}T(0)
\end{equation}
with $m_r=m/2$ the reduced mass of two atoms.

In the calculation of the fermionic scattering length it is important
to keep $\mu=0$ since the scattering proceeds in the vacuum. The
diagrams in Fig. \ref{fig:bubble} form a geometric series since each
loop is independent of the other loops. The $T$-matrix evaluated at
zero incoming and outgoing fermion four-momenta is
\begin{equation}
T(0) = \lambda+\lambda\Pi(0)\lambda+\lambda\Pi(0)\lambda\Pi(0)\lambda+\dots
= \frac{-\lambda}{1+\lambda\Pi(0)}.
\end{equation}
The bubble diagram $\Pi$ consists of two fermion propagators
integrated over the free four-momentum in the loop. That is,
\begin{equation}
  \Pi(0) = i\int \frac{d^4p}{(2\pi)^4}G_0(p)G_0(-p) = -\frac{m\Lambda}{2\pi^2}.
\end{equation}
Thus, in the one-channel model the atom-atom scattering length is
\begin{equation}
a = \left(-\frac{4\pi}{m\lambda}+\frac{2\Lambda}\pi\right)^{-1}.
\label{eq:aff}
\end{equation}

\begin{figure}[bt]
\begin{center}
\includegraphics[height=1.2 in]{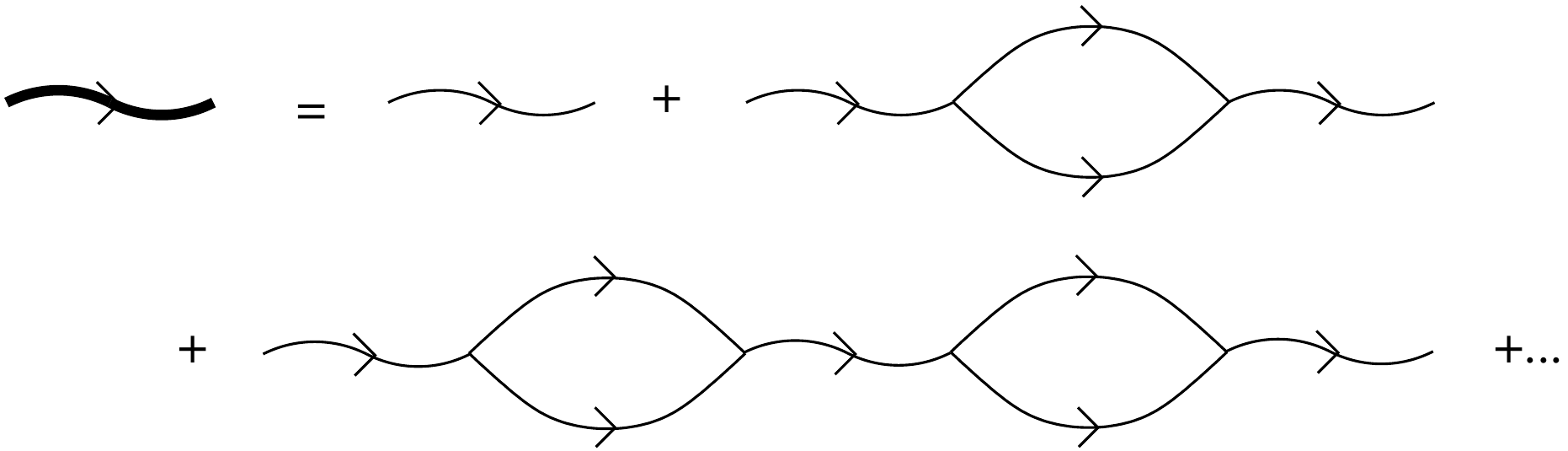}
\caption{The relation between the dressed molecular propagator (thick
  wavy line) and the bare molecular propagator (thin wavy line). The
  bare propagator equals $-\lambda$.}
\label{fig:bosonprop}
\end{center}
\end{figure}

In the one-channel model, the dynamics of the closed-channel molecule
is suppressed. Instead the propagator of the closed-channel model is
constant, equal to $-\lambda$. However, the molecule needs to be
``dressed'' by the self energy correction due to repeated interactions
between the two fermionic components. Fig. \ref{fig:bosonprop} shows
the propagator dressed by the self energy corrections, which consist
of fermionic bubble diagrams. The molecular propagator is then
\begin{equation}
D_0(p) = \frac{-\lambda}{1+\lambda\Pi(p)},
\end{equation}
where it has been used that the diagrams again form a geometric
series. The calculation now proceeds in the medium at a finite
chemical potential $\mu\leq0$. Then the fermionic loop takes the value
\begin{eqnarray}
\Pi(p,\omega) & = & i\int\frac{d^4q}{(2\pi)^4}G(p+q)G(-q) \nn \\
& = & -\frac{m\Lambda}{2\pi^2}+\frac{m^{3/2}}{4\pi}
\sqrt{-\omega+p^2/4m-2\mu-i0}.
\end{eqnarray}
The infinitesimal is derived from the fermionic propagators and is
such that also the molecular propagator is retarded. Note the
importance of having $\mu\leq0$. Had this not been the case, the
fermionic propagator would not have been retarded and hole propagation
would have occured.

Using the relation (\ref{eq:aff}) between the scattering length and
the cut-off, the molecular propagator becomes
\begin{equation}
  D_0(p,\omega) = \frac{4\pi}m\frac1{a^{-1}-
    \sqrt{m}\sqrt{-\omega+p^2/4m-2\mu-i0}}.
\label{eq:bosonprop}
\end{equation}

\section{BCS-BEC gap equation}
In order to derive the BCS-BEC gap equation (the equation for the
extremum of the action) from the one-channel model, it is convenient
to introduce a Hubbard-Stratonovich field $\Delta$. The one-channel
functional integral (\ref{eq:sf}) may then be written as
\begin{equation}
Z=\int \cD \bar \psi_\uparrow \cD \psi_\uparrow \cD \bar
\psi_\downarrow \cD \psi_\downarrow \cD \Delta \cD \bar
\Delta\,e^{iS_{\rm f}},
\end{equation}
with the action $S_{\rm int}$ replaced by
\begin{equation}
  S_{\rm HS}= -\int d^3x\,dt\,\left[ \frac{1}{\lambda} \bar \Delta
    \Delta +
    \left( \Delta \bar \psi_{\uparrow} \bar \psi_{\downarrow} + \bar
      \Delta \psi_{\downarrow} \psi_{\uparrow} \right)
  \right].
\label{eq:shs}
\end{equation}

The advantage of introducing the Hubbard-Stratonovich field is that
the action $S_{\rm f}$ becomes quadratic in fermion fields, and these may
then be integrated out. The resulting functional integral is
\begin{equation}
Z = \int \cD \Delta \cD \bar \Delta\,e^{iS_\Delta},
\end{equation}
with
\begin{equation}
S_{\Delta} = -i\,\tr \log \left( \matrix { \omega^+
-\frac{p^2}{2m}+\mu & -\Delta \cr - \bar \Delta & \omega^+ +
\frac{p^2}{2m}-\mu } \right) -\frac{1}{\lambda}
 \int d^3x\,dt\,\bar \Delta \Delta.
\label{eq:Seff1}
\end{equation}
Here $\omega^+=\omega+i0\, \sign~\omega$.

The gap equation is then derived by assuming that the extremum of the
action $S_\Delta$ occurs at a constant value $\Delta_0$ of
$\Delta$. This results in
\begin{equation}
  \frac{1}{\lambda} =\frac{1}{2} \int \frac{d^3 p}{(2 \pi)^3}
  \frac{1}{\sqrt{ \left(\frac{ p^2}{2 m} -  \mu \right)^2 + \Delta_0^2
    }}.
\label{eq:BCSBEC0}
\end{equation}
Using Eq. (\ref{eq:aff}) the divergence in the momentum integral of
Eq. (\ref{eq:BCSBEC0}) at large momenta may be regularized, and the
gap equation becomes
\begin{equation}
  -\frac{m}{4 \pi a} =\frac{1}{2} \int \frac{d^3p}{(2\pi)^3} \left [
    \frac{1}{\sqrt{ \left(\frac{p^2}{2 m} - \mu \right)^2 + \Delta_0^2 }}
    - \frac{2 m}{p^2} \right].
\label{eq:BCSBEC}
\end{equation}
%
%
%

\section{Equivalence of the one- and two-channel models}
The two-channel model described in section \ref{sec:2channel} contains
the dynamics of the molecular field. Compared with the one-channel
model, this model has an additional parameter, the detuning. This
parameter allows for the control of the effective range $r_0$
independently of the scattering length $a$. The precise relationships
are listed here again for convenience
\begin{eqnarray}
a & = & -\frac{mg^2}{4\pi\left(\epsilon_0-\frac{mg^2\Lambda}{2\pi^2}\right)},
\label{eq:scatintc}\\
r_0 & = & -\frac{8\pi}{mg^2}.
\end{eqnarray}
To ensure that a Feshbach resonance is wide, the effective range
should be vanishingly small. This means that the interaction strength
must be taken to $\infty$, while at the same time the detuning should
be adjusted in such a way that the scattering length remains finite.

This may be achieved as follows. Redefine the bosonic fields of the
two-channel model
\begin{equation}
\Delta = gb, \hspace{10mm}\bar\Delta = g\bar b.
\end{equation}
Using this definition, the interconversion part of the action,
Eq. (\ref{eq:interconversion}), becomes
\begin{equation}
S_{\rm bf} \to -\int d^3x\,dt\,\left[\Delta\bar\psi_\up\bar\psi_\down
+\bar\Delta\psi_\down\psi_\up\right].
\end{equation}
For the free bosonic action use
\begin{equation}
\epsilon_0 = -g^2\left(\frac m{4\pi a}-\frac{m\Lambda}{2\pi^2}\right)
= -\frac{g^2}\lambda,
\end{equation}
which follows from Eqs. (\ref{eq:aff}) and (\ref{eq:scatintc}). Then
in the limit $g\to\infty$ the free bosonic action
(\ref{eq:freebosonaction}) becomes
\begin{equation}
S_{0\rm b} = -\int d^3x\,dt\, \bar\Delta\Delta,
\end{equation}
and the sum $S_{0\rm b}+S_{\rm bf}$ exactly matches the
Hubbard-Stratonovich action $S_{\rm HS}$ of the one-channel model, see
Eq. (\ref{eq:shs}). Thus in the limit of wide resonances, the
one-channel model is equivalent to the two-channel model.

In order to ensure that the Feshbach resonance is wide, the
interaction strength is taken to the limit $g\to\infty$. Thus it is
again seen that it is not possible to treat $g$ as a small coupling
and construct a perturbative expansion in powers of $g$.

\section{The BEC regime of weakly bound molecules
  \label{sec:bcsbec}}

Having set up the perturbative framework in terms of the gas
parameter, the calculation of properties of the BEC of weakly bound
bosonic molecules proceeds in a fairly standard manner
\cite{AbrikosovBook,FetterBook}.

It is convenient to trade the expansion in powers of $na^3$ with a
diagrammatic expansion in the number of propagators beginning or
ending in the condensate. Each such line is assigned the value
$\Delta_0$, the expectation value of the field $\Delta$. $\Delta_0^2$
is not the density of particles in the condensate (the value assigned
to condensate lines in the standard dilute Bose gas) but is
proportional to it, with the relation given below. Since the density
of bosons in the condensate is lower than the total density, the
perturbative expansion in powers of the gas parameter may be viewed as
an expansion in the number of condensate lines in the diagrammatic
approach.

The presence of the condensate leads to the appearance of both normal
(one particle enters, one particle exits) and anomalous (two particles
enter and no particles exit or vice versa) propagators. These replace
the ``bare'' propagators $G_0$ and $D_0$ and will be denoted $G_{\rm
  n}$, $G_{\rm a}$ for the fermions and $D_{\rm n}$, $D_{\rm a}$ for
the bosons. The fermionic propagators are needed in order to properly
compute the particle number while the molecular propagators are needed
to determine parameters of the gas. In the systematic perturbative
treatment, the low-density approximations to the normal and anomalous
propagators must be constructed.

To this end it is convenient to construct the self energy corrections
to the propagators. The normal and anomalous fermionic self energies
are denoted $\Sigma_{\rm n}$ and $\Sigma_{\rm a}$ while $\Sigma_{11}$ and
$\Sigma_{20}$ are the normal and anomalous self energies for the
molecules. The perturbative expansion in the number of condensate
lines then consists in letting
$\Sigma=\Sigma^{(1)}+\Sigma^{(2)}+\dots$ where the superscript denotes
the number of condensate lines. However, as discovered by Lee, Huang,
and Yang \cite{Lee1957a,Lee1957b} non-trivial orders may be obtained
through the summation of certain classes of infinite numbers of
diagrams. This does not affect the lowest order results presented in
this chapter and is instead the subject of chapter \ref{chap:wu}.

\begin{figure}[bt]
\begin{center}
\includegraphics[height=1 in]{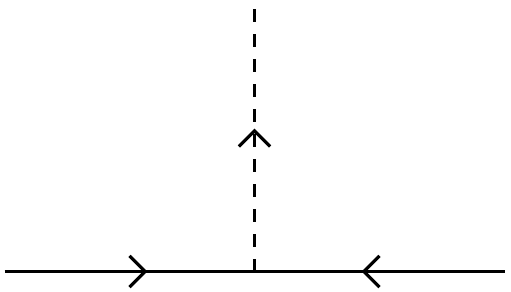}
\caption{Lowest order contribution in the low density approximation to
  the anomalous fermionic self energy, $\Sigma_{\rm a}$. Straight
  lines symbolize the fermionic atoms while the dashed line is a
  condensate line. The external fermion lines {\it do not} contribute
  to the self energy, they are present only to guide the eye.}
\label{fig:sigmaa}
\end{center}
\end{figure}

\begin{figure}[bt]
\begin{center}
\includegraphics[height=1 in]{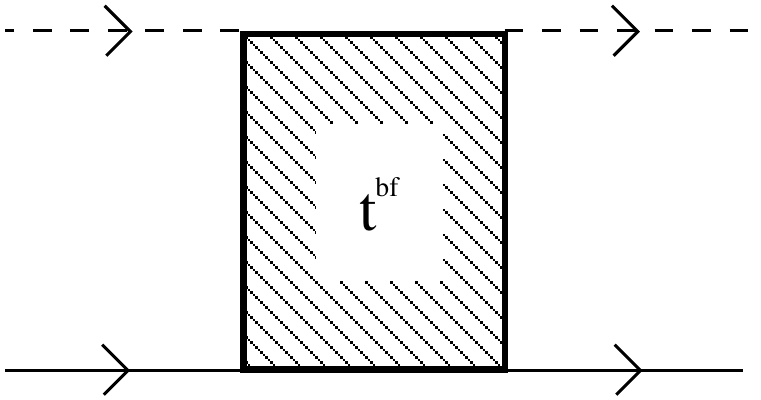}
\caption{Lowest non-vanishing contribution to the normal fermionic
  self energy, $\Sigma_{\rm n}$. The square block is the sum of all diagrams
  contributing to atom-molecule scattering.}
\label{fig:sigmaan}
\end{center}
\end{figure}

\begin{figure}[bt]
\begin{center}
\includegraphics[height=1.2 in]{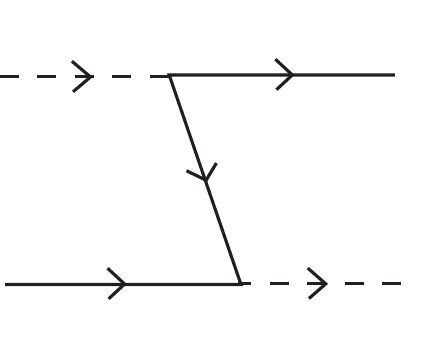}
\caption{The diagram which leads to the lowest order particle number
  (\ref{eq:particlenumber}). As opposed to the self energy diagrams,
  here the external lines must be included.}
\label{fig:particlenumber}
\end{center}
\end{figure}

No normal self energy for the fermions has only one condensate line
and thus
\begin{equation}
\Sigma_{\rm n}^{(1)}=0.
\end{equation}
The anomalous self energy in this order is given by the diagram
shown in Fig. \ref{fig:sigmaa}. This has the value
\begin{equation}
\Sigma_{\rm a}^{(1)}=\Delta_0.
\end{equation}
In the next order, the contributions to the normal self energy are as
shown in Fig. \ref{fig:sigmaan}. The square block shown in this
diagram equals the sum of all diagrams contributing to scattering of
an atom and a molecule.  There are an infinite number of such
diagrams.  The computation of such a sum will be discussed in detail
in section \ref{sec:3body} below. Denote by $t^{\rm bf}$ (the
atom-molecule scattering matrix) the result of the
summation.\footnote{The simplest such diagram consists of $\Sigma_{\rm
    a}^{(1)}$ repeated twice and should not be included (this is the
  diagram depicted in Fig. \ref{fig:particlenumber}, excluding
  external lines). This technical point will not be important for the
  following discussion.} Then
\begin{equation}
\Sigma_{\rm n}^{(2)}=\Delta_0^2t^{\rm bf}.
\end{equation}
No diagrams with two condensate lines contribute to the anomalous self
energy and thus
\begin{equation}
\Sigma_{\rm a}^{(2)}=0.
\end{equation}

\begin{figure}[bt]
\begin{center}
\includegraphics[height=1.9 in]{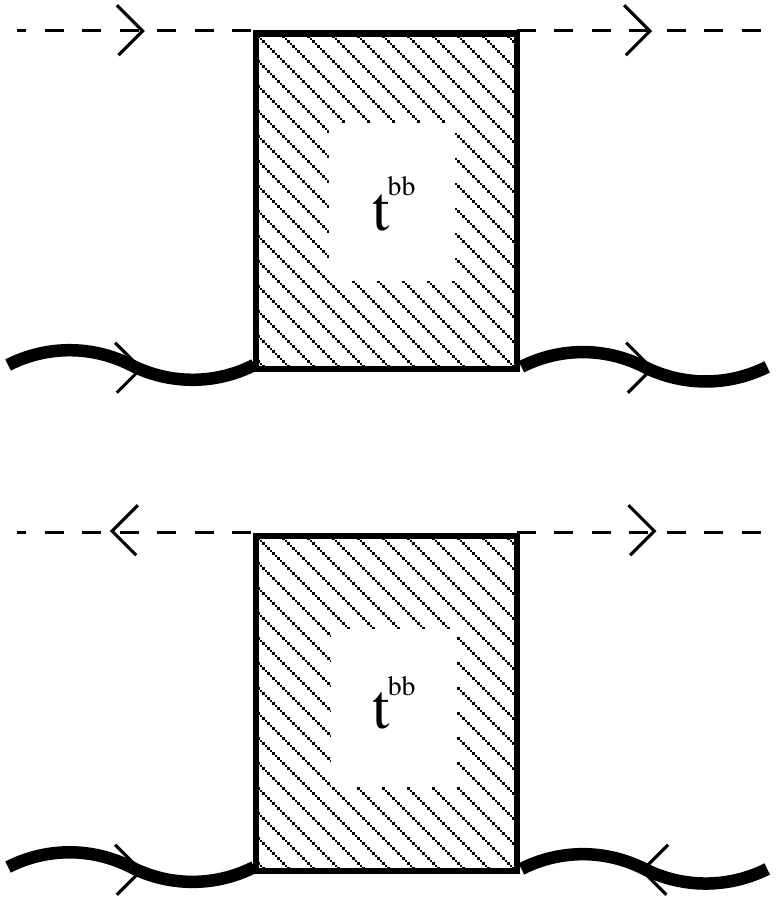}
\caption{Normal (top) and anomalous (bottom) self energy contributions
  to the bosonic propagator in the lowest non-vanishing order. Wavy
  lines are molecular propagators. The square blocks are the sum of
  all diagrams participating in molecule-molecule scattering. As with
  the fermion self energies, the external molecular propagators are
  not a part of the self energies. On the other hand, the condensate
  lines {\it do} contribute to the self energies.}
\label{fig:sigma}
\end{center}
\end{figure}

For the molecular self energies, no diagrams with only one condensate
line contribute and therefore
\begin{equation}
\Sigma_{11}^{(1)}=0,\hspace{1cm}\Sigma_{20}^{(1)}=0.
\end{equation}
In the next order, the self energies are shown in
Fig. \ref{fig:sigma}. The square block is the $t$-matrix of
molecule-molecule scattering calculated in section
\ref{sec:4body}. Again this consists of an infinite number of diagrams
which may be summed. The self energies are
\begin{eqnarray}
\Sigma_{11}^{(2)} & = & 2\Delta_0^2t^{\rm bb}, \\
\Sigma_{20}^{(2)} & = & \Delta_0^2t^{\rm bb}.
\end{eqnarray}
The factor 2 is due to the symmetry of the diagram.

Using the bosonic self energies $\Sigma_{11}$ and $\Sigma_{20}$, the
normal and anomalous propagators of the bosons may be computed. This
was first done by Beliaev \cite{Beliaev1958a,Beliaev1958b} with the
result
\begin{eqnarray}
D_{\rm n}(p,\omega) & = & \frac{D_0^{-1}(-p,-\omega)-
\Sigma_{11}(-p,-\omega)}{D(p,\omega)},
\label{eq:Dn}\\
D_{\rm a}(p,\omega) & = & \frac{\Sigma_{20}(p,\omega)}{D(p,\omega)}.
\label{eq:Da}
\end{eqnarray}
The denominator has the form
\begin{equation}
D(p,\omega) = -\left[\Sigma_{20}(p,\omega) \right]^2+ \left(
D_0^{-1}(p,\omega) -\Sigma_{11}(p,\omega)\right)
\left( D_0^{-1}(-p,-\omega) - \Sigma_{11}(-p,-\omega) \right).
\label{eq:denominator}
\end{equation}

In the BEC regime, it was shown by Hugenholtz and Pines
\cite{Hugenholtz1959} that the propagator of bosons must have a pole
as the four-momentum tends to zero, corresponding to the presence of a
gap-less sound mode in the condensate. This is the crucial observation
on which the results of this section depend, and in general the
Hugenholtz-Pines relation takes the form
\begin{equation}
D(0,0) = 0.
\label{eq:HP}
\end{equation}
In the lowest order in the expansion in powers of $\Delta_0$, the
propagator of bosons reduces to the form $D_0$ obtained in
Eq. (\ref{eq:bosonprop}). The chemical potential in the lowest order
is then
\begin{equation}
\mu = -\frac1{2ma^2}.
\label{eq:mulowest}
\end{equation}
Thus, in the lowest order of the BEC regime the chemical potential of
fermions equals half the binding energy of the molecular state as
expected.

It should be noted that the same result for the chemical potential
would have been obtained if in the gap equation (\ref{eq:BCSBEC}) the
expectation value $\Delta_0$ had been ignored compared with $\mu$.

The bosonic self energies should be related to the scattering length
of the molecules in vacuum, $a_{\rm b}$. The self energies appearing
in the Hugenholtz-Pines relation are on the other hand proportional to
the molecule-molecule scattering $t$-matrix in the presence of the
medium. In particular the $t$-matrix is to be evaluated at a finite
chemical potential $\mu$. However, as long as corrections to the
chemical potential are small, $t^{\rm bb}$ in the medium will
approximately equal $t^{\rm bb}$ evaluated in vacuum and will thus (as
the four-momentum tends to zero) be proportional to the vacuum
scattering length $a_{\rm b}$. The precise relationship is
\begin{equation}
t^{\rm bb} = \frac{2\pi}{mZ}a_{\rm b}.
\end{equation}
Here
\begin{equation}
Z=\frac{8\pi}{m^2a}
\label{eq:Z}
\end{equation}
is the residue of the molecular propagator at its pole and is needed
for proper normalization of external propagators \cite{LL4}.

The chemical potential of the bosonic molecules is defined as
\begin{equation}
\mu_{\rm b} = 2\mu+\frac1{ma^2},
\end{equation}
which vanishes in the lowest order of the low-density
approximation. Using this definition of the bosonic chemical
potential, the Hugenholtz-Pines relation (\ref{eq:HP}) results in
\begin{equation}
\mu_{\rm b} = Z\left[\Sigma_{11}(0)-\Sigma_{20}(0)\right]
-\frac{ma^2Z^2}4\left[\Sigma_{11}(0)-\Sigma_{20}(0)\right]^2.
\label{eq:HP2}
\end{equation}
Eq. (\ref{eq:HP}) will have one additional solution, however this
solution is unphysical \cite{AbrikosovBook,FetterBook}. In the lowest
order in density
\begin{eqnarray}
\mu_{\rm b} & \approx & Z\left[\Sigma_{11}(0)-\Sigma_{20}(0)\right]
 \nn \\
& = & \frac{aa_{\rm b}m\Delta_0^2}4,
\label{eq:mub}
\end{eqnarray}
and the relation becomes reminiscent of the Hugenholtz-Pines equation
in the standard dilute Bose gas \cite{Hugenholtz1959}.

The particle number may now be computed from the fermionic self
energies. The particle number equation is
\begin{eqnarray}
n & = & \frac1V\int d^3x\,\left[\left<\bar\psi_\up\psi_\up\right>+
\left<\bar\psi_\down\psi_\down\right>\right]
\nn \\
& = & -i\int_{\cal C}\frac{d\omega}{2\pi}\int\frac{d^3p}{(2\pi)^3}
\left[G_{n,\up}(p,\omega)+G_{n,\down}(p,\omega)\right],
\label{eq:particlenumberdef}
\end{eqnarray}
where ${\cal C}$ is the contour which includes the real axis and is
closed in the upper half plane ensuring that the field operators occur
in the proper order. The diagram which contributes to the particle
number in the lowest order is shown in
Fig. \ref{fig:particlenumber}. Thus the particle number is
\begin{eqnarray}
n & = & 2i\Delta_0^2\int_{\cal C}\frac{d\omega}{2\pi}\int\frac{d^3p}{(2\pi)^3}
\left[G_0(p,\omega)\right]^2G_0(-p,-\omega) \nn \\
& = & \frac{am^2\Delta_0^2}{4\pi}.
\label{eq:particlenumber}
\end{eqnarray}
The sign of the expression comes from fermionic anti-commutivity and
the factor 2 from the two species of fermions. In the evaluation of
the integral, the chemical potential has been set to its zero'th order
value, Eq. (\ref{eq:mulowest}). With the help of
Eq. (\ref{eq:particlenumber}) the bosonic chemical potential
Eq. (\ref{eq:mub}) becomes
\begin{equation}
  \mu_{\rm b} = \frac{\pi a_{\rm b} n}m =
  \frac{4\pi a_{\rm b} n_{\rm b}}{m_{\rm b}}.
\label{eq:mub2}
\end{equation}

The ground state energy $E_0$ (measured from the binding energy of the
molecules, i.e. from $-n_b/ma^2$) is calculated using the standard
relation \cite{Hugenholtz1959,FetterBook}
\begin{equation}
E_0/V-\frac12n_{\rm b} \mu_{\rm b} = \frac{i}{2Z}\int_{\cal C}\frac{d\omega}{2\pi}
\int\frac{d^3p}{(2\pi)^3}\left(\omega+p^2/4m\right)D_{\rm n}(p,\omega),
\label{eq:E0}
\end{equation}
where the contour ${\cal C}$ is again taken in the upper half
plane. In the lowest order, the normal molecular propagator reduces to
$D_0$ which is a retarded propagator. Thus with the help of
Eq. (\ref{eq:mub2}) the ground state energy becomes
\begin{equation}
E_0/V=\frac12n_{\rm b}\mu_{\rm b} = \frac{2\pi a_{\rm b} n_{\rm b}^2}{m_{\rm b}}.
\end{equation}
The effect of including the higher order molecular propagator
(\ref{eq:Dn}) in the equation for the ground state energy will be
discussed in chapter \ref{chap:wu}.

Even at zero temperature there is still a finite fraction of the
bosonic molecules not in the condensate. This phenomenon, caused by
the presence of interactions, is known as the condensate
depletion. The total density of bosons is
\begin{equation}
n_{\rm b} = n_{0{\rm ,b}} +\frac iZ \int_{\cal C}\frac{d\omega}{2\pi}
\int \frac{d^3p}{(2\pi)^3}D_{\rm n}(p,\omega),
\label{eq:conddep}
\end{equation}
with $n_{0{\rm ,b}}$ the density of bosons in the condensate. In the
lowest order, the density $n_{\rm b}=n_{0{\rm ,b}}$. Using
Eq. (\ref{eq:Dn}) for $D_{\rm n}(p,\omega)$ it is seen that the
frequency and momentum dependence of the molecular self energies
$\Sigma_{11}(p,\omega)$ and $\Sigma_{20}(p,\omega)$ is
needed. However, considering the integral in Eq. (\ref{eq:conddep}),
it is observed that the integration is dominated by frequencies of
order $\omega\sim \mu_{\rm b}\sim\Delta_0^2$ and momenta of order
$p^2/4m\sim\Delta_0^2$. For such low frequencies and momenta, the
underlying structure of the molecular propagator (\ref{eq:bosonprop})
is not important and it reduces to
\begin{equation}
D_0(p,\omega) \approx \frac Z{\omega-p^2/4m+\mu_{\rm b}+i0}.
\label{eq:D0approx}
\end{equation}
Also, $Z\Sigma_{11}(0)=2\mu_{\rm b}$ and $Z\Sigma_{20}(0)=\mu_{\rm b}$
coincide with $\Sigma_{11}(0)$ and $\Sigma_{20}(0)$ in the usual
dilute Bose gas. Hence, for small frequencies and momenta, the normal
molecular propagator divided by $Z$ coincides with the dilute Bose gas
normal propagator at this order in the expansion. The result of
evaluating the integral (\ref{eq:conddep}) then takes the usual form
\cite{AbrikosovBook,FetterBook}
\begin{equation}
n_{0,{\rm b}}=n_{\rm b}\left[1-\frac83\sqrt{\frac{n_{\rm b}a_{\rm b}^3}\pi}\right].
\end{equation}

Finally, the speed of sound wave propagation is evaluated by use of
the Hugenholtz-Pines relation, Eq. (\ref{eq:HP}), at finite
$p,\omega$. For small $\omega\lesssim \mu_{\rm b}$ and $p^2/4m\lesssim
\mu_{\rm b}$ the self energy terms approximately take the limiting
values $\Sigma_{11}^{(2)}(0)$ and $\Sigma_{20}^{(2)}(0)$ in the low
density expansion. Imposing the condition
\begin{equation}
D(p,\omega) = 0
\end{equation}
for the denominator, Eq. (\ref{eq:denominator}), of the molecular
propagators, it is found that
\begin{equation}
  \omega^2 = u^2 p^2, \hspace{10mm} u=\sqrt{\frac{4\pi n_{\rm b}
      a_{\rm b}}{m_{\rm b}^2}},
\label{eq:sof}
\end{equation}
where $u$ is the speed of sound.

The results of this section coincide with the results for the dilute
interacting Bose gas \cite{AbrikosovBook,FetterBook} with scattering
length $a_{\rm b}$.  In order to go beyond these results it is
necessary to modify the self energies in a systematic way. This will
be further investigated in chapter \ref{chap:wu}.

\section{Scattering of a molecule and an atom \label{sec:3body}}
In this section, the scattering problem of a fermionic atom with a
bound state of fermions is treated. The solution of this scattering
process has a long history. It was first solved in the approximation
of vanishing effective range by Skorniakov and Ter-Martirosian in 1956
\cite{Skorniakov1956} who studied the related problem of a neutron
scattering with a deuteron (a neutron-proton bound state). In the
context of cold two-component Fermi gases the scattering problem was
solved in Ref. \cite{Petrov2003} for any mass ratio of the fermionic
atoms below 13.6. The three-body calculation presented below is a
useful exercise before proceeding to more complicated few-body
calculations. It is important to note that the calculation below
relies on the approximation of a vanishing effective range,
i.e. $a\gg|r_0|$ and the results are only valid in this regime.

It should be mentioned that there is a fundamental difference between
the fermionic and the bosonic problem. The system of three identical
bosons with a large scattering length has an infinite number of bound
states, the effect known as the Efimov effect \cite{Efimov1971}. In
this case, to correctly describe the three-body physics, an additional
three-body parameter coming from short distance physics is required
\cite{Danilov1961,Bedaque1998,Braaten2006}. This is not the case in
the fermionic problem (for mass ratios below 13.6); here the
three-body problem is completely describable in terms of the two-body
scattering length and the fermionic masses \cite{Petrov2003}.

In the one channel model, the only length scale left is the scattering
length between the distinguishable atoms. Since no additional
parameters are required in order to describe the three-body problem,
the atom-molecule scattering length, $a_{\rm bf}$, must be
proportional to the atom-atom scattering length, $a$.

Apart from the three-body scattering length, it is also interesting to
note that the achievable densities for two-component Fermi gases are
limited by three-body recombination, a process governed by three-body
physics. Here two atoms form a deeply bound state with binding energy
of order $1/mR_e^2$ and a third atom escapes the trap as it carries
away the released binding energy as kinetic energy. The process
requires at least two identical fermionic atoms to approach each other
closely (at the order of the range of short distance physics, $R_e$)
and is suppressed by the Pauli principle. The loss rate depends on the
precise three-body physics and may be extracted from the three-body
calculations below. A precise discussion will be postponed for chapter
\ref{chap:mass} on mass-imbalanced two-component Fermi gases since the
loss-rate depends intimately on the mass ratio.

\begin{figure}[bt]
\begin{center}
\includegraphics[height=.6 in]{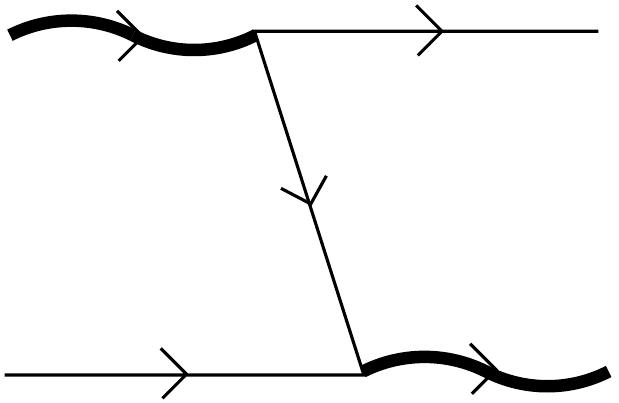}
\caption{The Born approximation to atom-molecule scattering.}
\label{fig:3bodyborn}
\end{center}
\end{figure}

\begin{figure}[bt]
\begin{center}
\includegraphics[height=.6 in]{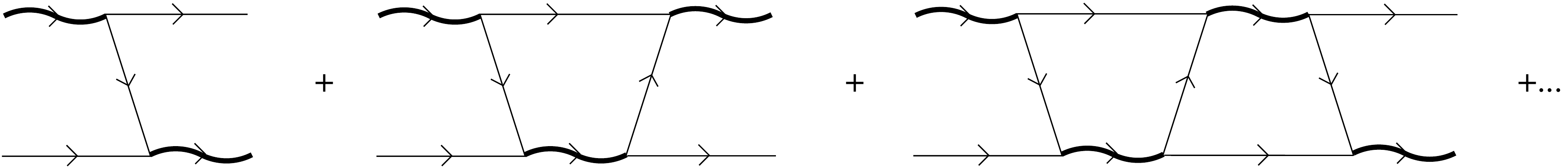}
\caption{The ladder type diagrams contributing to atom-molecule scattering.}
\label{fig:3bodydiagrams}
\end{center}
\end{figure}

This is the first of a number of few body problems treated using
integral equations and therefore this will be done very carefully. The
simplest process contributing to the atom-molecule scattering is that
the molecule breaks into its two constituent atoms, one of which forms
a bound state with the remaining atom. The process is illustrated in
Figure \ref{fig:3bodyborn}; had the scattering been weak, this would
have been the only contribution to the scattering and thus this
diagram will be called the Born contribution. However, as will be
discussed below, all of the diagrams contributing to the scattering
are of the ladder diagram type shown in Fig. \ref{fig:3bodydiagrams}
and these are all of the same order.

As pointed out above, the vacuum scattering diagrams are exactly those
needed for the normal fermionic self energy at next to leading order
in the low density expansion and the normal and anomalous molecular
self energy at lowest order.  That the type of diagrams shown in
Fig. \ref{fig:3bodydiagrams} is the only type contributing to the
scattering stems precisely from the fact that this is {\it vacuum}
scattering, i.e. all propagators are retarded and there is no hole
propagation. This in turn means diagrammatically that all propagators
specify a time direction and the possible diagrams are greatly
simplified. In fact, because of the simplicity of the diagrams it is
possible to sum all the possible diagrams by using an integral
equation of Lippmann-Schwinger type.

Since the scattering proceeds in the vacuum, the chemical potential
$\mu$ will be taken to vanish in this section.

The scattering $t$-matrix consists of all diagrams with external
incoming and outgoing lines being the scattering particles. The
external lines are not included in the $t$-matrix, however they do
affect the overall normalization. The first diagram in
Fig. \ref{fig:3bodydiagrams} thus contributes a single atomic
propagator, proportional to $ma^2$. In general, a diagram with $n$
loops will then contain $2n+1$ fermionic lines of order
$(ma^2)^{2n+1}$, $n$ bosonic lines of order $(a/m)^n$, and $n$
integrations over four-momenta $\sim m^{-n}a^{-5n}$. It follows that
each diagram will be of order $ma^2$.

\begin{figure}[bt]
\begin{center}
\includegraphics[height=.95 in]{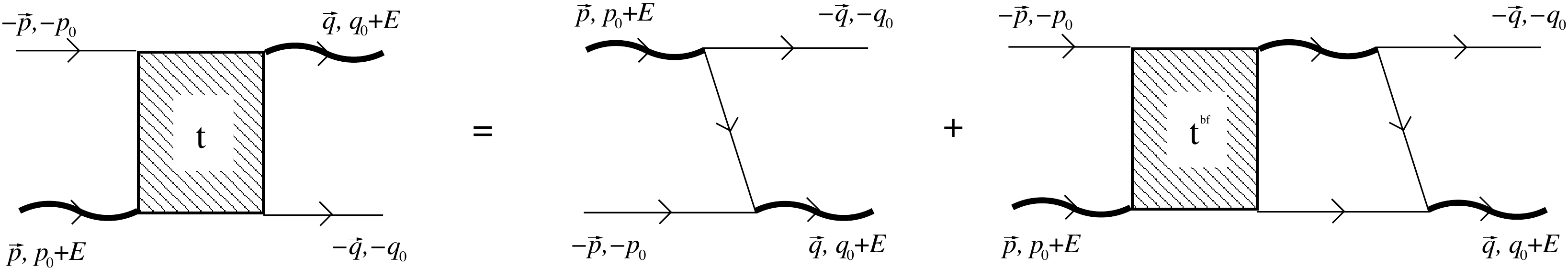}
  \caption{The integral equation for the $t$-matrix (shown as the
    square block) of atom-molecule scattering.}
\label{fig:inteq3body}
\end{center}
\end{figure}

Given the structure of the contributing diagrams it is possible to
write an integral equation for the sum of these. The integral equation
is shown in Fig. \ref{fig:inteq3body}.  In Eq. (\ref{eq:aff}) the
two-fermion bubble diagrams were summed to yield the atom-atom
scattering length by using a geometric series. The reason why the
three-body problem is an {\it integral equation} rather than a
geomatric series is that there will be an integral over the free
four-momentum in the loop of the iterated term.

Let the four-momenta be as shown in Fig. \ref{fig:inteq3body}. That
is, the incoming molecule and atom have four-momenta $(\vec p,p_0+E)$
and $(-\vec p,-p_0)$, respectively, while the outgoing molecule and
atom have four-momenta $(\vec q,q_0+E)$ and $(-\vec q,-q_0)$,
respectively. $E$ is the total energy going into the scattering and
since the initial and final states consist of a bound pair and an atom,
it is necessary to restrict the total energy to energies $E\le E_{\rm
  b}$. Otherwise, the final scattering state could be three free
particles. The scattering $t$-matrix with these kinematics is denoted
$t(\vec p,p_0;\vec q,q_0)$ (the dependence on total energy is
suppressed in this notation). Then the integral equation becomes
\begin{equation}
t(\vec p,p_0;\vec q,q_0) = -G(p+q+E)-i\int \frac{d^4Q}{(2\pi)^4}
G(Q+q+E)G(-Q)D(Q+E)t(\vec p,p_0;\vec Q,Q_0).
\label{eq:3bodygeneral0}
\end{equation}
Here, in a slightly abusive notation, the total energy $E$ is also
used as the total four-momentum $(\vec 0,E)$. To simplify notation
further, the four-momentum $(\vec p,p_0)$ is abbreviated as $p$. The
propagator of molecules is here given by the expression
(\ref{eq:bosonprop}) with the chemical potential taken to vanish.

The $t$-matrix depends on five variables; the two frequencies, the
amplitudes of the two momenta, and the angle between incoming and
outgoing momenta. However, as the dependence of $t$ on $p$ and $p_0$
is the same on both sides of the integral equation, the integral
equation for $t$ is only a three dimensional problem. Below, it will
be discussed how the frequencies $q_0$ and $Q_0$ may be integrated out
from Eq. (\ref{eq:3bodygeneral0}) and in chapter \ref{chap:mass} it
will be demonstrated how the integral equation may be projected onto
each spherical wave component, making the solution of each of these a
one-dimensional problem.

\begin{figure}[bt]
\begin{center}
\includegraphics[height=1.7 in]{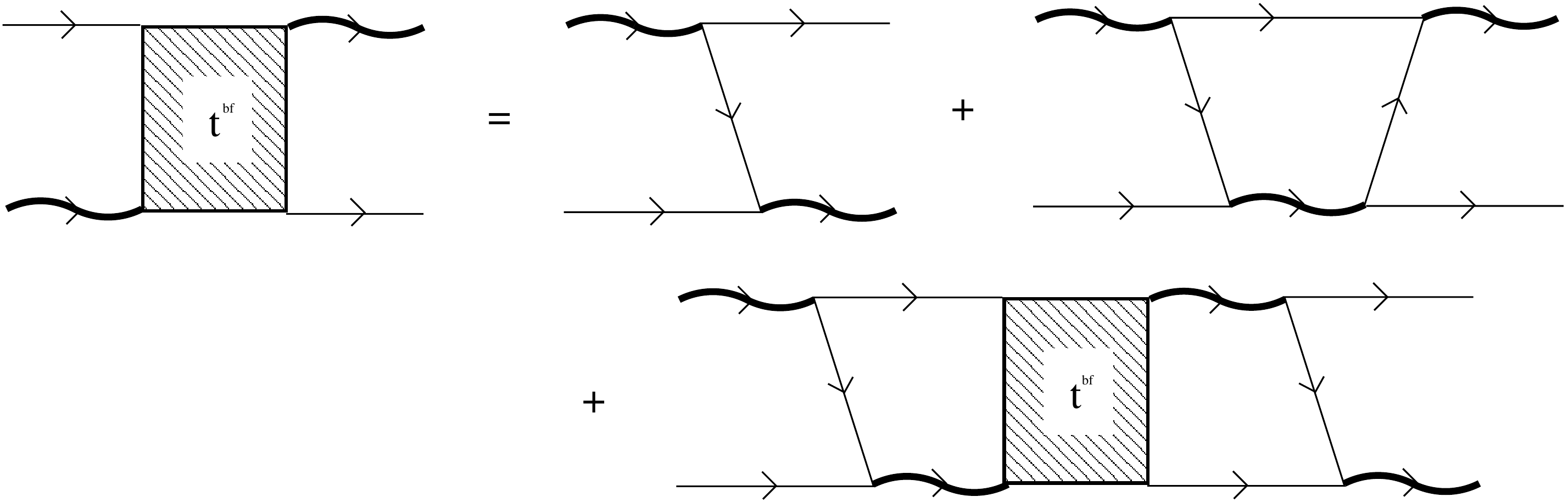}
\caption{The procedure for relating the on-shell $t$-matrix (the
  square block on the right hand side) to the $t$-matrix evaluated at
  a desired frequency (the left hand side).}
\label{fig:3bodyiteration}
\end{center}
\end{figure}

$t(\vec p,p_0;\vec q,q_0)$ is analytic in the upper half planes of
both $p_0$ and $q_0$. This may be seen from the ladder diagrams
contributing to the $t$-matrix or simply from
Eq. (\ref{eq:3bodygeneral0}). This in turn means that the integration
over $Q_0$ in Eq. (\ref{eq:3bodygeneral0}) may be carried out by
closing the contour in the upper half plane, with the only
contribution coming from the simple pole of $G(-Q)$, setting
$Q_0=-Q^2/2m$. The solution of the integral equation is then most
easily obtained by letting $q_0\to-q^2/2m$ which means that the
momentum dependence of the $t$-matrix is the same on both sides of the
integral equation. Should the value of the $t$-matrix be desired at a
different value of $q_0$, the result of solving the integral equation
at $q_0=-q^2/2m$ may simply be inserted on the right hand side of
Eq. (\ref{eq:3bodygeneral0}) and the integration carried out.  In this
sense the $t$-matrix with $p_0=-p^2/2m$ and $q_0=-q^2/2m$ is quite
natural. It is the result of attaching the external atomic propagators
to the $t$-matrix and integrating over their frequencies, closing the
contour in the upper half plane (where $t$ is analytic). This will be
called the ``on-shell'' $t$-matrix and satisfies the integral equation
\begin{eqnarray}
t(\vec p,-p^2/2m;\vec q,-q^2/2m)
& = & -G(p+q+E) \nn \\ && \hspace{-30mm}-\int \frac{d^3Q}{(2\pi)^3}
G(Q+q+E)D(Q+E)t(\vec p,-p^2/2m;\vec Q,-Q^2/2m).
\label{eq:onshell3body}
\end{eqnarray}
The iterative procedure described above for obtaining the solution to
the integral equation at the desired values of $p_0,q_0$ from the
on-shell $t$-matrix is illustrated in Fig. \ref{fig:3bodyiteration}.

To calculate the $s$-wave scattering length, the total energy should
be set to the two-body binding energy. The scattering length is then
found by solving the integral equation, averaging over angles, and
taking the limit of vanishing incoming and outgoing
four-momenta. Equivalently, the incoming momentum can be taken to
vanish prior to solving the integral equation after which there can be
no dependence on the angle between incoming and outgoing
momentum. Using this second approach yields the integral equation
\begin{equation}
t(p) = \frac{ma^2}
{1+p^2}+\frac1{\pi p}\int_0^\infty dq \frac{q\,t(q)}
{1-\sqrt{1+3q^2/4}}\log\frac{1+p^2+q^2+qp}{1+p^2+q^2-qp}.
\label{eq:inteq0}
\end{equation}
In this equation, momenta are measured in units of inverse scattering
length and the ``on-shell'' $s$-wave scattering matrix is $t(p)\equiv
t(\vec 0,0;\vec p,-p^2/2m)$.

\begin{figure}[bt]
\begin{center}
\includegraphics[height=2.5 in]{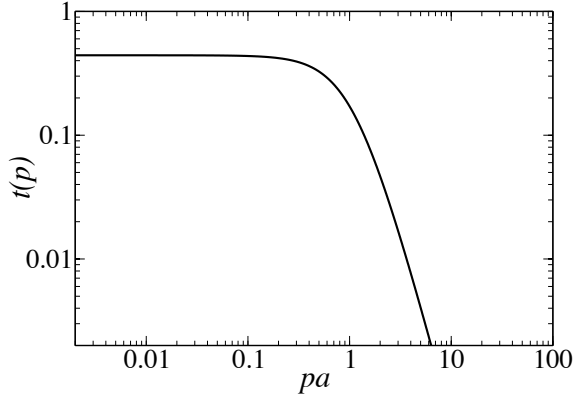}
\caption{The $t$-matrix $t(p)$ in units of $ma^2$ calculated by
  solving Eq. (\ref{eq:inteq0}). For $p\ll a^{-1}$ the $t$-matrix is
  approximately constant while for $p\gg a^{-1}$ the $t$-matrix
  satisfies a power-law behavior, decaying as $p^{-3.17}$
  \cite{Bedaque1998}.}
\label{fig:3bodyt}
\end{center}
\end{figure}

In general, the integral equations have a form which in matrix
notation is
\begin{equation}
t_i = v_i+K_{ij}t_j.
\label{eq:num01}
\end{equation}
The matrix equation has the formal solution
\begin{equation}
t_i=(I-K)^{-1}_{ij}v_j.
\label{eq:num02}
\end{equation}
The matrix $I$ is the identity matrix. The numerical solution of the
$t$-matrix is then a matter of choosing a discretization and solving
the matrix equation. Typically, for purposes of convergence, it is
advantageous to perform a change of variable such that the limits of
integration become finite \cite{numrecipes}. A convenient choice is
\begin{equation}
p = \frac2{z+1}-1, \hspace{1cm} z\in]-1,1].
\end{equation}
All integrals are then performed employing Gauss-Legendre quadrature
\cite{numrecipes}.  Using this method the $t$-matrix is found to take
the shape shown in Fig. \ref{fig:3bodyt}.

The atom-molecule scattering length $a_{\rm bf}$ is proportional to the
$t$-matrix of Eq. (\ref{eq:inteq0}) evaluated at vanishing momentum
and energy $E=E_{\rm b}$. To obtain the scattering amplitude from the
$t$-matrix it is necessary to renormalize the external
propagators. The fermionic propagators are already free propagators,
however the molecular propagators need to be renormalized by the
square root of the residue of the molecular propagator at the energy
of the bound state \cite{LL4}, i.e.
\begin{equation}
T(0) = Zt(0),
\end{equation}
with $Z$ given by Eq. (\ref{eq:Z}). The scattering amplitude is
related to the scattering length by
\begin{equation}
T(0) = \frac{2\pi}{m_{3{\rm r}}}a_{\rm bf}=\frac{3\pi}ma_{\rm bf},
\end{equation}
where $m_{3{\rm r}}$ is the three-body reduced mass. Using the method
described above it is found that
\begin{equation}
a_{\rm bf}\approx1.18a,
\end{equation}
in perfect agreement with the literature
\cite{Skorniakov1956,Petrov2003}.

Bound states in the three-body problem show up as poles of the
scattering amplitude. This corresponds to solutions of the homogenous
integral equation
\begin{equation}
t_i = K_{ij}t_j,
\end{equation}
or in other words to the integration kernel $K$ having a unit
eigenvalue. Bound three-body states will have $E<E_{\rm b}$, and writing
$E=(1+\epsilon_3)E_{\rm b}$ the homogenous equation becomes
\begin{equation}
  t(p) = \frac1{\pi p}\int_0^\infty dq \frac{q\,t(q)}
  {1-\sqrt{1+\epsilon_3+3q^2/4}}\log\frac{1+\epsilon_3+p^2+q^2+qp}
  {1+\epsilon_3+p^2+q^2-qp}.
\end{equation}
It is found that the three-body problem does not have any $s$-wave
bound states. Note that the validity of this statement is limited by
the zero effective range approximation. There may still be deeply
bound states with binding energies of order $1/mr_0^2$.

\section{Molecule-molecule scattering \label{sec:4body}}
\begin{figure}[bt]
\begin{center}
\includegraphics[height=1 in]{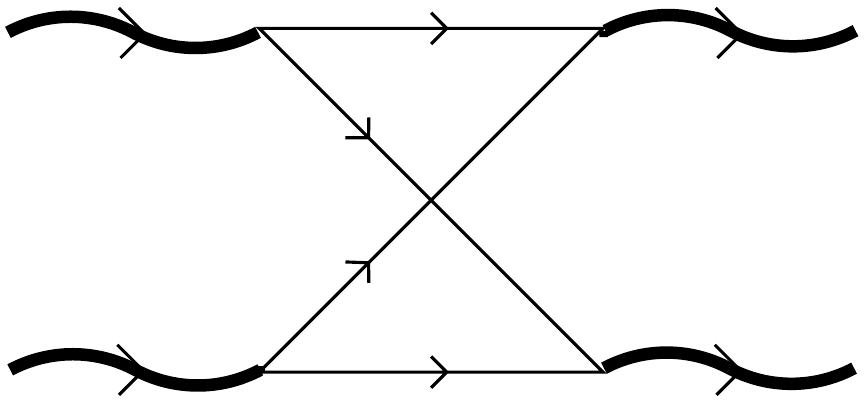}
  \caption{The simplest contribution to molecule-molecule scattering.}
\label{fig:4bodyborn}
\end{center}
\end{figure}

As discussed above, properties of the gas of weakly bound molecules
depend intimately on the molecule-molecule scattering length, $a_{\rm
  b}$.  Mean field theory gives the result $a_{\rm b}=2a$. This result
implicitly assumes that the coupling is weak and that only the diagram
shown in Fig. \ref{fig:4bodyborn} contributes to the scattering. An
approximate diagrammatic method was developed by Pieri and Strinati in
Ref. \cite{Pieri2000} who found $a_{\rm b}\approx 0.75 a$. Later the
problem was solved exactly by Petrov, Salomon, and Shlyapnikov
\cite{Petrov2005a} by solving the quantum mechanical four-body problem
using the Bethe-Peierl's method. In this work it was found that
$a_{\rm b}\approx0.6a$. This last result is confirmed below by the use
of diagrammatic methods. While the work presented in this section and
published in Ref. \cite{Levinsen2006} was in progress, the problem was
solved by Brodsky {\it et al} in Ref. \cite{Brodsky2005} by
essentially the same diagrammatic method.

The scattering $t$-matrix consists of all possible diagrams with two
incoming and two outgoing molecules, excluding the external
propagators. In the atom-molecule scattering problem treated above, it
was discussed how diagrams contributing to the molecule-molecule
scattering do not contain any internal condensate lines nor any hole
propagation. This is again the case, since the process considered is
scattering in the vacuum. This observation greatly simplifies the
diagrams possible in the scattering process. Since all propagators can
be assigned a definite time direction, at any time in the scattering
of two molecules there will be either two molecules, one molecule and
two atoms, or four atoms.

\begin{figure}[bt]
\begin{center}
\includegraphics[height=2.4 in]{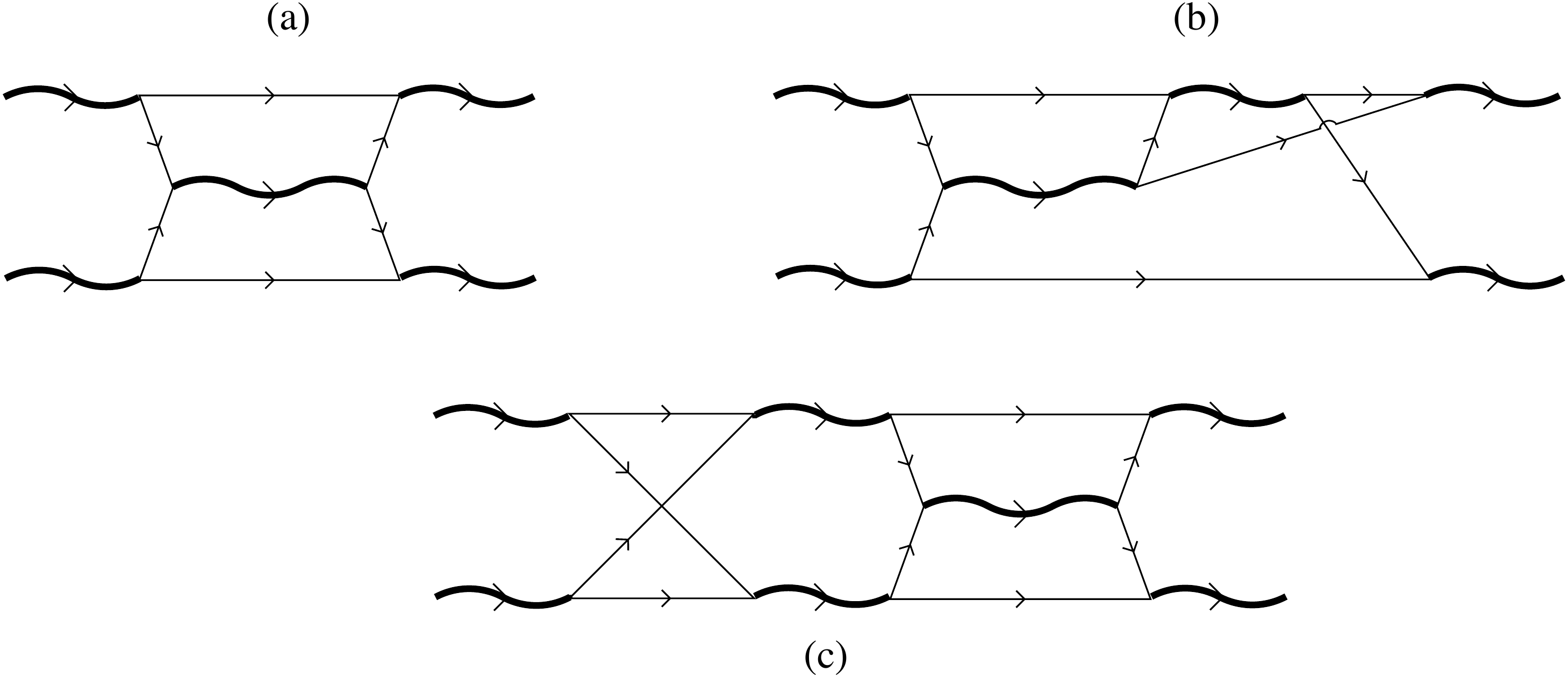}
  \caption{Diagrams which contribute to molecule-molecule scattering.}
\label{fig:4bodydiagrams}
\end{center}
\end{figure}

With this observation in mind, consider the different diagrams which
contribute to the scattering $t$-matrix. The simplest possible diagram
is the Born diagram shown in Fig. \ref{fig:4bodyborn} where the two
molecules exchange their constituent fermionic atoms. Other possible
diagrams contributing to the scattering are shown in
Fig. \ref{fig:4bodydiagrams}. Fig. \ref{fig:4bodydiagrams}a shows the
propagation of an intermediate molecule, Fig. \ref{fig:4bodydiagrams}b
a slightly more complicated diagram, and \ref{fig:4bodydiagrams}c a
diagram where a pair of molecules propagate at an intermediate
step. The crucial observation is that all of these diagrams are of the
same order, namely they are all proportinal to $m^3a^3$. Indeed this
is true for all possible vacuum scattering diagrams. Thus the problem
of molecule-molecule scattering is not amenable to a perturbative
treatment. Instead of considering only the Born diagram,
Fig. \ref{fig:4bodyborn}, an infinite number of diagrams need to be
taken into account.

\begin{figure}[bt]
\begin{center}
\includegraphics[height=.9 in]{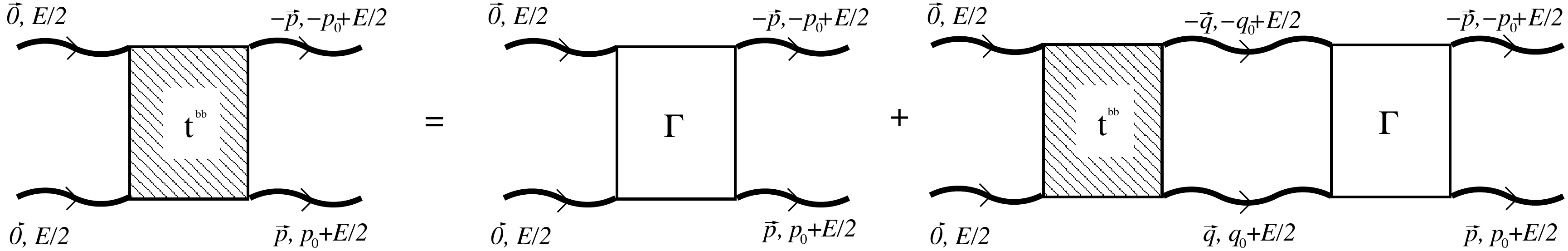}
\caption{Integral equation for the $t$-matrix (shown as the square
  block) of molecule-molecule scattering. Here, for simplicity $E$ has
  been used as the four-momentum $(\vec 0,E)$.}
\label{fig:4bodyinteq}
\end{center}
\end{figure}

Fortunately it is possible to perform the summation of all diagrams by
using an integral equation. To this end, it is useful to define the
concept of a {\it two-boson irreducible diagram} as a diagram which
cannot be cut in two by cutting two molecular lines only. For
instance, the diagrams in Fig. \ref{fig:4bodydiagrams}a and b are
two-boson irreducible while the diagram in
Fig. \ref{fig:4bodydiagrams}c is not. The scattering $t$-matrix may
then be viewed as the repeated propagation and scattering of two
molecules, each pair of molecular propagators separated by the sum of
all two boson irreducible diagrams. In this sense, the sum of
two-boson irreducible diagrams is similar to the bubble diagram, $\Pi$
(see Fig. \ref{fig:bubble}) occurring in the repeated atom-atom
scattering. However, there will be an integral over a free
four-momentum and thus the sum does not form a geometric series.
Defining $\Gamma$ as the sum of all two-boson irreducible diagrams,
$t$ satisfies an integral equation as shown in
Fig. \ref{fig:4bodyinteq}.

Each diagram contributing to $\Gamma$ will have a number, $n$, of
molecular propagators, $4+2n$ fermionic propagators, and $n+1$
integrations. The total order of such a diagram is $\left(\frac
  am\right)^n\left(ma^2\right)^{4+2n}\left(\frac1{ma^5}\right)^{n+1}=m^3a^3$. Any
diagram contributing to $t$ will then have $n+1$ insertions of
$\Gamma$, $2n$ molecular propagators and $n$ integrations and will be
proportional to $m^3a^3$.

The integral equation for the $t$-matrix with kinematics as shown in
Fig. \ref{fig:4bodyinteq} is
\begin{equation}
t(p,p_0)=\Gamma(0,0;p,p_0)+\frac{i}{4\pi^3}\hspace{-1mm}
\int \hspace{-1mm}q^2dq\,dq_0D(q,q_0+E/2)
D(q,-q_0+E/2)\Gamma(q,q_0;p,p_0)t(q,q_0).
\label{eq:4bodyinteq}
\end{equation}
The incoming momentum has been taken to zero while the outgoing
four-momenta of the molecules are $(\pm \vec p,E/2\pm p_0)$. This
choice of kinematics is convenient since ultimately the quantity of
interest is the $s$-wave scattering length, proportional to
$t(0,0)$. The scalar $t(\vec p,p_0)$ cannot depend on the direction of
$\vec p$, which is why it is possible to average the sum of two-boson
irreducible diagrams over the angle between incoming and outgoing
momentum. This average is denoted $\Gamma(p,p_0;q,q_0)$. To determine
the scattering length, let $E=2E_B$, while to search for bound states
look for solutions of the homogenous integral equation with
$E<-|2E_{\rm b}|$.

Because of the symmetry of the problem, it is convenient to slightly
redefine what is meant by $G$ and $D$. Below, {\it in this section
  only}, the convention will be used that all atomic propagators have
their frequency shifted by a quarter of the energy going into the
scattering, while the molecules have their frequency shifted by half
the energy. That is
\begin{eqnarray}
G(p,p_0) & \equiv & \frac1{p_0-p^2/2m+E/4+i0}, \\
D(p,p_0) & \equiv & \frac{4\pi}{m}\frac1{a^{-1}-\sqrt{m}\sqrt{-p_0-E/2
+p^2/4m-i0}}.
\end{eqnarray}
For the purpose of calculating the molecular scattering length $a_{\rm
  b}$, the total energy is $2E_{\rm b}$ and in this case each molecule
will carry the molecular binding energy.

\begin{figure}[bt]
\begin{center}
\includegraphics[height=1.9 in]{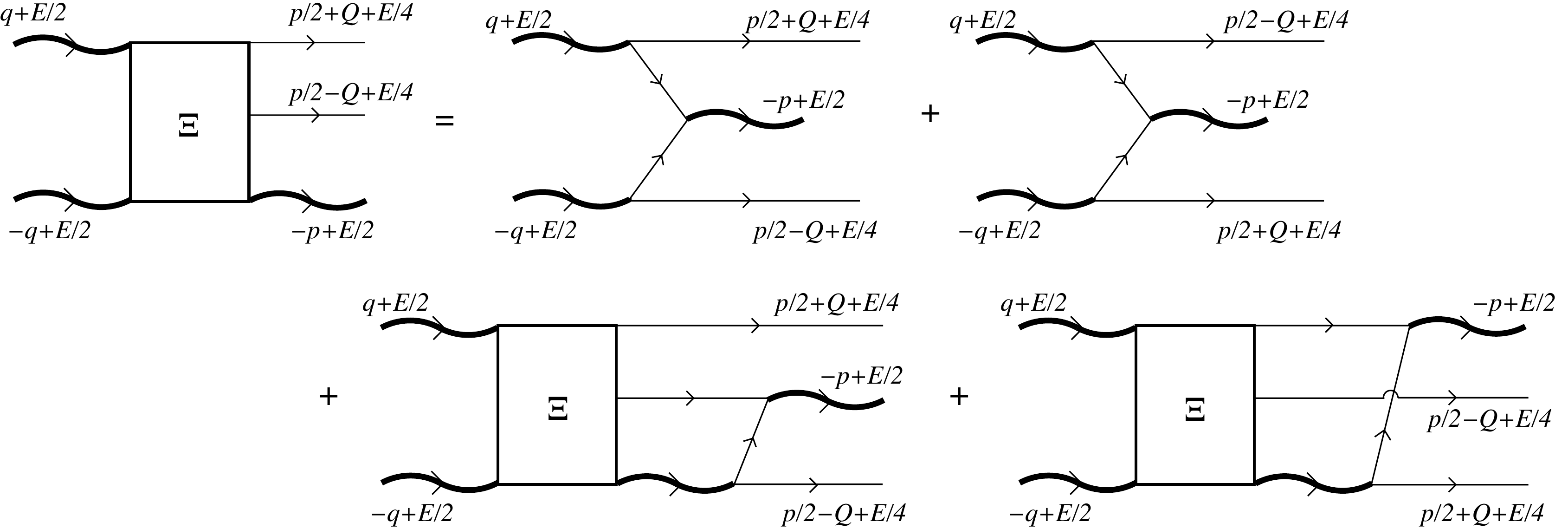}
\caption{Integral equation for the vertex $\Xi$. Here the notation $E$
  has been used to denote the four-momentum $(\vec 0,E)$.}
\label{fig:inteqxi}
\end{center}
\end{figure}

The integral equation (\ref{eq:4bodyinteq}) is not sufficient to
compute the $t$-matrix. The sum of all two-boson irreducible diagrams
still needs to be computed and this sum may again be obtained through
the solution of an integral equation. It is not possible to directly
write down an integral equation satisfied by $\Gamma$. However, it
is possible to relate $\Gamma$ to the solution of an integral equation
for the sum, $\Xi$, of two-boson irreducible diagrams with two incoming
molecules, one outgoing molecule and two outgoing fermions. The
integral equation to be solved is depicted in
Fig. \ref{fig:inteqxi}. This integral equation is quite similar to the
equation for atom-molecule scattering, see Fig. \ref{fig:inteq3body}.

Let the kinematics be as shown in Fig. \ref{fig:inteqxi}. $\Xi$ with
incoming four-momenta of $(\pm \vec q,E/2\pm q_0)$ for the molecules,
$(-\vec p,E/2-p_0)$ for the outgoing molecule, and $(\vec p/2\pm\vec
Q,E/4+p_0/2\pm Q_0)$ for the outgoing fermions is denoted
$\Xi(q;p/2+Q;p/2-Q)$. The integral equation for $\Xi$ may then be
written as\footnote{The calculation of the vertex $\Xi$ presented here
  differs slightly from the formalism in Ref. \cite{Levinsen2006}. The
  present formalism is better suited to an extension to treating the
  mass imbalanced problem investigated in chapter \ref{chap:mass}.}
\begin{eqnarray}
  \Xi(q;p/2+Q;p/2-Q) & = & -\int\frac{d\Omega_{\vec q}}{4\pi}\left[
    G(-p/2-q+Q)G(-p/2+q-Q)+\left(q\leftrightarrow -q\right)
\right]
  \nn \\ &&
\hspace{-48mm}
-i\int\frac{d^4Q'}{(2\pi)^4}\left\{G(p/2-Q')
    G(-3p/2+Q')D(-p-Q+Q')\Xi(q;p/2+Q;p/2-Q')
\right. \nn \\ &&
\hspace{-31mm} \left.
+G(p/2+Q')
    G(-3p/2-Q')D(-p+Q-Q')\Xi(q;p/2+Q';p/2-Q)
  \right\}\hspace{-.9mm}.
\label{eq:inteqxi0}
\end{eqnarray}
The integration over directions of $\vec q$ (with notation $\int
d\Omega_{\vec q}$) is the averaging over directions of incoming
momentum, i.e. the projection onto the $s$-wave. Minus signs in front
of both the Born terms and the iterated terms all follow from
anti-commutivity of the fermionic atoms.

Eq. (\ref{eq:inteqxi0}) has to be solved at every $|\vec q|$ and
$q_0$. It is an integral equation in 5 variables, the length of $\vec
p$ and $\vec Q$, the angle between these vectors, and the
corresponding frequencies. As will be shown below, it is possible to
integrate out the two frequencies such that the resulting integral
equation becomes three-dimensional.

\begin{figure}[bt]
\begin{center}
\includegraphics[height=1.1 in]{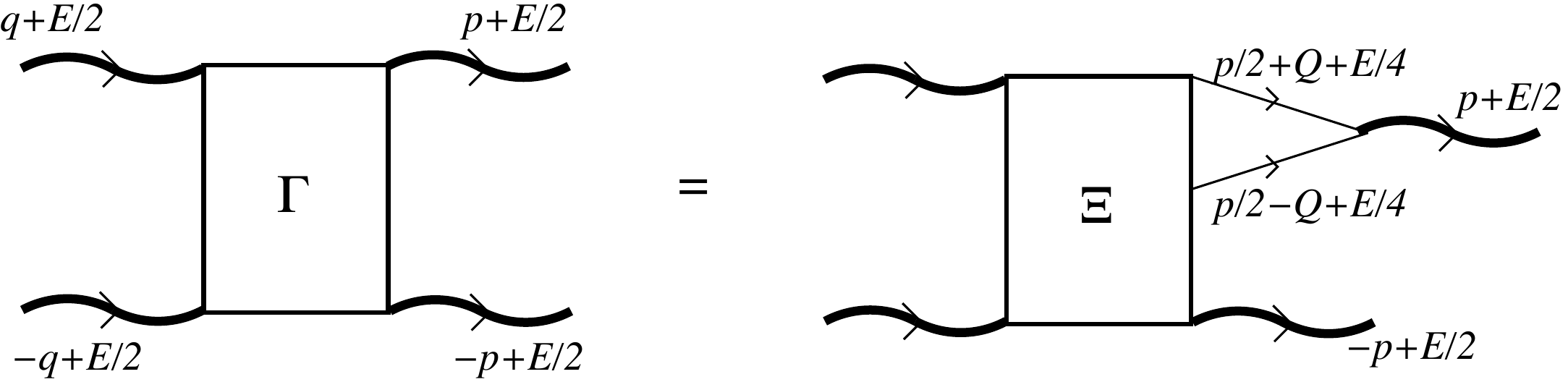}
  \caption{The relation between $\Gamma$ and $\Xi$.}
\label{fig:gammaxi}
\end{center}
\end{figure}

Finally, the two-boson irreducible vertex $\Gamma$ is related to $\Xi$
by joining the two external fermionic propagators into a molecule as
shown in Fig. \ref{fig:gammaxi}. This relation may be written as
\begin{eqnarray}
\Gamma(q,q_0;p,p_0) = \frac i2\int\frac{d^4Q}{(2\pi)^4}
G(p/2+Q)G(p/2-Q)\Xi(q;p/2+Q;p/2-Q).
\label{eq:gammaxi0}
\end{eqnarray}
The factor $\frac12$ ensures that all diagrams enter $\Gamma$ with the
correct weights.

It is advantageous to split $\Xi(q;p/2+Q;p/2-Q)$ into parts
$\Xi^+(q;p/2+Q;p/2-Q)$ [$\Xi^-(q;p/2+Q;p/2-Q)$] analytic in the upper
[lower] half planes of $Q_0$, with
\begin{equation}
\Xi(q;p/2+Q;p/2-Q)=\Xi^+(q;p/2+Q;p/2-Q)+\Xi^-(q;p/2+Q;p/2-Q).
\end{equation}
To this end, note that the iterated terms are already thus split. To
split up the Born terms, use the relation
\begin{equation}
G(p)G(q)=\frac{G(p)+G(q)}{G(p)^{-1}+G(q)^{-1}}.
\label{eq:gsquaresplit}
\end{equation}
Then Eq. (\ref{eq:gammaxi0}) becomes
\begin{eqnarray}
\hspace{-4mm}\Gamma(q,q_0;p,p_0) & = & \nn \\ && \hspace{-28mm}
\frac12\int\frac{d^3Q}{(2\pi)^3}
\left[G(p/2-Q)\left.\Xi^-(q;p/2+Q;p/2-Q)\right|_{Q_0=
-p_0/2+(p/2+Q)^2/2m-E/4}\right. \nn \\ && \left.
\hspace{-12mm}+G(p/2+Q)\left.\Xi^+(q;p/2+Q;p/2-Q)\right|_{Q_0=
p_0/2-(p/2-Q)^2/2m+E/4}\right].
\label{eq:gammaxi20}
\end{eqnarray}

In each of the two last terms of Eq. (\ref{eq:inteqxi0}) it is
possible to integrate out $Q_0'$ by closing around the single pole in
the lower half plane. The two parts of $\Xi$ appearing in
Eq. (\ref{eq:gammaxi20}) satisfy the coupled integral equations
\begin{eqnarray}
&&  \hspace{-10mm}\Xi^-\left(q;\vec p/2+\vec Q,(\vec p/2+\vec Q)^2/2m-E/4;
\vec p/2-\vec Q,p_0-(\vec p/2+\vec Q)^2/2m+E/4\right) 
\nn \\
& = & 
-\int\frac{d\Omega_{\vec q}}{4\pi}
\left[\frac{G(-p/2+q-Q)}{G^{-1}(-p/2-q+Q)+G^{-1}(-p/2+q-Q)}
+(q\leftrightarrow-q)\right]
\nn \\ &&
-\int\frac{d^3Q'}{(2\pi)^3}G(-3\vec p/2+\vec Q',-p_0-
(p/2-Q')^2/2m+E/4)D(-p-Q+Q')
\nn \\ && \hspace{12mm}\times
 \Xi\left(q;\vec p/2+\vec Q,(p/2+Q)^2/2m-E/4;\vec p/2-\vec Q'
,(p/2-Q')^2/2m-E/4\right),
\nn \\ &&
 \hspace{-10mm}\Xi^+\left(q;\vec p/2+\vec Q,p_0-(p/2-Q)^2/2m+E/4;
\vec p/2-\vec Q,(p/2-Q)^2/2m-E/4\right)
\nn \\ & = &
-\int\frac{d\Omega_{\vec q}}{4\pi}
\left[\frac{G(-p/2-q+Q)}{G^{-1}(-p/2-q+Q)+G^{-1}(-p/2+q-Q)}
+(q\leftrightarrow-q)\right]
\nn \\ &&
-\int\frac{d^3Q'}{(2\pi)^3}G(-3\vec p/2-\vec Q',-p_0-
(p/2+Q')^2/2m+E/4)D(-p+Q-Q')
\nn \\ && \hspace{12mm}\times
\Xi\left(q;\vec p/2+\vec Q',(p/2+Q')^2/2m-E/4;\vec p/2-\vec Q
,(p/2-Q)^2/2m-E/4\right). \nn \\ &&
\label{eq:xisplit0}
\end{eqnarray}
In the first of these equations $Q_0=-p_0/2+(p/2+Q)^2/2m-E/4$ and
$Q_0'=p_0/2-(p/2-Q')^2/2m+E/4$. In the second equation
$Q_0=p_0/2-(p/2-Q)^2/2m+E/4$ and
$Q_0'=-p_0/2+(p/2+Q')^2/2m-E/4$.

On the right hand side of Eqs. (\ref{eq:xisplit0}), the vertex $\Xi$
only appears in the form
\begin{equation}
  \Xi(q;\vec p_1,p_1^2/2m-E/4;\vec p_2,p_2^2/2m-E/4) \equiv
\chi(q;\vec p_1,\vec p_2).
\end{equation}
This will be called the ``on-shell'' vertex. The reason for this
terminology is clear; had the propagators $G(p_1)$ and
$G(p_2)$ been attached to the vertex $\Xi(q;p_1;p_2)$ and the
frequencies been integrated over using the poles of the propagators,
then these on-shell values of the frequencies would have been
obtained. In this sense, the on-shell vertex is equivalent to the
atom-molecule scattering vertex $t^{\rm bf}(\vec p,-p^2/2m;\vec
q,-q^2/2m)$ discussed above, see Eq. (\ref{eq:onshell3body}).

Solving the integral equations is now possible with the
substitution
\begin{equation}
p_0\to (\vec p/2+\vec Q)^2/2m+(\vec p/2-\vec Q)^2/2m-E/2,
\end{equation}
at which value also the left hand sides of Eqs. (\ref{eq:xisplit0})
will be on-shell. Adding the expressions in Eqs. (\ref{eq:xisplit0})
at this value of $p_0$ results in an integral equation in three
variables for the total on-shell vertex which may subsequently be
inserted on the right hand side of Eqs. (\ref{eq:xisplit0}) to find
the value of the vertex $\Xi$ at any value of $p_0$.

After a change of variables to $\vec p_1=\vec p/2+\vec Q$ and $\vec
p_2=\vec p/2-\vec Q$, the equation for the total on-shell vertex takes
the form
\begin{eqnarray}
\hspace{-4mm}
\chi(q;\vec p_1,\vec p_2) & = & -\int\frac{d\Omega_{\vec q}}{4\pi}\left[
G(q-p_1)G(-q-p_2)+(q\leftrightarrow -q)\right]
\nn \\ && \hspace{-15mm}
-\int\frac{d^3Q}{(2\pi)^3}\left\{\left.
G(-p_1-p_2-Q)D(-Q-p_1)\chi(q;\vec p_1,\vec Q)\right|_{Q_0=Q^2/2m-E/4}
\right.\nn \\ && \left.
\hspace{0mm}+\left.G(-p_1-p_2-Q)D(-Q-p_2)\chi(q;\vec Q,\vec p_2)
\right|_{Q_0=Q^2/2m-E/4}\right\},
\label{eq:onshellchi0}
\end{eqnarray}
where $(p_1)_0=p_1^2/2m-E/4$ and $(p_2)_0=p_2^2/2m-E/4$. 

\begin{figure}[bt]
\begin{center}
\includegraphics[height=.75 in]{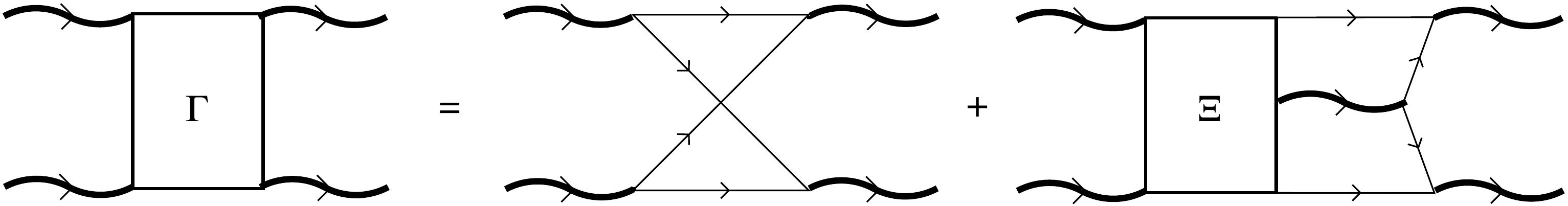}
\caption{An illustration of Eq. (\ref{eq:gammaxi30}).}
\label{fig:gammaxiiterate}
\end{center}
\end{figure}

Inserting Eq. (\ref{eq:onshellchi0}) into the right hand sides of
Eqs. (\ref{eq:xisplit0}) and the resulting expression into the right
hand side of Eq. (\ref{eq:gammaxi20}), the vertex $\Gamma$ finally
becomes
\begin{eqnarray}
\hspace{-10mm}\Gamma(q,q_0;p,p_0) & = & \Gamma^{(0)}(q,q_0;p,p_0)
\nn \\ && \hspace{-28mm}
-\frac12\int\frac{d^3p_1}{(2\pi)^3} \frac{d^3p_2}{(2\pi)^3}
\left[G(p-p_1)G(-p-p_2)
+(p\leftrightarrow -p)\right]
D(-p_1-p_2)\chi(q;\vec p_1,\vec p_2).
\label{eq:gammaxi30}
\end{eqnarray}
Here again $(p_1)_0=p_1^2/2m-E/4$ and $(p_2)_0=p_1^2/2m-E/4$ while it
is important to note that $p_0$ is a free parameter. This method
(\ref{eq:gammaxi30}) of constructing the two-boson irreducible vertex
corresponds diagrammatically to the situation depicted in
Fig. \ref{fig:gammaxiiterate}.
$\Gamma^{(0)}$ is the result of calculating the Born diagram, shown in
Fig. (\ref{fig:4bodyborn}),
\begin{eqnarray}
\Gamma^{(0)}(q,q_0;p,p_0) & = & -i\int\frac{d\Omega_{\vec q}}{4\pi}
\int\frac{d^4Q}{(2\pi)^4}
G(Q+p/2+q/2)G(Q-p/2-q/2) \nn\\ &&\hspace{20mm}\times
G(-Q-p/2+q/2)G(-Q+p/2-q/2)
\label{eq:4bodybornterm}\\
& = & -2\int\frac{d\Omega_{\vec q}}{4\pi}
\int\frac{d^3Q}{(2\pi)^3}\frac{A}{(A^2-B^2)(A^2-C^2)},
\end{eqnarray}
with
\begin{eqnarray}
A & = & E/2-Q^2/m-p^2/4m-q^2/4m, \nn \\
B & = & p_0-\vec Q\cdot\vec q/m, \nn \\
C & = & q_0-\vec Q\cdot\vec p/m.
\end{eqnarray}
The angular integrations may be performed independently by first
integrating over directions of $\vec q$ and then over directions of
$\vec Q$. The result is
\begin{eqnarray}
\hspace{-10mm}
\Gamma^{(0)}(q,q_0;p,p_0) & = & \frac{m^2}{16\pi^2pq}\int_0^\infty \frac{dQ}
{-E/2+Q^2/m+p^2/4m+q^2/4m}
\nn \\ && 
\times
\log\frac{(-E/2+Q^2/m+p^2/4m+q^2/4m+Qq/2m)^2-p_0^2}
{(-E/2+Q^2/m+p^2/4m+q^2/4m-Qq/2m)^2-p_0^2}
\nn \\ && 
\times
\log\frac{(-E/2+Q^2/m+p^2/4m+q^2/4m+Qp/2m)^2-q_0^2}
{(-E/2+Q^2/m+p^2/4m+q^2/4m-Qp/2m)^2-q_0^2}.
\end{eqnarray}

The procedure for solving the set of integral equations is now as
follows. The integral equation (\ref{eq:onshellchi0}) for the on-shell
$\chi$ has to be solved at every $(\vec q,q_0)$. However, the kernel
of this integral equation does not depend on $(\vec q,q_0)$, an
observation which is of great help in the numerical solution, since
this means that the identity matrix minus the integration kernel needs
to be inverted only once [see Eqs. (\ref{eq:num01}) and
(\ref{eq:num02})]. With the on-shell $\chi$ determined, this should be
inserted in Eq. (\ref{eq:gammaxi30}) to find the two-boson irreducible
vertex $\Gamma$. Finally, $\Gamma$ may be inserted in
Eq. (\ref{eq:4bodyinteq}) to find the scattering $t$-matrix.

\begin{figure}[bt]
\begin{center}
\includegraphics[height=2 in]{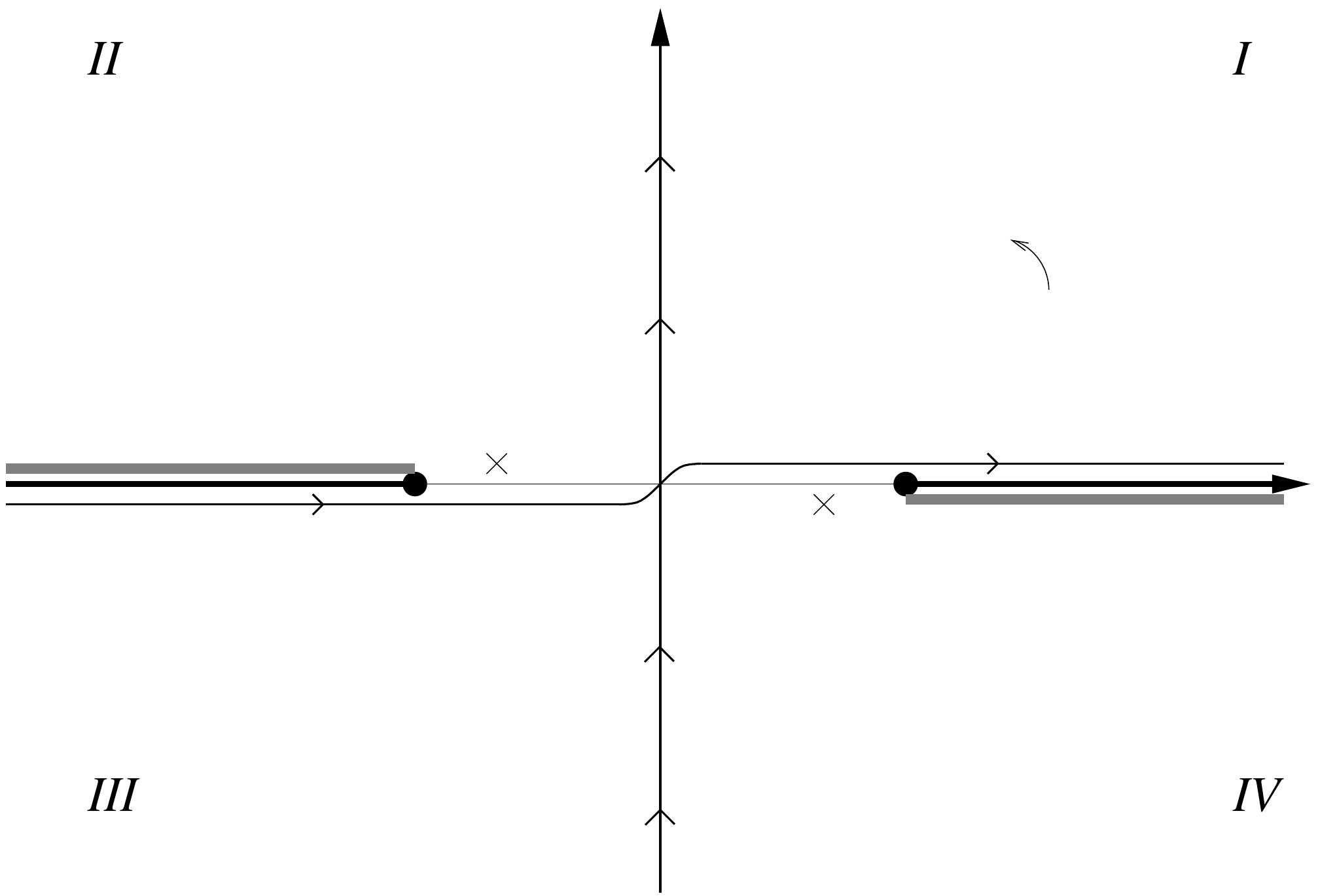}
\caption{The original and rotated contour of frequency integration
  used in Eq. (\ref{eq:4bodyinteq}). The crosses correspond to poles
  of the molecular propagators while the black lines correspond to the
  square root branch cuts. The gray boxes are the areas in which the
  two-boson irreducible vertex $\Gamma$ may have non-analytic
  structure.}
\label{fig:wick}
\end{center}
\end{figure}

A bit of care needs to be taken in peforming the integration over
frequency in Eq. (\ref{eq:4bodyinteq}). Due to the square-root
structure of the molecular propagators there will be branch cuts
starting at
\begin{equation}
q_0=-E/2+q^2/4m\ge0,\hspace{2mm}\mbox{and}\hspace{2mm} q_0=E/2-q^2/4m\le0
\end{equation}
and continuing towards $\pm\infty$, respectively. The contour of
integration is such that for $\mbox{Re}(q_0)>0$ [$\mbox{Re}(q_0)<0$]
the integration is above [below] the branch cut, which follows from
the infinitesimal in the molecular propagator,
Eq. (\ref{eq:bosonprop}). Furthermore, the molecular propagators have
poles at $q_0=\pm(q^2/4m+E_{\rm b}-E/2-i0)$. It is thus advantageous
to perform a Wick rotation as illustrated in Fig. \ref{fig:wick}. This
removes the problem of integrating close to the branch cuts and
poles. In order to perform such a Wick rotation, the vertex
$\Gamma(p,p_0;q,q_0)$ needs to be analytic in quadrants $I$ and
$III$. Eqs. (\ref{eq:onshellchi0}, \ref{eq:gammaxi30},
\ref{eq:4bodybornterm}) show that this is indeed the case and that any
non-analytic structure of $\Gamma$ must be restricted to the areas
marked as gray boxes in Fig. \ref{fig:wick}.

When computing the scattering length, the total energy should be set
to $E=2E_{\rm b}$. Then the simple poles of the molecular propagators
present in Eq. (\ref{eq:4bodyinteq}) occur at $q_0=\pm(q^2/4m-i0)$,
i.e. close to both the original and the Wick rotated integration
contour. These simple poles are integrable and may be treated by a
change of variables to ``polar'' coordinates,
\begin{equation}
  q^2/4m = R^2\sin\theta, \hspace{1cm}
  q_0 = R^2\cos\theta.
\end{equation}
Here $R\in[0,\infty[$ and $\theta\in[0,\pi]$.

To search for bound states in the four-fermion problem, let $E<2E_{\rm
  b}$ and look for solutions to the homogenous equation. In a matrix
notation this corresponds to the existence of a solution to the
equation $t_i=K_{ij}t_j$ or, in other words, the existence of a bound
state implies that the kernel of the integral equation, $K$, has an
eigenvalue equal to 1. No bound states are found by using this
method. As the one-channel model is employed here, the effective range
$r_0$ has been taken to zero. Thus the validity of this method is only
for bound states with binding energies $|E|\ll\frac{1}{mr_0^2}$ and
the result does not exclude deeply bound states.

\begin{figure}[bt]
\begin{center}
\includegraphics[height=2.5 in]{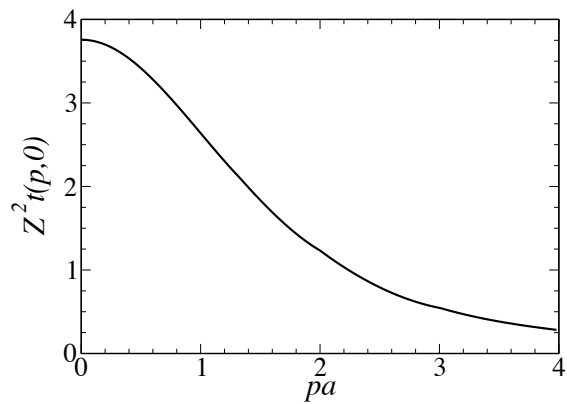}
\caption{The scattering amplitude $Z^2t(p,p_0)$ in unites of $a/m$
  evaluated at $p_0=0$.}
\label{fig:4bodyscat}
\end{center}
\end{figure}

The scattering amplitude is calculated on-shell, with the incoming and
outgoing molecules all having four-momentum $(\vec 0,E_{\rm b})$. Thus
$E=2E_{\rm b}$. As in the three-body problem above, to calculate the
scattering amplitude each external molecular propagator needs to be
renormalized by $\sqrt Z$, and the relation between the scattering
amplitude and the scattering matrix is
\begin{equation}
T(0) = Z^2t(0,0).
\end{equation}
The scattering length is then found through
\begin{equation}
T(0)=\frac{2\pi}{M_{\rm r}}a_{\rm b}
\end{equation}
with $M_{\rm r}=m$ the molecule-molecule reduced mass. By solving the set of
equations described above, it is found that
\begin{equation}
a_{\rm b}\approx 0.60a
\end{equation}
in complete agreement with
Refs. \cite{Brodsky2005,Petrov2005a}. Fig. \ref{fig:4bodyscat} shows
the computed $t$-matrix, from which $t(0,0)$ may be extracted.

\chapter{Quantum corrections \label{chap:wu}}
In chapter \ref{chap:bcsbec} properties of the BEC regime of the
dilute Bose gas consisting of weakly bound molecules were
calculated. The calculations were performed in the low density limit
and it was seen how this limit corresponds to a diagrammatic expansion
in terms of lines beginning or ending in the condensate. The results
obtained were seen to exactly match the results of the standard dilute
Bose gas \cite{AbrikosovBook,FetterBook} with the effect of the
fermionic atoms being that the boson-boson scattering length is
related to the fermion-fermion scattering length by $a_{\rm
  b}\approx0.60a$.

\begin{figure}[bt]
\begin{center}
\includegraphics[height=1.5 in]{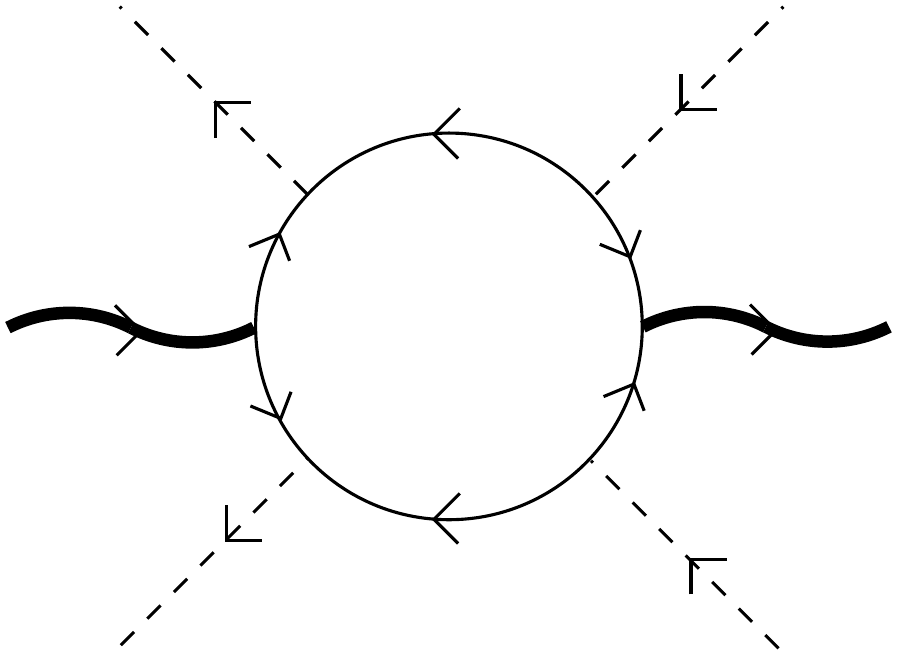}
\caption{A high order contribution to the molecular self energy.}
\label{fig:highorder}
\end{center}
\end{figure}

It is possible to extend the low density expansion in order to obtain
the chemical potential and ground state energy in higher orders. This
involves the determination of the fermionic and bosonic self energies
in higher orders. Somewhat surprisingly, the next order self energies
for the bosons {\it do not} involve diagrams such as in
Fig. \ref{fig:highorder} which only contributes at even higher
order. Rather, they arise due to non-trivial resummations of diagrams
occuring because of infrared divergences of the molecular
propagator. Lee, Huang, and Yang computed the first correction to the
ground state energy \cite{Lee1957a,Lee1957b}. Later, results were
obtained in the next order by Wu \cite{Wu1959} and Sawada
\cite{Sawada1959} and the results of these authors confirmed by
Hugenholtz and Pines \cite{Hugenholtz1959}.

The results in the dilute Bose gas are the following:
\begin{eqnarray}
\mu_{\rm b} & = & \frac{4\pi a_{\rm b}n_{\rm b}}{m_{\rm b}}\left[1+\frac{32}3\sqrt{
\frac{n_{\rm b}a_{\rm b}^3}\pi}
+12\pi\left(\frac43-\frac{\sqrt3}\pi\right)n_{\rm b}a_{\rm b}^3
\log n_{\rm b}a_{\rm b}^3
\right],\\
\frac{E_0}V & = & \frac{2\pi a_{\rm b}n_{\rm b}^2}{m_{\rm b}}\left[1+\frac{128}{15}\sqrt{
\frac{a_{\rm b}^3n_{\rm b}}\pi}+8\pi\left(\frac43-\frac{\sqrt3}\pi
\right)n_{\rm b}a_{\rm b}^3\log n_{\rm b}a_{\rm b}^3\right].
\end{eqnarray}
Below, it will be shown that these results are also correct in the gas
of weakly bound molecules, with the scattering length $a_{\rm b}$
related to the underlying fermion scattering length $a$.

The first two corrections to the ground state energy and chemical
potential are universal in the sense that they only depend on the
scattering length $a_{\rm b}$, the density, and the mass of
bosons. Higher order corrections will depend on the shape of the
interatomic potential, in particular on the effective range
$r_0$. These will not be considered here.

\section{The molecular propagator at low densities and momenta}
Define the chemical potential and particle number as an expansion in
$\Delta_0$:
\begin{eqnarray}
\mu_{\rm b} & = & \mu_{\rm b}^{(2)}+\mu_{\rm b}^{(3)}+\dots \\
n_{\rm b} & = & n_{\rm b}^{(2)}+n_{\rm b}^{(3)}+\dots
\end{eqnarray}
The superscript denotes powers of $\Delta_0$. In chapter
\ref{chap:bcsbec} the results
\begin{eqnarray}
\mu_{\rm b}^{(2)} & = & \frac{aa_{\rm b}m\Delta_0^2}4, \\
n_{\rm b}^{(2)} & = & \frac{am^2\Delta_0^2}{8\pi},
\end{eqnarray}
were found. Below, the next two orders in the chemical potential will
be calculated.

It is useful to consider the molecular propagator at frequencies
$\omega\ll 1/ma^2$ and momenta $p\ll a^{-1}$. In the zero density
approximation it becomes a free propagator
\begin{equation}
D_0(p,\omega) \approx \frac Z{\omega-p^2/2m_{\rm b}+i0},
\hspace{10mm}\left(p\ll a^{-1}, \omega\ll1/ma^2\right).
\end{equation}
In the first non-trivial order of the low density expansion, the
normal and anomalous propagators were similarly seen in section
\ref{sec:bcsbec} to be approximately $Z$ times the usual dilute Bose
gas normal and anomalous propagators. The poles of the iterated
propagators at low frequencies and momenta may then be separated as
\cite{Hugenholtz1959}
\begin{eqnarray}
\hspace{-2mm}
D_{\rm n}(p,\omega) & \approx & Z\frac{u_p^2}{\omega-\epsilon(p)+i0}
-Z\frac{v_p^2}{\omega+\epsilon(p)-i0},
\hspace{10mm}\left(p\ll a^{-1}, \omega\ll1/ma^2\right),
\label{eq:dn2}
\\
\hspace{-2mm}
D_{\rm a}(p,\omega) & \approx & Z\frac{-u_pv_p}{\omega-\epsilon(p)+i0}
+Z\frac{u_pv_p}{\omega+\epsilon(p)-i0},
\hspace{10mm}\left(p\ll a^{-1}, \omega\ll1/ma^2\right),
\label{eq:da2}
\end{eqnarray}
with
\begin{eqnarray}
u_p^2 & = & \frac{\frac{p^2}{2m_{\rm b}}+\mu_{\rm b}^{(2)}
+\epsilon(p)}{2\epsilon(p)}, \\
v_p^2 & = & \frac{\frac{p^2}{2m_{\rm b}}+\mu_{\rm b}^{(2)}
-\epsilon(p)}{2\epsilon(p)}.
\end{eqnarray}
Here
\begin{equation}
\epsilon(p) = \sqrt{\left(\frac{p^2}{2m_{\rm b}}\right)^2+\frac{p^2}{m_{\rm b}}
\mu_{\rm b}^{(2)}}
\end{equation}
is the spectrum of low-lying excitations found by solving the
Hugenholtz-Pines relation (\ref{eq:HP}) at low frequencies and
momenta.

For the purpose of computing the vacuum scattering lengths $a_{\rm
  bf}$ and $a_{\rm b}$ in chapter \ref{chap:bcsbec} there was only one
relevant length scale, namely the atom-atom scattering length
$a$. However, in the BEC regime another length scale is given by the
average interparticle spacing $n^{-1/3}$ and $na^3$ is a dimensionless
parameter. It is now seen how another important length scale appears
of order $\frac1{\sqrt{na}}\gg a$. This scale, known as the healing
length, defines a crossover in the excitation spectrum of low-lying
excitations; for $p\ll \sqrt{na}$ these behave as sound waves with a
constant velocity while for $\sqrt{na}\ll p\ll a^{-1}$ they are almost
free particles. Indeed, for small momenta
\begin{equation}
\epsilon(p) \approx p\sqrt{\frac{\mu_{\rm b}^{(2)}}{m_{\rm b}}},
\hspace{10mm}p\ll \sqrt{na}
\label{eq:epssmall}
\end{equation}
as found in Eq. (\ref{eq:sof}), while for higher momenta
\begin{equation}
\epsilon(p) \approx \frac{p^2}{2m_{\rm b}}+\mu_{\rm b}^{(2)}
-\left(\mu_{\rm b}^{(2)}\right)^2\frac{m_{\rm b}}{p^2},
\hspace{10mm}\sqrt{na}\ll p\ll a^{-1}.
\label{eq:epslarge}
\end{equation}

Thus there are three regimes of momenta relevant for the problem,
small momenta $p\ll \sqrt{na}$, intermediate momenta $\sqrt{na}\ll
p\ll a^{-1}$, and large momenta $p\gg a^{-1}$. In the first two
\cite{Hugenholtz1959}
\begin{eqnarray}
u_p^2 \approx v_p^2 \approx
\frac{\sqrt{\mu_{\rm b}^{(2)}}}{2p},
\hspace{1cm} & p\ll\sqrt{na}, & \\
u_p^2\approx 1,\hspace{5mm} v_p^2=\frac{\left(
\mu_{\rm b}^{(2)}
\right)^2}{p^4}, & \hspace{10mm} \sqrt{na}\ll p \ll a^{-1}. &
\end{eqnarray}
In the limit of large frequencies and momenta, the square root
structure of the molecular propagator dominates, and in this limit it
will no longer be possible to separate the poles.

\section{Corrections to the chemical potential and ground state
  energy \label{sec:lee}}
In this section the first corrections to the chemical potential and
ground state energy will be calculated in the low density
expansion. This will be done by a very direct method, utilizing the
particle number equation for the fermionic atoms,
Eq. (\ref{eq:particlenumberdef}), and the Hugenholtz-Pines relation,
Eq. (\ref{eq:HP2}).

The chemical potential of the bosonic molecules was found in
Eq. (\ref{eq:HP2}) by requiring a gap-less sound mode in the BEC. The
result was
\begin{equation}
\mu_{\rm b} = Z\left[\Sigma_{11}(0)-\Sigma_{20}(0)\right]
-\frac{ma^2Z^2}4\left[\Sigma_{11}(0)-\Sigma_{20}(0)\right]^2.
\label{eq:HP3}
\end{equation}
The lowest order contribution to $\mu_{\rm b}$ is of order
$\Delta_0^2$. The two non-trivial corrections to the chemical
potential considered in this chapter are of order $\Delta_0^3$ and
$\Delta_0^4\log\Delta_0$. Thus it is seen that the last term in
Eq. (\ref{eq:HP3}) does not influence these corrections to the
chemical potential and the Hugenholtz-Pines relation becomes
\begin{equation}
\mu_{\rm b} \approx Z\left[\Sigma_{11}(0)-\Sigma_{20}(0)\right].
\end{equation}

\begin{figure}[bt]
\begin{center}
\includegraphics[height=1.3 in]{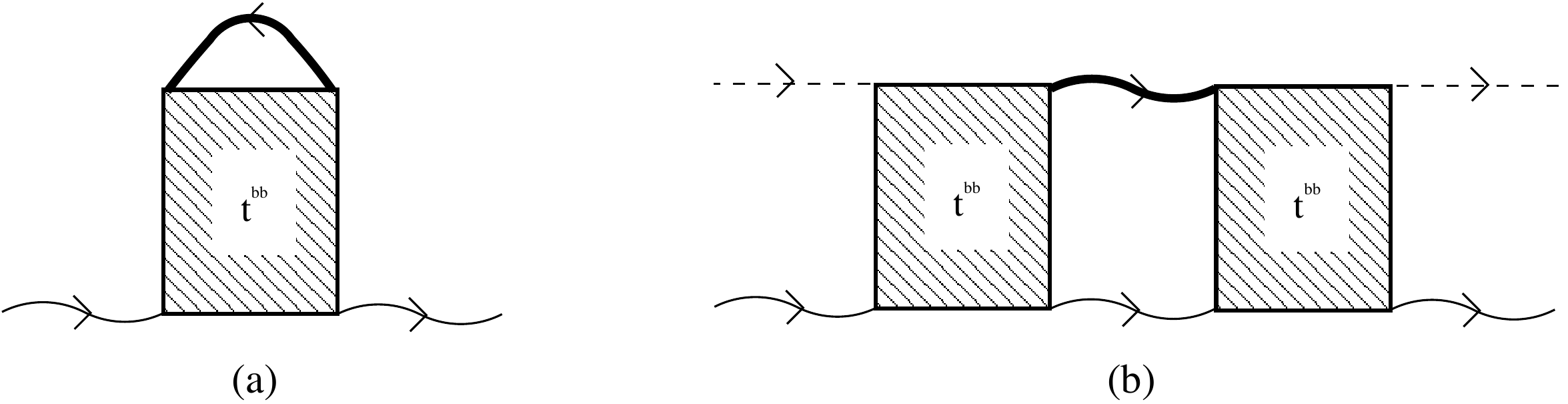}
\caption{The normal self energy at order $\Delta_0^3$. In this
  chapter, the thin wavy lines are molecular propagators $D_0$, while
  the thick wavy lines are the iterated molecular propagators $D_{\rm
    n}$ and $D_{\rm a}$ [Eqs. (\ref{eq:Dn}) and (\ref{eq:Da})],
  distinguished by the directions of arrows.}
\label{fig:normal3}
\end{center}
\end{figure}

The difference between the normal and anomalous self energy of the
bosons is needed at order $\Delta_0^3$. The diagrams contributing to
the normal self energy in this order are depicted in
Fig. \ref{fig:normal3}. The diagram in Fig. \ref{fig:normal3}a has no
analogue in $\Sigma_{11}^{(2)}$. Its value is
\begin{equation}
2i\lim_{\eta\to 0^+}\int \frac{d\omega}{2\pi}e^{i\eta\omega}
\int\frac{d^3p}{(2\pi)^3}t^{\rm bb}
D_{\rm n}(p,\omega),
\end{equation}
where the factor 2 appears as a combinatorial factor and the
exponential factor ensures the correct order of field operators. The
four-momentum dependence of $t^{\rm bb}$ has not been made
explicit. The $t$-matrix has the property that it is roughly constant
at $\omega\ll1/ma^2$ and $p\ll a^{-1}$, while it drops quickly to zero
for large frequencies and momenta. The value of the integral may then
be estimated by assuming that the main contribution is from
frequencies and momenta small compared with $|E_b|$ and
$a^{-1}$. Using Eq. (\ref{eq:dn2}) for the normal propagator, the
frequency integration is performed by closing the contour in the upper
half plane. The integration picks out the part of the propagator
proportional to $v_p^2$ which is suppressed at momenta $p\gg
\sqrt{na}$, and the value of the diagram is approximately
\begin{equation}
t^{\rm bb}\int\frac{d^3p}{(2\pi)^3}\frac{-\epsilon(p)+\frac{p^2}{2m_{\rm b}}
+\mu_{\rm b}^{(2)}}{\epsilon(p)}.
\end{equation}
Using the asymptotic forms (\ref{eq:epssmall}) and (\ref{eq:epslarge})
of the excitation spectrum $\epsilon(p)$, the integrand is seen to go
as $p$ for $p\ll \sqrt{na}$ and as $p^{-2}$ for $\sqrt{na}\ll
a^{-1}$. This means that the integral is dominated by momenta of order
$\sqrt{na}\ll a^{-1}$ and the assumption that $t^{\rm bb}$ could be taken
constant is justified. The diagram in Fig. \ref{fig:normal3}a then has
the value
\begin{equation}
\frac{16}3Z^{-1}\frac{aa_{\rm b}m\Delta_0^2}4
\sqrt{\frac{a_{\rm b}^3}\pi\frac{am^2\Delta_0^2}{8\pi}}.
\label{eq:normalsediag}
\end{equation}

The diagram shown in Fig. \ref{fig:normal3}b is already contained in
$\Sigma_{11}^{(2)}$ if the iterated normal propagator is replaced by
$D_0$. To compute the contribution to $\Sigma_{11}^{(3)}$ from the
diagram, the part proportional to $\Delta_0^2$ must be subtracted and
the value of the diagram is
\begin{eqnarray}
&& 4i\Delta_0^2\int \frac{d\omega}{2\pi}
\int\frac{d^3p}{(2\pi)^3}t^{\rm bb}
\left[D_{\rm n}(p,\omega)-D_0(p,\omega)\right]D_0(-p,-\omega) \nn \\
& = & 16Z^{-1}\frac{aa_{\rm b}m\Delta_0^2}4
\sqrt{\frac{a_{\rm b}^3}\pi\frac{am^2\Delta_0^2}{8\pi}}.
\end{eqnarray}
Again the integration is dominated by momenta of order
$p\sim\sqrt{na}$. The sum of the normal self energies gives
\begin{equation}
Z\Sigma_{11}^{(3)}(0)=\left(\frac{16}3+16\right)\frac{aa_{\rm b}m\Delta_0^2}4
\sqrt{\frac{a_{\rm b}^3}\pi\frac{am^2\Delta_0^2}{8\pi}}.
\end{equation}

\begin{figure}[bt]
\begin{center}
\includegraphics[height=1 in]{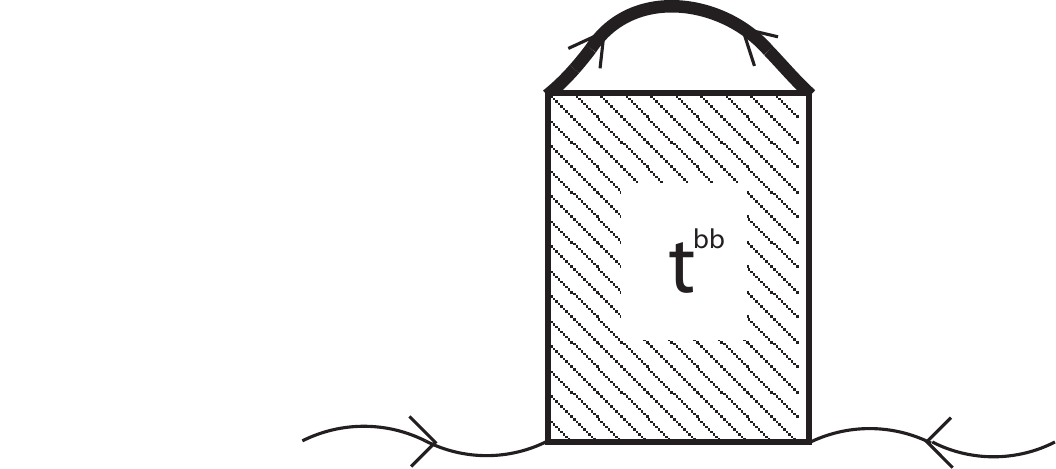}
\caption{The anomalous self energy for the bosons at order $\Delta_0^3$.}
\label{fig:anomalous3}
\end{center}
\end{figure}

\begin{figure}[bt]
\begin{center}
\includegraphics[height=1 in]{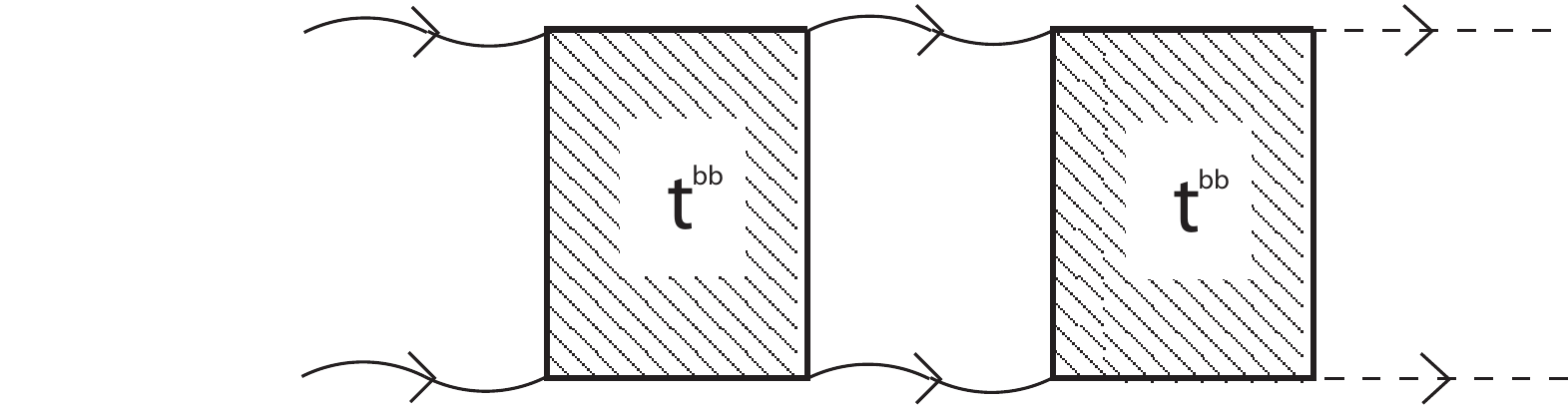}
\caption{A contribution to the anomalous self energy already included
  at order $\Delta_0^2$.}
\label{fig:anomalous3cancel}
\end{center}
\end{figure}

The only contribution to the anomalous self energy $\Sigma_{11}^{(3)}$ in
this order is given by the diagram shown in Fig. \ref{fig:anomalous3}.
\begin{equation}
i\int\frac{d\omega}{2\pi}\int\frac{d^3p}{(2\pi)^3}t^{\rm bb}D_{\rm a}(p,\omega).
\end{equation}
Assuming that the integration is dominated by small $p,\omega$ results
in
\begin{equation}
-\frac{Zt^{\rm bb}\mu_{\rm b}^{(2)}}2\int\frac{d^3p}{(2\pi)^3}
\frac1{\epsilon(p)},
\end{equation}
which is formally divergent at large $p$. However, the divergence is
proportional to $\Delta_0^2$ and arises because the diagram shown in
Fig. \ref{fig:anomalous3cancel}, and belonging to
$\Sigma_{11}^{(2)}$, is included in the diagram in
Fig. \ref{fig:anomalous3}. The diagram should be subtracted and the
anomalous self energy at order $\Delta_0^3$ becomes
\begin{eqnarray}
Z\Sigma_{20}^{(3)}(0) & = & -\frac{Z^2t^{\rm bb}
\mu_{\rm b}^{(2)}}2\int\frac{d^3p}{(2\pi)^3}
\left[\frac1{\epsilon(p)}-\frac{2m_{\rm b}}{p^2}\right] \nn \\
& = & 8\frac{aa_{\rm b}m\Delta_0^2}4
\sqrt{\frac{a_{\rm b}^3}\pi\frac{am^2\Delta_0^2}{8\pi}}.
\end{eqnarray}
With the regularization, the diagram is dominated by $p\sim\sqrt{na}$.

The chemical potential up to order $\Delta_0^3$ is found to be
\begin{eqnarray}
\mu_{\rm b} & = & Z\left[\Sigma_{11}(0)-\Sigma_{20}(0)\right] \nn \\
& = & \frac{aa_{\rm b}m\Delta_0^2}4\left(1+
\frac{40}3\sqrt{\frac{a_{\rm b}^3}\pi\frac{am^2\Delta_0^2}{8\pi}}\right).
\label{eq:mub3}
\end{eqnarray}

\begin{figure}[bt]
\begin{center}
\includegraphics[height=1 in]{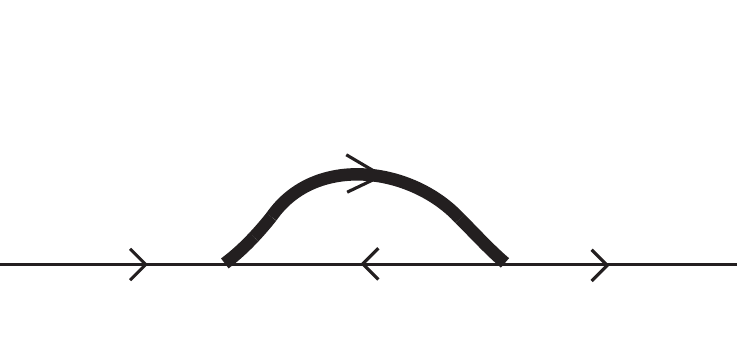}
\caption{The diagram giving the particle number at order $\Delta_0^3$.}
\label{fig:nb3}
\end{center}
\end{figure}

The particle number must be determined in the same order. $n_{\rm b}^{(3)}$
originates from only one diagram, shown in Fig. \ref{fig:nb3}, with
the value
\begin{eqnarray}
\hspace{-3mm}
n_{\rm b}^{(3)} & = & \frac122i^2\lim_{\eta\to0^+}
\int\frac{dp_0}{2\pi}e^{i\eta p_0}\int\frac{d^3p}{(2\pi)^3}
\int\frac{d^4q}{(2\pi^4)}\left[G(p)\right]^2G(-p+q)D_{\rm n}(q) \nn \\
& \approx & -2Z\lim_{\eta\to0^+}
\int\frac{dp_0}{2\pi}e^{i\eta p_0}\int\frac{d^3p}{(2\pi)^3}
\left(\frac1{p_0-p^2/2m-1/2ma^2+i0}\right)^2
\nn \\ && \hspace{-10mm}\times
\int\frac{d^4q}{(2\pi^4)}
\frac1{-p_0+q_0-(-\vec p+\vec q)^2/2m-1/2ma^2+i0}
\frac{q_0+q^2/2m_{\rm b}+\mu_{\rm b}^{(2)}}{q_0^2-\epsilon(q)^2+i0},
\end{eqnarray}
where it has again been assumed that the integration over $q$ is
dominated by $q\ll a^{-1}$. The factor $\frac12$ comes from $n_{\rm b}=n/2$
while the factor 2 is from the sum on fermion spins. Integrating over
frequencies and shifting the momentum $\vec p\to \vec p+\vec q/2$ to
get rid of angular dependence, the expression reduces to
\begin{eqnarray}
n_{\rm b}^{(3)} & = & Z\int\frac{d^3p}{(2\pi)^3}\frac{d^3q}{(2\pi)^3}
\left(\frac1{\epsilon(q)+p^2/m+q^2/2m_{\rm b}+1/ma^2}\right)^2
\frac{-\epsilon(q)+q^2/2m_{\rm b}+\mu_{\rm b}^{(2)}}{\epsilon(q)} \nn \\
& \approx & \frac{m^3a^{3/2}a_{\rm b}^{3/2}\Delta_0^3}{3\sqrt2\pi^2}.
\end{eqnarray}
The $q$ integration is indeed dominated by momenta $q$ of order
$\sqrt{na}$. The integration also results in terms of order
$\Delta_0^4$ and higher which have been ignored.

The particle number to order $\Delta_0^3$ is thus found to be
\begin{equation}
n_{\rm b} = \frac{am^2\Delta_0^2}{4\pi}\left(1+\frac83\sqrt{
\frac{a_{\rm b}^3}\pi\frac{am^2\Delta_0^2}{8\pi}}\right).
\label{eq:nopartic}
\end{equation}
Combining this result with the result for the chemical potential,
Eq. (\ref{eq:mub3}), results in
\begin{equation}
\mu_{\rm b} = \frac{4\pi a_{\rm b}n_{\rm b}}{m_{\rm b}}\left(1+\frac{32}3\sqrt{
\frac{a_{\rm b}^3n_{\rm b}}\pi}\right),
\end{equation}
exactly coinciding with the result in the standard dilute Bose gas
\cite{Lee1957a}.

The chemical potential is related to the ground state energy by
\begin{equation}
\mu_{\rm b} = \frac1V\frac{\partial E_0}{\partial n_{\rm b}}.
\label{eq:mubfrome0}
\end{equation}
Assuming a form $E_0=\alpha n_{\rm b}^2+\beta n_{\rm b}^{5/2}$ the ground state
energy is found to be
\begin{equation}
  \frac{E_0}V = \frac{2\pi a_{\rm b}n_{\rm b}^2}
  {m_{\rm b}}\left(1+\frac{128}{15}\sqrt{
      \frac{a_{\rm b}^3n_{\rm b}}\pi}\right).
\label{eq:e0lhy}
\end{equation}
As this result is computed from the chemical potential it of course
also matches the usual dilute Bose gas result.

\section{Next order corrections}
An elegant derivation of the chemical potential and ground state
energy in the standard dilute Bose gas was presented by Hugenholtz and
Pines in Ref. \cite{Hugenholtz1959}. These authors noticed that the
two quantities are linked by two equations. The first of these was
given as Eq. (\ref{eq:E0}) and is
\begin{equation}
\frac{E_0}V-\frac12n_{\rm b} \mu_{\rm b} =
\frac{i}{2Z}\lim_{\eta\to0^+}\int\frac{d\omega}{2\pi}e^{i\eta\omega}
\int\frac{d^3p}{(2\pi)^3}\left(\omega+p^2/2m_{\rm b}\right)D_{\rm n}(p,\omega),
\label{eq:E02}
\end{equation}
while the second equation is Eq. (\ref{eq:mubfrome0}). The crucial
observation was that given the normal bosonic propagator in some order
in the low density expansion, when inserted in Eq. (\ref{eq:E02}) the
resulting relation between the chemical potential and the ground state
energy will be an equation in the next order. This will be
demonstrated below.

The method may be used to compute the results of section \ref{sec:lee}
above. Inserting the propagator $D_{\rm n}$ from Eq. (\ref{eq:dn2}), correct
up to order $\Delta_0^2$ in Eq. (\ref{eq:E02}), results in
\begin{equation}
\frac{E_0}V-\frac12n_{\rm b} \mu_{\rm b} =
-\frac{i}{2}\lim_{\eta\to0^+}\int\frac{d\omega}{2\pi}e^{i\eta\omega}
\int\frac{d^3p}{(2\pi)^3}\left(\omega+p^2/2m_{\rm b}\right)
\frac{v_p^2}{\omega+\epsilon(p)-i0},
\end{equation}
where it has been assumed that only small $p,\omega$ contribute to the
integral. Carrying out the integration, the assumption is seen to be
valid, and the result is
\begin{eqnarray}
  \frac{E_0}V-\frac12n_{\rm b} \mu_{\rm b} & = & -\frac{2}{15\pi^2m_{\rm b}}
  (4\pi a_{\rm b})^{5/2}
  \left(\frac{am_{\rm b}^2\Delta_0^2}{32\pi}\right)^{5/2} \nn \\
  & = & -\frac{2}{15\pi^2m_{\rm b}}(4\pi a_{\rm b})^{5/2}n_{\rm b}^{5/2}.
\label{eq:e0andmub}
\end{eqnarray}
In the last step, $\Delta_0^2$ has been traded for $n_{\rm b}$ using
Eq. (\ref{eq:particlenumber}). This is correct at the current order in
$na^3$. Now assume the following forms for the ground state energy and
chemical potential
\begin{eqnarray}
  \frac{E_0}V & = & \frac{2\pi a_{\rm b}n_{\rm b}^2}{m_{\rm b}}+\alpha
\frac{a_{\rm b}^{5/2}n_{\rm b}^{5/2}}{m_{\rm b}}, \\
  \mu_{\rm b} & = & \frac{4\pi a_{\rm b}n_{\rm b}}{m_{\rm b}}+\beta 
\frac{a_{\rm b}^{5/2}n_{\rm b}^{3/2}}{m_{\rm b}}.
\end{eqnarray}
Performing the differentiation in Eq. (\ref{eq:mubfrome0}) gives
\begin{equation}
\mu_{\rm b} = \frac{4\pi a_{\rm b}n_{\rm b}}{m_{\rm b}}+\frac52
\alpha\frac{ a_{\rm b}^{5/2}n_{\rm b}^{3/2}}{m_{\rm b}},
\end{equation}
from which it is seen that $\beta=\frac52\alpha$. A second linear
equation is provided by comparing powers of $n_{\rm b}$ in
Eq. (\ref{eq:e0andmub}). Solving the equation gives
$\alpha=\frac{256\sqrt\pi}{15}$, and inserting $\alpha$ and $\beta$ in
the equations for $\mu_{\rm b}$ and $E_0$, the resulting ground state
energy and chemical potential are seen to exactly match the results in
Sec. \ref{sec:lee} above.

\subsection{The logarithmic contributions}
In the low density expansion, the next order contribution (computed by
Wu and Sawada \cite{Wu1959,Sawada1959} and denoted below by WS) to the
chemical potential and ground state energy is logarithmic,
proportional to $\log\frac{a^{-1}}{\sqrt{na}}$. This in turn means
that the correction arises from momenta in the range $\sqrt{na}\ll
p\ll a^{-1}$. It should thus not come as a great surprise that the BEC
of weakly bound molecules also at this order coincides with the
standard dilute Bose gas since the precise structure of $t^{\rm bb}$
will not be probed by these momenta.

As above, the normal molecular propagator is needed. Define
\begin{equation}
D_{\rm n}(p,\omega) = D_{\rm n}^{(1)}(p,\omega)+D_{\rm n}^{(2)}(p,\omega)+\dots
\end{equation}
The propagator $D_{\rm n}^{(1)}(p,\omega)$ is the one considered above
in calculating the Lee-Huang-Yang (LHY) ground state energy
(\ref{eq:e0lhy}), and is correct to order $\Delta_0^2$. $D_{\rm
  n}^{(2)}(p,\omega)$ is the part of the molecular propagator leading
to the logarithmic corrections while the dots correspond to even
higher order parts of the normal propagator. The anomalous propagator
$D_{\rm a}^{(2)}(p,\omega)$ is defined in the same manner.

\begin{figure}[bt]
\begin{center}
\includegraphics[height=.35 in]{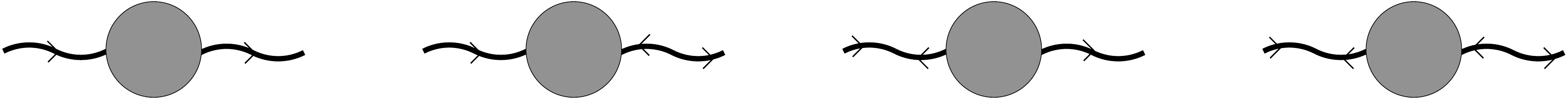}
\caption{The expansion of the normal propagator,
  Eq. (\ref{eq:Dn2expansion}). The circles are the effective self
  energies $\Sigma'_{11}$ and $\Sigma'_{20}$.}
\label{fig:Dn2expansion}
\end{center}
\end{figure}

In Eqs. (\ref{eq:Dn}) and (\ref{eq:Da}) the normal and anomalous
molecular propagators were related to the ``bare'' propagator $D_0$
and the exact self energies, the result of Beliaev
\cite{Beliaev1958a}. It is possible to derive an analogous expression
relating the exact propagators not to the bare propagator but to
$D_{\rm n}^{(1)}(p,\omega)$ and $D_{\rm a}^{(1)}(p,\omega)$
\cite{Hugenholtz1959}. This requires the definition of new effective
self energies
\begin{eqnarray}
\Sigma'_{11}(p,\omega) & \equiv &
\Sigma_{11}(p,\omega)-\Sigma^{(2)}_{11}(0,0), \\
\Sigma'_{20}(p,\omega) & \equiv &
\Sigma_{20}(p,\omega)-\Sigma^{(2)}_{20}(0,0),
\end{eqnarray}
since the self energies (at zero four-momentum) $\Sigma^{(2)}_{11}$
and $\Sigma^{(2)}_{20}$ are already included in the LHY propagators
$D_{\rm n}^{(1)}$ and $D_{\rm a}^{(1)}$. In order to compute the WS order, it is
sufficient to approximate the normal molecular propagator by
\cite{Hugenholtz1959}
\begin{eqnarray}
D_{\rm n}^{(2)}(p,\omega) & = &  D_{\rm n}^{(1)}(p,\omega)\Sigma'_{11}(p,\omega)
D_{\rm n}^{(1)}(p,\omega)
+D_{\rm n}^{(1)}(p,\omega)\Sigma'_{20}(p,\omega)D_{\rm a}^{(1)}(p,\omega)
\nn \\ && \hspace{-13mm}
+D_{\rm a}^{(1)}(p,\omega)\Sigma'_{20}(p,\omega)D_{\rm n}^{(1)}(p,\omega)
+D_{\rm a}^{(1)}(p,\omega)\Sigma'_{11}(-p,-\omega)D_{\rm a}^{(1)}(p,\omega).
\label{eq:Dn2expansion}
\end{eqnarray}
Diagrammatically, the expansion is shown in
Fig. \ref{fig:Dn2expansion}. Because of the range of momenta relevant
to the problem ($p\gg\sqrt{na}$), the propagators may be expanded in
powers of the density and become
\begin{eqnarray}
&& D_{\rm n}^{(1)}(p,\omega) \approx D_0(p,\omega) \approx 
\frac{Z}{\omega-p^2/2m_{\rm b}+i0}, \nn \\
&& D_{\rm a}^{(1)}(p,\omega) \approx \Delta_0^2t^{\rm bb}D_0(p,\omega)
D_0(-p,-\omega) \approx 
\frac{Z\mu_{\rm b}^{(2)}}{\omega-p^2/2m_{\rm b}+i0}
\frac1{-\omega-p^2/2m_{\rm b}+i0}.
\nn\\ &&
\end{eqnarray}

Inserting the resulting normal molecular propagator $D_{\rm n}^{(2)}$ in
Eq. (\ref{eq:E02}) results in an equation between the ground state
energy and the chemical potential in WS order
\begin{eqnarray}
\hspace{-4mm}  \frac{\tilde E_0}V-\frac12n_{\rm b}\tilde\mu_{\rm b}
  & = & \frac{iZ}2\lim_{\eta\to0^+}\int\frac{d\omega}{2\pi}
  e^{i\eta\omega}
  \int\frac{d^3p}{(2\pi)^3}\left[\frac{\Sigma_{11}(p,\omega)
      (\omega+p^2/2m_{\rm b})}{\omega-p^2/2m_{\rm b}+i0} 
\right.\nn \\ && \hspace{-12mm}\left.
    -\frac{2\Sigma'_{20}(p,\omega)\mu_{\rm b}^{(2)}}{(\omega-p^2/2m_{\rm b}+i0)^2}
    +\frac{\Sigma'_{11}(-p,-\omega)\mu_{\rm b}^{(2)}}{(\omega-p^2/2m_{\rm b}+i0)^2
      (\omega+p^2/2m_{\rm b}+i0)}\right].
\label{eq:e0aa}
\end{eqnarray}
The quantity in square brackets is $(\omega+p^2/2m_{\rm
  b})D_{\rm n}^{(2)}(p,\omega)$ and the normal propagator has the property
that any poles must lie slightly above the negative real $\omega$ axis
or slightly below the positive real $\omega$ axis. Looking at the
$\omega$ integration, which must be closed in the upper half plane, it
is then concluded that the following contributions to the self
energies need to be considered:
\begin{itemize}
\item{poles of $\Sigma_{20}'(p,\omega)$ above the negative real axis}
\item{poles of $\Sigma_{11}'(p,\omega)$ above the negative real axis,
    and}
\item{poles of $\Sigma_{11}'(-p,-\omega)$ above the negative real axis
    as well as the momentum dependence of
    $\Sigma_{11}^{(2)}(p,p^2/2)$.}
\end{itemize}

\begin{figure}[bt]
\begin{center}
\includegraphics[height=1 in]{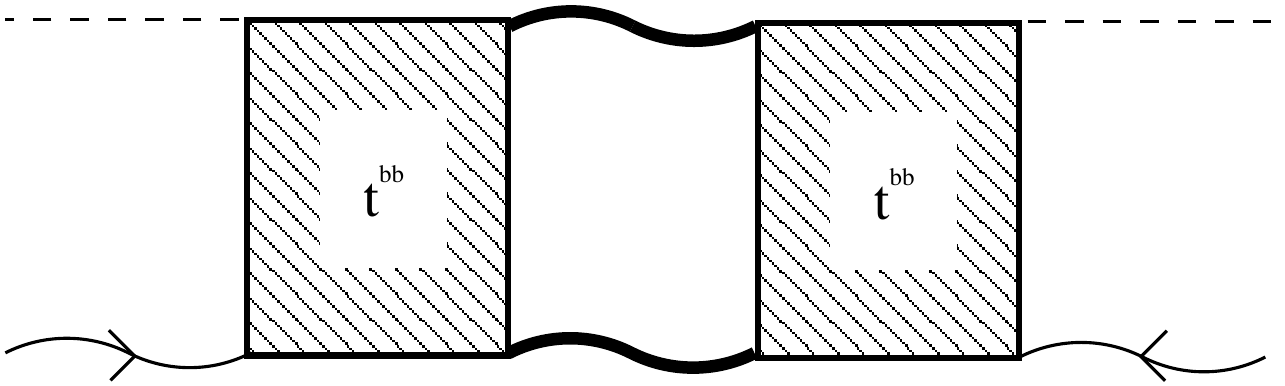}
\caption{The class of diagrams which must be considered in the
  anomalous self energy.}
\label{fig:s20diagrams}
\end{center}
\end{figure}

\begin{figure}[bt]
\begin{center}
\includegraphics[height=1.05 in]{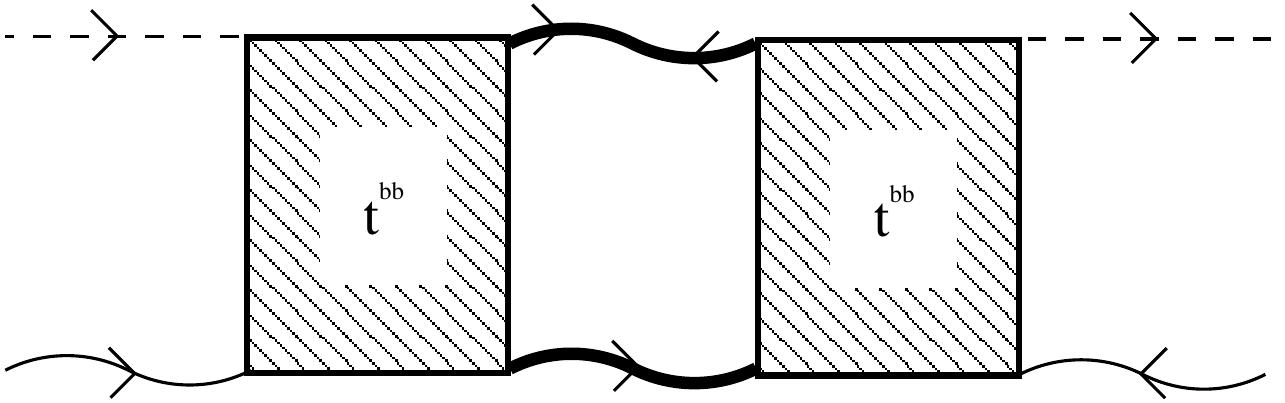}
\caption{The diagram belonging to $\Sigma'_{20}(p,\omega)$ which gives
  the leading contribution in Eq. (\ref{eq:e0aa}).}
\label{fig:s20leading}
\end{center}
\end{figure}

The simplest class of diagrams entering the anomalous self energy sum
to give the molecule-molecule scattering $t$-matrix. This reduces to a
constant at $p\ll a^{-1}$ and this class of diagrams does not have any
poles at relevant momenta. Poles above the negative real exis become
possible if, in the molecule-molecule scattering problem, a pair of bare
molecular propagators are replaced by the iterated propagators
$D^{(1)}$, thus all diagrams of the form depicted in
Fig. \ref{fig:s20diagrams} must be taken into account. The leading of
these diagrams is shown in Fig. \ref{fig:s20leading}. Carrying out the
frequency integration, anticipating that the main contribution will be
from $p\ll a^{-1}$ and $\omega\ll 1/ma^2$, results in
\begin{equation}
\tilde \Sigma'_{20}(p,\omega) =
4\Delta_0^2Z^2\mu_{\rm b}^{(2)}\left(t^{\rm bb}\right)^2\int
\frac{d^3q}{(2\pi)^3}\frac{m_{\rm b}}
{q^2\left(\omega+q^2/2m_{\rm b}+k^2/2m_{\rm b}
-i0\right)}.
\label{eq:s20a}
\end{equation}
For simplicity $\vec k\equiv \vec p+\vec q$. The tilde signifies that
this is the only relevant part of the self energy at WS order. In the
prefactor, the factor 4 comes from combinatorics, $Z^2\mu_{\rm
  b}^{(2)}$ from propagators, $\Delta_0^2$ from condensate lines, and
$\left(t^{\rm bb}\right)^2$ from the two $t$-matrices evaluated at low
momenta.

\begin{figure}[bt]
\begin{center}
\includegraphics[height=1.05 in]{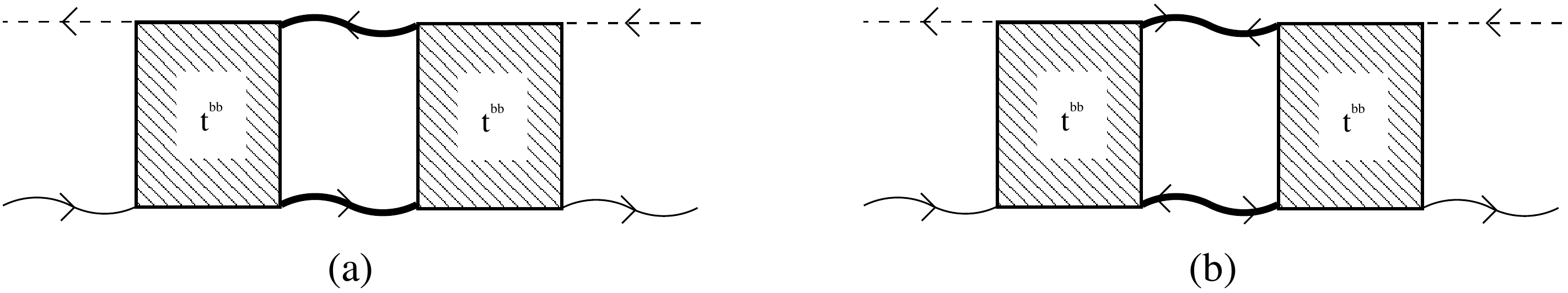}
\caption{The diagrams belonging to $\Sigma'_{11}(p,\omega)$ which
  result in the leading terms in the low density expansion of
  Eq. (\ref{eq:e0aa}).}
\label{fig:s11diagrams}
\end{center}
\end{figure}

The normal self energy is constructed in the same way. The two leading
diagrams relevant for computing the poles of $\Sigma'_{11}(p,\omega)$
above the negative real $\omega$ axis are shown in
Fig. \ref{fig:s11diagrams}. The sum of these diagrams is
\begin{equation}
\tilde \Sigma'_{11}(p,\omega)=
-4\Delta_0^2Z^2\left(\mu_{\rm b}^{(2)}\right)^2\left(t^{\rm bb}\right)^2
\int\frac{d^3q}{(2\pi)^3}\left(\frac1
 {q^2k^2}+\frac1{k^4}\right)\frac{m_{\rm b}^2}
{\omega+q^2/2m_{\rm b}+k^2/2m_{\rm b}-i0},
\label{eq:s11a}
\end{equation}
where it is again anticipated that only momenta small compared with
the inverse scattering length are relevant.

\begin{figure}[bt]
\begin{center}
\includegraphics[height=1.0 in]{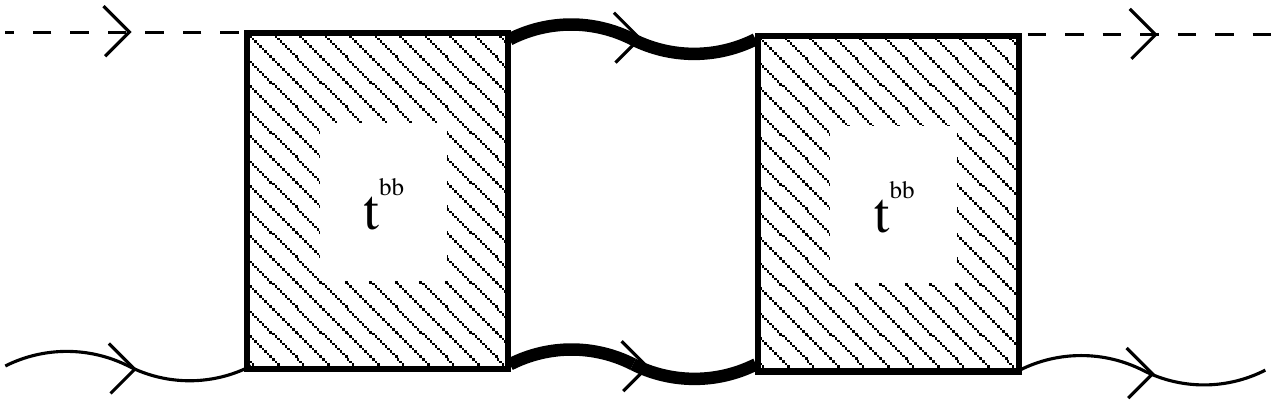}
\caption{A contribution from $\Sigma'_{11}(-p,-\omega)$.}
\label{fig:s11diagram2}
\end{center}
\end{figure}

The final contribution from molecular self energies comes from
$\Sigma'_{11}(-p,-\omega)$ and here it is important to include the
momentum dependence of the self energy. This may be included by
considering the relation between the normal self energy and the
scattering amplitude. This relation was first derived by Beliaev
\cite{Beliaev1958a,Beliaev1958b} (and similar results for fermion
scattering by Galitskii \cite{Galitskii1958}) and at relevant momenta
it takes the approximate form \cite{Hugenholtz1959}
\begin{eqnarray}
\hspace{-10mm}\Sigma_{11}(p,\omega) & \approx &
2\Delta_0^2\mbox{Im}f_s(p/2,p/2) \nn \\ && \hspace{-24mm}+
2Z^2\Delta_0^2\left(t^{\rm bb}\right)^2\int\frac{d^3q}{(2\pi)^3}
\left(\frac1{\omega-p^2/4m_{\rm b}-q^2/m_{\rm b}+i0}+
\frac1{q^2/m_{\rm b}-p^2/4m_{\rm b}-i0}\right).
\label{eq:kin}
\end{eqnarray}
Here $f_s(p,q)\equiv \left(f(p,q)+f(-q,p)\right)/2$ is a symmetrized
scattering amplitude and reduces at small momenta to $t^{\rm bb}$. Note
how the first term under the integral simply corresponds to the
diagram of Fig. \ref{fig:s11diagram2} while the second term under the
integral is a kinematic term. The reasoning behind Eq. (\ref{eq:kin})
is that while the scattering amplitude between two particles contains
the possibility of multiple scattering, it is still an intrinsic
two-body quantity in the sense that it is assumed that after the
scattering the two particles are in the vacuum and on-shell which is
certainly not true in the many-body case. The proof by Beliaev of this
equation is for a dilute gas of bosons but it goes through in exactly
the same way in the problem of the weakly bound bosonic molecules,
since the assumption is that the potential between the bosons (which
in the case of the fermion problem is replaced by the two-boson
irreducible diagram $\Gamma$, see chapter \ref{chap:bcsbec}) reduces
to a constant at momenta $p\ll a^{-1}$.

The imaginary part in the first term of Eq. (\ref{eq:kin}) may be
written in a different way, as also found by Beliaev,
\begin{equation}
\mbox{Im} f_s(p/2,p/2) = -Z^2\left(t^{\rm bb}\right)^2\mbox{Im}\int\frac{d^3q}
     {(2\pi)^3}\frac1{q^2/m_{\rm b}-p^2/4m_{\rm b}-i0}.
\end{equation}
The real part of the first term in Eq. (\ref{eq:kin}) is simply the
scattering matrix $t^{\rm bb}$ which reduces to a constant and does not
contribute to the ground state energy.

\begin{figure}[bt]
\begin{center}
\includegraphics[height=1.3 in]{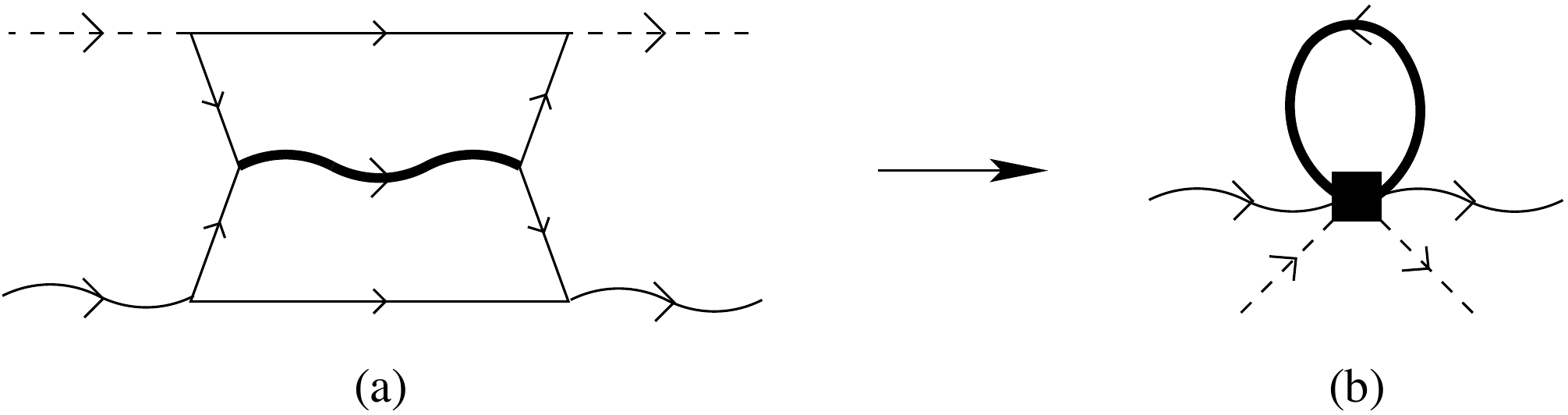}
\caption{An illustration of how fermions may be thought of as
  constants at low momenta. Fermions lines are drawn as straight thin
  lines.}
\label{fig:fermionexample}
\end{center}
\end{figure}

At this point it is important to investigate whether the presence of
the fermionic atoms affects self energies at the current order. The
effect of replacing a bare molecular propagator insde the two-boson
irreducible diagram by an iterated propagator $D_{\rm n}^{(1)}$ must
be examined. As an example, consider the diagrams in
Fig. \ref{fig:fermionexample}. The fermionic propagators in this
diagram all have chemical potential $\mu=-1/2ma^2$ in the BEC regime
of the gas, see Eq. (\ref{eq:mulowest}). Equivalently, when a molecule
in the vacuum breaks into two atoms they equally share the binding
energy $-1/ma^2$. For momenta small compared with the inverse
scattering length, the fermionic propagators are thus approximately
constant and the diagram shown in Fig. \ref{fig:fermionexample}a
effectively shrinks to that of Fig. \ref{fig:fermionexample}b. The
integration over the normal propagator closed on itself was performed
in Eq. (\ref{eq:normalsediag}) and goes as $\Delta_0^3$. Thus the
diagram in Fig. \ref{fig:fermionexample}a goes as $\Delta_0^5$ and is
of higher order than the WS order. The same argument goes through for
all two-boson irreducible diagrams with molecular propagators replaced
by iterated propagators.

The fermionic atoms are also modified by their self energies. The
anomalous fermion self energy at lowest order is simply two fermions
coming in and a condensate line going out. Such anomalous self
energies must come in pairs and thus modifying the two-boson
irreducible diagram by the simplest two such anomalous self energies
simply corresponds to adding the possibility of three boson
scattering. This in turn only contributes at order $\Delta_0^4$. The
normal fermion self energy consists of atom-molecule
scattering. Replacing a fermion propagator with the atom-molecule
scattering process inside the two-boson irreducible diagram again
results in a diagram belonging to the class of three boson
scattering. Furthermore, the atom-molecule diagram reduces to a
constant at $p\ll a^{-1}$. Thus the presence of this self energy does
not change the pole structure of the two-boson irreducible diagram in
the range of momenta and energies under consideration.

Each two-boson irreducible diagram will contain exactly one fermion
loop. The external four-momentum going into the two-boson irreducible
diagram may be contained completely inside the loop and since the
fermionic propagators are constant at $p\ll a^{-1}$, the pole structure
of the molecular self energies at small momenta is completely
unchanged by the presence of fermions in the problem.

The conclusion to this discussion is that the presence of fermionic
atoms does not change the chemical potential and ground state energy
from the results of the standard dilute Bose gas beyond replacing the
scattering length $a_{\rm b}$ with $0.60a$.

The rest is mathematics. Inserting the self energies,
Eqs. (\ref{eq:s20a}), (\ref{eq:s11a}), and (\ref{eq:kin}) in
Eq. (\ref{eq:e0aa}) and performing the integrals, the result is
\cite{Hugenholtz1959}
\begin{equation}
\frac{\tilde E_0}V-\frac12n_{\rm b}\tilde\mu_{\rm b} = Z\Delta_0^2
\left(\mu_{\rm b}^{(2)}\right)^2\frac1{32\pi^2}\left(
\frac43-\frac{\sqrt3}\pi\right)\int \frac {dp}p.
\end{equation}
The range of integration is determined by $p$ being in the range
$\sqrt{na}\ll p \ll a^{-1}$ and with logarithmic accuracy $\sqrt{na}$
and $a^{-1}$ may be used as the actual limits of integration.

Using the relation (\ref{eq:mubfrome0}) between the ground state
energy and the chemical potential as well as the particle number
equation (\ref{eq:nopartic}) the result of the usual dilute Bose gas
\cite{Wu1959,Sawada1959} is reproduced
\begin{eqnarray}
\frac{\tilde E_0}V & = & 16\pi^2\frac{n_{\rm b}^3a_{\rm b}^4}{m_{\rm b}}\left(
\frac43-\frac{\sqrt3}\pi\right)\log(n_{\rm b}a_{\rm b}^3),\\
\tilde \mu_{\rm b} & = & 48\pi^2\frac{n_{\rm b}^2a_{\rm b}^4}{m_{\rm b}}\left(
\frac43-\frac{\sqrt3}\pi\right)\log(n_{\rm b}a_{\rm b}^3).
\end{eqnarray}
It has been assumed that $n_{\rm b}$ does not contain terms of order
$\Delta_0^4\log\Delta_0$. Arguments similar to those given above show
that this assumption is indeed correct.

In conclusion, the LHY and WS terms in the low density expansion arise
from non-trivial renormalizations due to the infrared divergences of
molecular propagators. At low momenta the fermionic atoms in the BEC
regime are nearly constants and their dynamics do not affect the
low-density expansion of the chemical potential and ground state
energy at the order considered here. However, from the above
discussion, and also from the Hugenholtz-Pines relation
(\ref{eq:HP2}), it is obvious that the presence of fermions will
affect next order quantities in a non-trivial way.

\chapter{Two-component Fermi gases with a mass
  imbalance \label{chap:mass}}

Since the experimental discovery of the BCS-BEC crossover in ultracold
mixtures of atoms in two different hyperfine states of e.g. $^{40}$K
\cite{Ticknor2004}, a major effort has gone into creating other more
exotic superfluids. An example is a system with a mass imbalance,
where instead of studying two different hyperfine states of the same
fermionic atom, a mixture of two different atomic species is
used. While experiments in these mixtures are still in their infancy,
these systems have received considerable theoretical attention
\cite{Petrov2003,Petrov2005a,Iskin2007}. The interest in these systems
has several reasons. For instance, the advent of the use of optical
lattices in ultracold gases has opened up the possibility of
continuously changing the effective mass of one of the species
independently of the other, since the resonant frequencies of the
atomic species will be different.

One might naively expect that the presence of a mass imbalance would
serve only to modify the few-body coupling constants computed in
chapter \ref{chap:bcsbec}. However, it turns out that beyond a certain
mass ratio, the physical picture becomes quite different. It was
discovered in Ref. \cite{Petrov2004a} that, as the mass imbalance is
increased, collisional losses of the Fermi gas also increase. For
small mass imbalances, the gas becomes more stable as a Feshbach
resonance is approached. Conversely, beyond a mass ratio of 12.33 the
system becomes less stable as a Feshbach resonance is approached and
finally, beyond the mass ratio 13.6, the three body system was found
to display bound states in a manner very similar to the Efimov
phenomenon in systems of three bosons \cite{Efimov1971}.

For moderate mass ratios, the main effect of introducing a mass
imbalance is indeed to change the few-body coupling constants. Results
for the three and four fermion problems, calculated in coordinate
space using the Bethe-Peirls method, have been obtained in
Refs. \cite{Petrov2003,Petrov2005b}. The results for four fermions
have been confirmed by Monte Carlo methods in
Ref. \cite{Vonstecher2007}. Thus, the results in this chapter for
three- and four-body scattering calculated using the momentum space
formalism are not new, rather they confirm already known results by
the use of a complimentary method.

Even though the Fermi energies will be different, in heteronuclear
mixtures with equal densities of the two atomic species, the Fermi
momenta will be equal. This means that the usual Cooper pairing is
still taking place on the BCS side and the picture of a BCS-BEC
crossover is valid. In this chapter the BEC regime with equal
densities will be examined.

\section{Molecular propagator}
The formalism presented here is very similar to that presented in
chapter \ref{chap:bcsbec}. Consider two species of fermionic atoms
interacting close to a heteronuclear Feshbach resonance. For
simplicity of notation, even though the particles are not different
hyperfine states of the same isotope, the particles will be called
$\up$ and $\down$. Let the particles have masses
\begin{equation}
m_\uparrow\equiv \gamma m, \hspace{1cm}
m_\downarrow\equiv m
\end{equation}
and chemical potentials $\mu_\uparrow, \mu_\downarrow<0$.

To calculate the molecular propagator, the polarization bubble which
appears in the repeated scattering between the two species of atoms is
needed (see Fig. \ref{fig:bubble}). The polarization bubble has the
value
\begin{eqnarray}
\Pi(p,\omega) & = & i\int\frac{d^4q}{(2\pi)^4}G_\uparrow(q+p/2)G_\downarrow
(-q+p/2) \nn \\
& = & \int\frac{d^3q}{(2\pi)^3}\frac1{\omega-(q+p/2)^2/2m_\uparrow-
(q-p/2)^2/2m_\downarrow+\mu_\uparrow+\mu_\downarrow+i0} \nn \\
& = & -\frac{m_{\rm r}\Lambda^2}{\pi^2}+\frac{m_{\rm r}^{3/2}}{\sqrt2\pi}
\sqrt{-\omega+p^2/2m_{\rm b}-\mu_\uparrow-\mu_\downarrow-i0},
\end{eqnarray}
where it has been used that $q$ can be shifted such that the angular
integration becomes trivial (let $q\to
q+\frac12\frac{m_\uparrow-m_\downarrow}{m_\uparrow+m_\downarrow}p$). Here
\begin{equation}
m_{\rm r} = \frac{m_\uparrow m_\downarrow}{m_\uparrow+m_\downarrow}
\end{equation}
is the reduced mass and $m_{\rm b}=m_\uparrow+m_\downarrow$ is the
mass of the molecule.

The scattering length between the two species of atoms is related to
the {\it vacuum} polarization bubble (for which
$\mu_\uparrow=\mu_\downarrow=0$) by
\begin{equation}
  a=\frac{m_{\rm r}}{2\pi}T(0)=\frac{m_{\rm r}}{2\pi}\frac{-\lambda}
{1+\lambda\Pi(0)} =\left(-\frac{2\pi}{m_{\rm r}\lambda}+\frac{2\Lambda}\pi
\right)^{-1}.
\end{equation}
The molecular propagator is then given by
\begin{equation}
D(p,\omega) = \frac{-\lambda}{1+\lambda\Pi(p,\omega)}
=\frac{2\pi}{m_{\rm r}}\frac1{a^{-1}-\sqrt{2m_{\rm r}}\sqrt{-\omega
+p^2/2m_{\rm b}-\mu_\uparrow-\mu_\downarrow-i0}},
\end{equation}
which reduces to the usual form (\ref{eq:bosonprop}) in the limit of
equal masses. According to Hugenholtz and Pines \cite{Hugenholtz1959},
the molecular propagator should have a pole at $p,\omega=0$. This
leads to the relation
\begin{equation}
E_{\rm b}=\mu_\uparrow+\mu_\downarrow=-\frac1{2m_{\rm r}a^2}
\end{equation}
for the binding energy of the molecule.

In the limit of small momenta and energies, $p\ll a^{-1}$ and
$\omega\ll 1/2m_{\rm r}a^2$, the underlying structure of the molecule is not
probed. In this limit the propagator reduces to
\begin{equation}
D(p,\omega)\approx\frac Z{\omega-p^2/2m_{\rm b}+i0}.
\end{equation}
$Z=\frac{2\pi}{m_{\rm r}^2a}$ is the residue and the remainder is the free
propagator of a particle with mass $m_{\rm b}=m_\uparrow+m_\downarrow$ as
expected.

\section{Three-body problem}
\begin{figure}[bt]
\begin{center}
\includegraphics[height=.95 in]{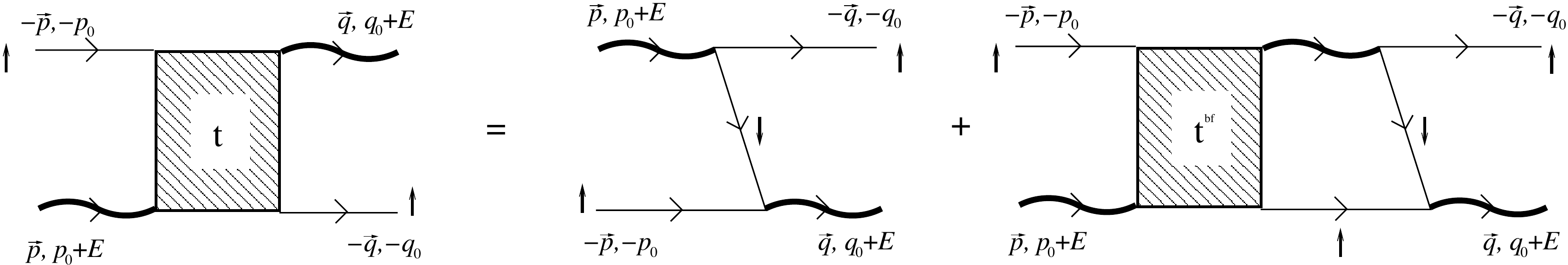}
  \caption{The integral equation satisfied by the atom-molecule
    scattering $t$-matrix.}
\label{fig:fbmass}
\end{center}
\end{figure}
Assume an atom of type $\up$ scatters off the bound molecular
state. The general integral equation for the atom-molecule scattering
matrix with kinematics as shown in Fig. \ref{fig:fbmass} is given by
\begin{equation}
  t(\vec p,p_0;\vec q,q_0) = -G_\down(p+q+E)-i\int \frac{d^4Q}{(2\pi)^4}
G_\down(Q+q+E)G_\up(-Q)D(Q+E)t(\vec p,p_0;\vec Q,Q_0).
\label{eq:3bodygeneral}
\end{equation}
$E$ is here used both as the total energy and as the total
four-momentum $(\vec 0,E)$. The symmetry factors are the same as in
the mass-balanced case.

As in chapter \ref{chap:bcsbec}, in the notation for the $t$-matrix
the dependence on the total energy has been suppressed. Also, $E\leq
E_{\rm b}$ in order for processes in which the molecule and atom scatters
into a final state of three free atoms to be forbidden.

The frequency $Q_0$ can be integrated out by closing the contour in
the upper half plane, where the only pole is the pole of
$G_\up(-Q)$. This results in
\begin{equation}
  t(\vec p,p_0;\vec q,q_0) = -G_\down(p+q+E)-\int \frac{d^3Q}{(2\pi)^3}
  G_\down(Q+q+E)D(Q+E)t(\vec p,p_0;\vec Q,Q_0),
\end{equation}
where $Q_0=-Q^2/2m_\up$.

Consider first the $s$-wave scattering length which is proportional to
$t(\vec 0,0;\vec 0,0)$ evaluated at $E=E_{\rm b}$. The incoming momentum and
frequency may be taken to vanish but the outgoing should be kept
finite in order to have a solvable integral equation. The result is
the equation
\begin{eqnarray}
t(p) & = & \frac{2m_{\rm r}a^2}
{1+p^2}
\nn \\ && \hspace{-10mm}
+\frac{1+\gamma^{-1}}{2\pi p}\int_0^\infty dq \frac{q\,t(q)}
{1-\sqrt{1+q^2\left[1-\gamma^2/(1+\gamma)^2\right]}}
\log\frac{1+p^2+q^2+2qp/\left(1+\gamma^{-1}\right)}
{1+p^2+q^2-2qp/\left(1+\gamma^{-1}\right)}, \nn \\ &&
\label{eq:inteqmass}
\end{eqnarray}
where momenta are measure in units of inverse scattering length. Here
the integral equation for $t(p)\equiv t(\vec 0,0;\vec p,-p^2/2m_\up)$ is
solved with $p_0=-p^2/2m_\up$ to make the dependence on momentum the
same on both sides of the equation. This point has been discussed in
detail in section \ref{sec:3body}.

\begin{figure}[bt]
\begin{center}
\includegraphics[height=2.4 in]{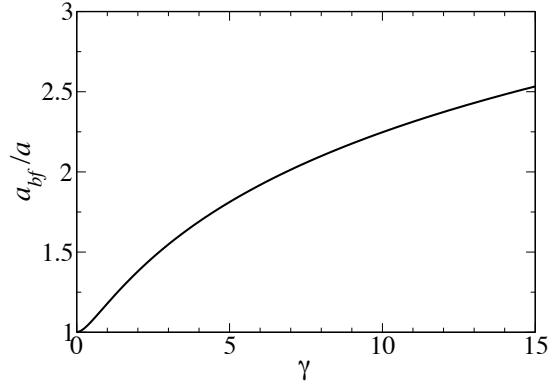}
  \caption{$a_{\rm bf}/a$ as a function of
    $\gamma\equiv m_\uparrow/m_\downarrow$.}
\label{fig:abfmass}
\end{center}
\end{figure}

Having solved Eq. (\ref{eq:inteqmass}) at a certain mass ratio, to
calculate the scattering amplitude each external molecular propagator
has to be renormalized by the square root of the residue of the pole
of the molecular propagator. The scattering amplitude at vanishing
momentum is thus
\begin{equation}
T(0) = Z\,t(0).
\end{equation}
The relationship between the scattering amplitude and the scattering
length is
\begin{equation}
T(0) = \frac{2\pi}{m_{3\rm r}}a_{\rm bf},
\end{equation}
with $m_{3\rm r}=\frac{m_{\rm b}m_\up}{m_{\rm b}+m_\up}$ being the
three-body reduced mass. The three-body scattering length is found to
be
\begin{equation}
a_{\rm bf}=\frac{m_{3\rm r}}{m_{\rm r}}\frac{t(0)}{m_{\rm r}a^2}a.
\end{equation}
The result is shown in Fig. \ref{fig:abfmass} and exactly matches
the result of Petrov \cite{Petrov2003}.

\subsection{$p$-wave scattering and collisional
  relaxation \label{sec:relaxation}}
Up to now the only bound molecular two-atom state considered has been
the weakly bound and large state of size comparable with the
scattering length $a$. This state is the state appearing at a Feshbach
resonance and is the highest ro-vibrational state. Additionally, in
the two body problem there are tightly bound states of size $R_e$, the
scale at which short distance physics becomes important. Through
collisions the weakly bound molecules can fall into these deeply lying
states and in the process a binding energy of order $1/(mR_e^2)$ is
converted into kinetic energy. The released energy is quite large and
the colliding atoms will escape the system. This process is called
collisional relaxation and is the process which determines the
lifetime of the Bose gas of weakly bound molecules \cite{Petrov2004a}.

The rate at which the collisional relaxation occurs is conventionally
written as either
\begin{equation}
\dot n_{\rm b} = -\alpha_{\rm rel}n_{\rm b}^2
\label{eq:relax}
\end{equation}
or
\begin{equation}
\dot n_{\rm b} = -\beta_{\rm rel}n_{\rm b}.
\end{equation}
Both versions are natural; for the first, two molecules need to
approach each other closely for the relaxation process to occur and
$\alpha_{\rm rel}$ is the probability for this to occur. For
the second $\beta_{\rm rel}^{-1}$ is the characteristic
relaxation time. Below, the result will be presented in the form
(\ref{eq:relax}).

For equal masses of the constituent fermions the relaxation rate
$\alpha_{\rm rel}$ decreases with increasing $a$
as \cite{Petrov2004a}
\begin{equation}
  \alpha_{\rm rel} \propto \frac{R_e}{m}
  \left(\frac{R_e}a\right)^{2.55}.
\end{equation}
This ensures the stability of the Fermi gas in the vicinity of a
Feshbach resonance as opposed to the case of Bose gases which become
unstable close to a Feshbach resonance. The reason for the stability
of the Fermi gas is that the relaxation process requires three
particles to approach each other within a distance of order $R_e$,
while the remaining particle in the molecule-molecule scattering will
typically be at a distance of order $a$. The probability for three
fermions to approach each other is suppressed because of the Pauli
principle, since two of the fermions will necessarily be identical.

\begin{figure}[bt]
\begin{center}
\includegraphics[height=1 in]{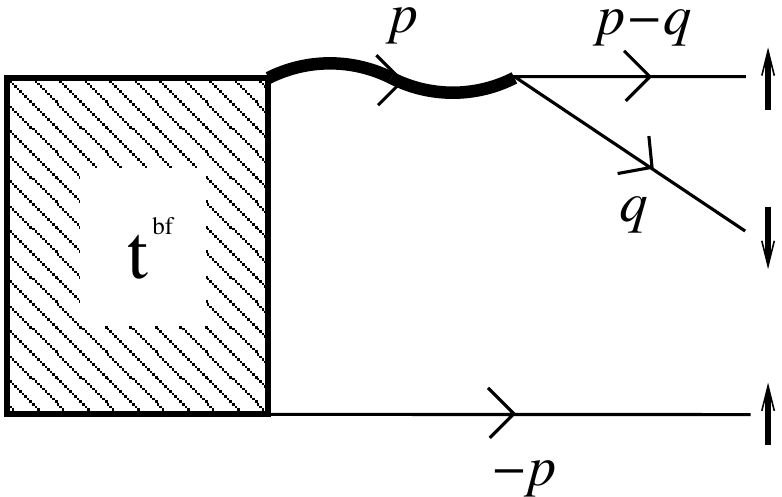}
  \caption{Relation between the scattering matrix and the scattered
    wave function.}
\label{fig:ttowave}
\end{center}
\end{figure}

It was shown in Ref. \cite{Petrov2003} that the leading relaxation
channel at large $a$ is scattering through the $p$-wave channel. In
general the relaxation rate may be written as
\begin{equation}
  \alpha_{\rm rel} \propto \frac{R_e}{m}
  \left(\frac{R_e}a\right)^s.
\label{eq:relaxrate}
\end{equation}
The power $s$ is a function of the mass imbalance. $s$ is related to
the power-law behavior $\psi\sim \rho^{\nu-1}$ satisfied by (the
$p$-wave part of) the three-body wave function in the range $R_e\ll
\rho \ll a$ by $s=2\nu+1$.  This can be seen from estimating the
probability for the three particles to be within a distance of $R_e$
of each other. This probability is
\begin{equation}
  \frac{\int_0^{R_e}d^3\rho \,|\psi(\rho)|^2}
  {\int_0^{a}d^3\rho \,|\psi(\rho)|^2}=\left(\frac{R_e}a\right)^{2\nu+1}.
\end{equation}
The relation between the outgoing part of the
wavefunction and (the $p$-wave part of) the $t$-matrix is
\begin{eqnarray}
\psi_{\rm out}(\vec r_1,\vec r_2,\vec r_3) & = & \int\frac{d^4p}{(2\pi)^4}
\int\frac{d^4q}{(2\pi)^4}D(p+E)t\left(\vec 0,0;\vec p,p_0\right)
\nn \\ && \hspace{0mm} \times
G_\up(p-q+E)G_\up(-p)G_\down(q)
e^{i\vec r_1\cdot (\vec p_1+\vec p_2)+i\vec r_2\cdot \vec p_2-
i\vec r_3\cdot \vec p_1}.
\end{eqnarray}
This relation is illustrated in Fig. \ref{fig:ttowave}.
If the $t$-matrix scales as $t(p)\propto p^\delta$ at large
momenta, $p\gg a^{-1}$, then
\begin{equation}
s = -3-2\delta.
\end{equation}

The collisional relaxation rate may be calculated by noting that the
scattering $t$-matrix can be written as the sum over partial waves
\begin{equation}
  t(\vec p,p_0;\vec q,q_0)=\sum_{\ell=0}^{\infty}(2\ell+1)
  P_\ell(\cos\theta)t^{(\ell)}(p,p_0;q,q_0).
\end{equation}
Here $\theta$ is the angle between $\vec p$ and $\vec
q$. $P_\ell(\cos\theta)$ are the Legendre polynomials of which the first
few are
\begin{eqnarray}
P_0(x) & = & 1, \nn \\
P_1(x) & = & x, \nn \\
P_2(x) & = & \frac12(3x^2-1). \nn
\end{eqnarray}
These satisfy the orthogonality relation
\begin{equation}
\int_{-1}^1P_{\ell'}(x)P_\ell(x)\,dx=\frac{2}{2\ell+1}\delta_{\ell'\ell}.
\label{eq:ortholegendre}
\end{equation}

To see that the integral equations for the partial waves decouple,
consider the general integral equation, Eq.
(\ref{eq:3bodygeneral}). Let $\theta$ denote the angle between $\vec
p$ and $\vec q$ and likewise $\phi$ the angle between $\vec q$ and
$\vec Q$. Then the angle $\gamma$ between $\vec p$ and $\vec Q$
satisfies
$\cos\gamma=\cos\theta\cos\phi+\sin\theta\sin\phi\cos\lambda$. With
this definition of angles, only the $t$-matrix under the integral
depends on the azimuthal angle and the corresponding integration may
be carried out as
\begin{equation}
\int_0^{2\pi}\frac{d\lambda}{2\pi}\,t(\vec p,p_0;\vec Q,Q_0)=
\sum_{\ell=0}^{\infty}P_\ell(\cos\theta)P_\ell(\cos\phi)t^{(\ell)}(p,p_0;Q,Q_0).
\end{equation}
Here the addition theorem for Legendre polynomials \cite{LL}
\begin{equation}
P_\ell(\cos\gamma)=P_\ell(\cos\theta)P_\ell(\cos\phi)+\sum_{m=1}^\ell2
\frac{(\ell-m)!}{(\ell+m)!}
P_\ell^m(\cos\theta)P_\ell^m(\cos\phi)\cos m\lambda
\end{equation}
has been used. The integral equations are now decoupled by using the
orthogonality relation \ref{eq:ortholegendre} with the result
\begin{eqnarray}
\hspace{-10mm}  t^{(\ell)}(p,p_0;q,q_0) & = & -\frac12
\int_{-1}^1d(\cos\theta)\, P_\ell(\cos\theta)
G_\down(p+q+E)
\nn \\ && \hspace{-33mm}
-\frac1{4\pi^2}\int_0^\infty Q^2dQ
t^{(\ell)}(p,p_0;Q,Q_0)D(Q+E) 
\int_{-1}^1d(\cos\phi)\,P_\ell(\cos\phi)
G_\down(q+Q+E),
\label{eq:3decouple}
\end{eqnarray}
where $Q_0=-Q^2/2m_\up$.

\begin{figure}[bt]
\begin{center}
\includegraphics[height=2.4 in]{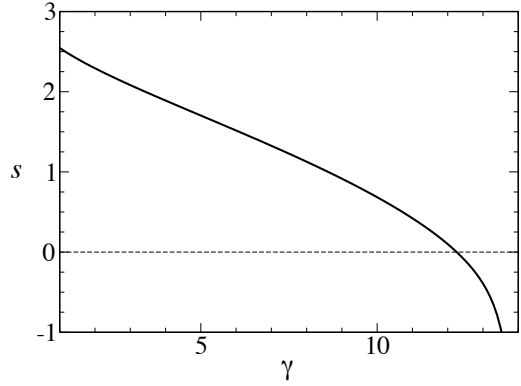}
\caption{The power $s$ appearing in the relaxation rate,
  Eq. (\ref{eq:relaxrate}), as a function of $\gamma$.}
\label{fig:powers}
\end{center}
\end{figure}

To solve the integral equation for the $p$-wave contribution, let
$q_0\to -q^2/2m_\up$. For simplicity let also $p_0\to -p^2/2m_\up$.
The $p$-wave contribution to the $t$-matrix satisfies
\begin{eqnarray}
&& \hspace{-18mm}t^p(p,-p^2/2m_\up;q,-q^2/2m_\up)
\nn \\ &
= &ma^2\left(\frac1{pq}
-\frac{(1+\gamma^{-1})(1+p^2+q^2)}{4p^2q^2}
\log\frac{1+p^2+q^2+2pq/(1+\gamma^{-1})}{1+p^2+q^2-2pq/(1+\gamma^{-1})}\right)
\nn \\ &&
+\frac{1+\gamma^{-1}}\pi\int_0^\infty Q^2dQ\frac{t(p,-p^2/2m_\up;Q,-Q^2/2m_\up)}
{1-\sqrt{1+Q^2\left[1-\gamma^2/(1+\gamma)^2\right]}}
\nn \\ && \times
\left(\frac1{qQ}
-\frac{(1+\gamma^{-1})(1+q^2+Q^2)}{4q^2Q^2}
\log\frac{1+q^2+Q^2+2qQ/(1+\gamma^{-1})}{1+q^2+Q^2-2qQ/(1+\gamma^{-1})}\right)
\label{eq:pwaveofswave}
\end{eqnarray}
at $E=E_{\rm b}$. Fig. \ref{fig:powers} shows the dependence of $s$ on the mass
imbalance found by solving Eq. (\ref{eq:pwaveofswave}). The curve
matches the result of Ref. \cite{Petrov2005b}.

It is seen that above a mass ratio of $m_\up/m_\down\approx12.3$ the
relaxation rate {\it increases} with increasing scattering
length. This is nicely explained in the Born-Oppenheimer picture
\cite{Petrov2005b}; in the case of a large mass ratio, if the two
heavy atoms are separated by a distance much smaller than $a$, then the
presence of the light atom will mediate an effective attractive
$\frac1{r^2}$ potential between the heavy atoms. On the other hand,
the Pauli principle results in the centrifugal barrier, which is a
repulsive $\frac1{r^2}$ potential. The mass imbalance of 12.3
corresponds to the cancellation of these potentials in the
Born-Oppenheimer picture. As the mass ratio is further increased, the
attractive potential is also increased and at a mass ratio of
approximately 13.6 one encounters the phenomenon of the fall to the
centre \cite{LL}: The system displays a behavior characteristic of the
Efimov phenomenon\cite{Efimov1971} where a hierarchy of three-body
bound states appear \cite{Petrov2005b}. The short distance physics
becomes very important and a short range three-body parameter is
needed in order to accurately describe the physics, see
e.g. Refs. \cite{Bedaque1998,Braaten2006}.

\section{Scattering of composite molecules}
Consider now the 4-body problem with a mass imbalance. The present
calculation follows closely the mass-balanced molecule-molecule
scattering calculation presented in section \ref{sec:4body}. Thus some
intermediate steps in the calculation will be skipped.

The governing integral equation is formally identical to the problem
of equal masses, Eq. (\ref{eq:4bodyinteq}), that is
\begin{equation}
t(p,p_0)
\hspace{-1mm}
=
\hspace{-1mm}
\Gamma(0,0;p,p_0)+\frac{i}{4\pi^3}
\hspace{-1mm}
\int q^2dq\,dq_0D(q,q_0+E/2)
D(q,-q_0+E/2)\Gamma(q,q_0;p,p_0)t(q,q_0).
\end{equation}
The difference from the previous calculation is that now both the
molecular propagator and the two-boson irreducible vertex, $\Gamma$,
depend on the mass-ratio.

Below, the convention is used that all atomic propagators have their
frequency shifted by a quarter of the energy going into the
scattering, while the molecules have their frequency shifted by half
the energy. That is
\begin{eqnarray}
G_{\up,\down}(p,p_0) & \equiv & \frac1{p_0-p^2/2m_{\up,\down}+E/4+i0}, \\
D(p,p_0) & = & \frac{2\pi}{m_{\rm r}}
\frac1{a^{-1}-\sqrt{2m_{\rm r}}\sqrt{-p_0-E/2
+p^2/2m_{\rm b}-i0}}.
\end{eqnarray}

The relation between $\Gamma$ and $\Xi$ is
\begin{eqnarray}
\Gamma(q,q_0;p,p_0) = \frac i2\int\frac{d^4Q}{(2\pi)^4}
G_\uparrow(p/2+Q)G_\downarrow(p/2-Q)\Xi(q;p/2+Q;p/2-Q),
\label{eq:gammaxi}
\end{eqnarray}
and the vertex $\Xi$ satisfies
\begin{eqnarray}
  \Xi(q;p/2+Q;p/2-Q) & = &\hspace{-2mm}
-\int\frac{d\Omega_{\vec q}}{4\pi}\left[
    G_\up(-p/2-q+Q)G_\down(-p/2+q-Q)+\left(q\leftrightarrow -q\right)
\right]
  \nn \\ && \hspace{-40mm}
-i\int\frac{d^4Q'}{(2\pi)^4}\left\{G_\down(p/2-Q')
    G_\up(-3p/2+Q')D(-p-Q+Q')\Xi(q;p/2+Q;p/2-Q')
\right. \nn \\ &&  \hspace{-35mm}\left.
+G_\up(p/2+Q')
    G_\down(-3p/2-Q')D(-p+Q-Q')\Xi(q;p/2+Q';p/2-Q)
  \right\}.
\end{eqnarray}

The vertex $\Xi(q;p/2+Q;p/2-Q)$ may again be split into parts
$\Xi^+(q;p/2+Q;p/2-Q)$ [$\Xi^-(q;p/2+Q;p/2-Q)$] analytic in the upper
[lower] half planes of $Q_0$, with
\begin{equation}
\Xi(q;p/2+Q;p/2-Q)=\Xi^+(q;p/2+Q;p/2-Q)+\Xi^-(q;p/2+Q;p/2-Q).
\label{eq:xisplitmass}
\end{equation}
To achieve this use Eq. (\ref{eq:gsquaresplit}). Now insert
Eq. (\ref{eq:xisplitmass}) into Eq. (\ref{eq:gammaxi}) and perform the
frequency integrations over simple poles (fermionic propagators). The
iterated terms in the integral equations for $\Xi^\pm$ evaluated at
the values of frequency dictated by the simple poles will then only
depend on an on-shell vertex
\begin{equation}
  \Xi(q;\vec p_1,p_1^2/2m_\up-E/4;\vec p_2,p_2^2/2m_\down-E/4) \equiv
\chi(q;\vec p_1,\vec p_2).
\end{equation}
Again, as in section \ref{sec:4body}, the on-shell frequencies would
have been obtained had $G_\up(p_1)$ and $G_\down(p_2)$ been attached
to the vertex $\Xi(q;p_1;p_2)$ and their frequencies been integrated
out using a contour around their simple poles.

The solution of the integral equation for the on-shell vertex $\chi$
is now possible with the substitution
\begin{equation}
p_0\to (\vec p/2+\vec Q)^2/2m_\up+(\vec p/2-\vec Q)^2/2m_\down-E/2.
\end{equation}
The equation for the on-shell vertex is
\begin{eqnarray}
\hspace{-7mm}
\chi(q;\vec p_1,\vec p_2) & = & -\int\frac{d\Omega_{\vec q}}{4\pi}\left[
G_\down(q-p_1)G_\up(-q-p_2)+(q\leftrightarrow -q)\right]
\nn \\ && 
\hspace{-17mm}
-\int\frac{d^3Q}{(2\pi)^3}\left\{\left.
G_\up(-p_1-p_2-Q)D(-Q-p_1)\chi(q;\vec p_1,\vec Q)\right|_{Q_0=Q^2/2m_\down-E/4}
\right.\nn \\ && \left.
\hspace{-1mm}+\left.G_\down(-p_1-p_2-Q)D(-Q-p_2)\chi(q;\vec Q,\vec p_2)
\right|_{Q_0=Q^2/2m_\up-E/4}\right\},
\label{eq:onshellchi}
\end{eqnarray}
where $(p_1)_0=p_1^2/2m_\up-E/4$ and $(p_2)_0=p_2^2/2m_\down-E/4$. 

The relation between $\Gamma$ and $\chi$ is
\begin{eqnarray}
\hspace{-10mm}
\Gamma(q,q_0;p,p_0) & = & \Gamma^{(0)}(q,q_0;p,p_0)
\nn \\ && \hspace{-30mm}
-\frac12\int\frac{d^3p_1}{(2\pi)^3} \frac{d^3p_2}{(2\pi)^3}
\left[G_\down(p-p_1)G_\up(-p-p_2)
+(p\leftrightarrow -p)\right]
D(-p_1-p_2)\chi(q;\vec p_1,\vec p_2).
\end{eqnarray}
Again $(p_1)_0=p_1^2/2m_\up-E/4$ and $(p_2)_0=p_1^2/2m_\down-E/4$ and
$p_0$ is a free parameter. $\Gamma^{(0)}$ is the result of calculating
the Born diagram,
\begin{eqnarray}
\Gamma^{(0)}(q,q_0;p,p_0) & = & -i\int\frac{d\Omega_{\vec q}}{4\pi}
\int\frac{d^4Q}{(2\pi)^4}
G_\up(Q+p/2+q/2)G_\up(Q-p/2-q/2)
\nn \\ && \hspace{20mm}\times
G_\down(-Q-p/2+q/2)G_\down(-Q+p/2-q/2)
\nn \\
& = & -2\int\frac{d\Omega_{\vec q}}{4\pi}
\int\frac{d^3Q}{(2\pi)^3}\frac{A}{(A^2-B^2)(A^2-C^2)},
\end{eqnarray}
with
\begin{eqnarray}
A & = & E/2-Q^2/2m_{\rm r}-p^2/8m_{\rm r}-q^2/8m_{\rm r}-\vec p\cdot \vec q/4m_\up+
\vec p\cdot\vec q/4m_\down, \\
B & = & p_0-\vec Q\cdot\vec p/2m_\up+\vec Q\cdot\vec p/2m_\down
-\vec Q\cdot\vec q/2m_{\rm r}, \\
C & = & q_0-\vec Q\cdot\vec q/2m_\up+\vec Q\cdot\vec q/2m_\down
-\vec Q\cdot\vec p/2m_{\rm r}.
\end{eqnarray}

\begin{figure}[bt]
\begin{center}
\includegraphics[height=2.4 in]{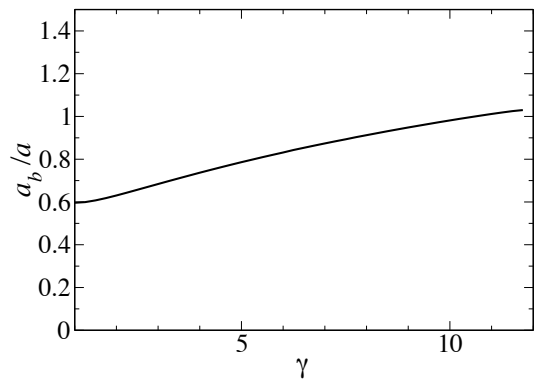}
\caption{The scattering length $a_{\rm b}$ as a function of
  mass-imbalance, $\gamma=m_\up/m_\down$. Above a mass ratio of about
  $\gamma\approx10$ convergence becomes very slow and numerical errors
  quite large. The curve is thus only plotted to $\gamma=12$.}
\label{fig:4bodymass}
\end{center}
\end{figure}

Finally, to relate the scattering amplitude to the scattering
$t$-matrix at vanishing incoming energy and momentum use
\begin{equation}
T(0)=Z^2t(0),
\end{equation}
while the scattering amplitude is related to the scattering length by
\begin{equation}
T(0)=\frac{2\pi}{M_{\rm r}}a_b.
\end{equation}
$M_{\rm r}=\frac{1+\gamma}2m$ is the reduced molecular-molecular mass. The
result is shown in Fig. \ref{fig:4bodymass} and matches the result of
Refs. \cite{Petrov2005b,Vonstecher2007}.

\chapter{Superfluidity close to a $p$-wave resonance
  \label{chap:pwavesetup}}

Attention will now be turned to atomic Fermi gases interacting close
to a $p$-wave Feshbach resonance. At low energies, because of the
angular momentum barrier, typically $s$-wave scattering will
dominate. However, in gases consisting of identical fermionic atoms,
the Pauli exclusion principle prevents $s$-wave scattering. At low
temperatures in these systems, $p$-wave scattering will dominate over
higher angular momentum scattering, and the study of $p$-wave
resonantly paired Fermi gases is therefore of fundamental interest.

The $p$-wave resonantly coupled superfluid state has not yet been
observed experimentally in degenerate atomic gases, but $p$-wave
Feshbach resonances have been observed in both $^{40}$K
\cite{Ticknor2004} and $^6$Li \cite{Schunck2005}. Considerable
theoretical attention has been given to the subject
\cite{Ho2005,Botelho2005,Gurarie2005,Cheng2005,Gurarie2007a},
particularly since it was pointed out \cite{Gurarie2005} that the
two-dimensional $p$-wave superfluids may display phases which are
candidates for quantum computing systems. Below, only
three-dimensional systems will be considered.

The gas of $p$-wave resonantly paired fermions is in many ways a
richer system than the $s$-wave system considered in the previous
chapters. In the $s$-wave paired superfluid, the BCS and BEC phases
were merely different limits of the same phase, but this is no longer
true in $p$-wave systems. The order parameters now correspond to
different projections of angular momentum and distinct
symmetries. This allows for the possibility of genuine phase
transitions which in some cases can even be topological
\cite{VolovikBook,Read2000,Botelho2005,Gurarie2005} and the phase
diagram becomes much richer.

An interesting theoretical feature of the $p$-wave Feshbach resonances
is that they are naturally narrow, whereas most $s$-wave resonances
studied in experiments are wide. This means that a quantitatively
correct perturbative approach may be used across the whole crossover
\cite{Gurarie2007a}. However, as discussed below, the Feshbach
resonances fall into categories of weak and strong, and while the weak
resonances are amenable to a mean field treatment, as performed in
Ref. \cite{Gurarie2007a}, the strong resonances must be treated more
carefully. In particular, it will be shown in chapter \ref{chap:pwave}
that the system close to a strong resonance will contain three-body
bound states (trimers), an effect first noticed by Y.~Castin and
co-workers \cite{CastinPrivate}.

In this chapter, the two-channel model for $p$-wave Feshbach
resonances will be introduced and its parameters related to those of
basic scattering theory. Next the important distinctions between wide
and narrow and between weak and strong Feshbach resonances will be
discussed.

\section{A two-channel model for $p$-wave resonances}
Consider a system of $N$ identical (atoms in the same hyperfine state)
spin-less fermions of mass $m$. At low energies, the scattering will
proceed entirely in the $p$-wave channel, since $s$-wave scattering is
excluded due to the Pauli principle and higher order scattering is
suppressed. As in the $s$-wave problem, it is convenient to describe
the system by a two-channel model. For the $p$-wave resonantly coupled
superfluid the two-channel Hamiltonian is given by
\cite{Timmermans1999,Timmermans2001,Holland2001,Gurarie2005,Cheng2005,Gurarie2007a}
\begin{equation}
H = \sum_{p} \frac{p^2}{2m} \hat a^\dagger_{\vec p}
\,\hat a_{\vec p} + \sum_{{\vec q}, \mu} \left(\epsilon_0 +
{\frac{q^2}{4m}} \right)
\hat b_{\mu,\,{\vec q}}^\dagger\, \hat b_{\mu,\, {\vec q}}
+\sum_{{\vec p},{\vec q},\mu} ~{g ( \left| \vec p\right|)
\over \sqrt{V}} \left( \hat b_{\mu,\,
{\vec q}} ~p_\mu~ \hat a^\dagger_{{\vec q\over 2}+\vec p}
~\hat a^\dagger_{{\vec q
\over 2}-\vec p} + h. c. \right).
\label{eq:pwavehamiltonian}
\end{equation}
The operators $\hat a_{\vec p}^\dagger$, $\hat a_{\vec p}$ are
creation and annihilation operators of the fermionic atoms of momentum
${\vec p}$. Similarly, $\hat b_{\mu,\, {\vec p}}^\dagger\,$, $\hat
b_{\mu,\, {\vec p}}$ are creation and annihilation operators of the
bosonic spin 1 molecules with the vector index $\mu$ being the spin
projection on some axis. The coupling $g(|{\vec p}|)$ appears as the
amplitude for the transition from a pair of identical fermions with
relative momentum $\vec p$ and orbital angular momentum 1 into the
bare closed-channel molecular state with internal angular momentum 1.

The superfluid described by the two-channel Hamiltonian
(\ref{eq:pwavehamiltonian}) is controlled by four parameters. The
first of these is the bare detuning $\epsilon_0$. This may in
principle have a directional dependence, however for simplicity it
will be taken constant in this thesis. The second parameter is the
particle number $N$, which is the expectation value of the operator
\begin{equation}
\hat N = \sum_p \hat a^\dagger_{\vec p}\, \hat a_{\vec p} +
2 \sum_{\mu, {\vec q}} \hat b^\dagger_{\mu,\, \vec q}\, \hat b_{\mu,\, \vec q}.
\end{equation}
It is convenient to trade the particle number for the energy scale
\begin{equation}
\epsilon_F=\left(\frac{6\pi^2N}V\right)^{2/3}/2m.
\end{equation}
In the absence of the Feshbach molecules this would have been the
Fermi energy. The remaining two parameters are contained in the
coupling constant $g(|{\vec p}|)$. Taking $g$ constant would imply that
the interactions take place at a point in space, as if the molecules
were infinitesimally small. However, in practice the molecules have a
finite size set by the range of the interactions, $R_e$. $g(|{\vec
  p}|)$ is in turn proportional to the wave function of the molecule
in momentum space and should be taken approximately constant for $p\ll
\Lambda\sim\frac1{R_e}$ while quickly dropping to zero for $p\gg
\Lambda$. Thus define a cut-off function
\begin{equation}
g(|{\vec p}|) \equiv g \,\xi(|{\vec p}|).
\label{eq:hardcut}
\end{equation}
In this introductory chapter to the physics of $p$-wave resonantly
coupled superfluids, the simplest cut-off function, a step function,
will be used. This choice is
\begin{equation}
\xi(|{\vec p}|)=\Theta(1-|{\vec
  p}|/\Lambda),
\label{eq:cutsimple}
\end{equation}
which is 1 for $|{\vec p}|<\Lambda$ and 0 above. In
chapter \ref{chap:pwave} both this and a ``softer'' cut-off will be
investigated.

\subsection{The diagrammatic approach}
The propagator of fermionic atoms is
\begin{equation}
G(p,\omega) = \frac1{\omega-p^2/2m+i0}.
\end{equation}
From the Hamiltonian (\ref{eq:pwavehamiltonian}), the propagator of
the bare closed-channel molecule may be read off to be
\begin{equation}
D_{0,\mu\nu}(p,\omega) = \frac{\delta_{\mu\nu}}{\omega-\epsilon_0-p^2/4m+i0}.
\end{equation}
As in chapter \ref{chap:bcsbec}, the bare molecular propagator needs
to be corrected by self energy insertions as shown in
Fig. \ref{fig:bosonprop}. The physical dressed propagator is then
\begin{equation}
D_{\mu\nu}(p,\omega) = \left(\frac1{D_0^{-1}(p,\omega)-\Sigma(p,\omega)}
\right)_{\mu\nu}.
\label{eq:dressedp}
\end{equation}
The value of the self energy bubble is
\begin{eqnarray}
\Sigma_{\mu\nu}(p,\omega) & = & \nn \\ && \hspace{-22mm}
2ig^2\int\frac{dq_0}{2\pi}\frac{d^3q}{(2\pi)^3}
\frac{q_\mu
  q_\nu\,\xi^2(|\vec q|)}{\left[\omega/2+q_0-(q+p/2)^2/2m+i0\right]
\left[\omega/2-q_0-(q-p/2)^2/2m+i0\right]}. \nn \\ &&
\end{eqnarray}
The factor 2 appears as the appropriate combinatorial factor; in the
$s$-wave problem, the operators giving the propagators of the two
distinguishable fermionic atoms inside the bubble could be contracted
in only one way, in the present case there are two possible
ways. Notice how the physical molecular propagator will in general
depend on the chosen method of cut-off. To proceed, the cut-off
function needs to be specified. Using the simplest cut-off,
Eq. (\ref{eq:cutsimple}), the self energy becomes
\begin{eqnarray}
\hspace{-5mm}\Sigma_{\mu\nu}(p,\omega)
& = & \frac23g^2\delta_{\mu\nu}\frac1{2\pi^2}\int_0^\Lambda
  dq\frac{q^4}{\omega-q^2/m-p^2/4m+i0} \nn \\
& = & \delta_{\mu\nu}\left\{-c_1+c_2(q^2/4m-q_0-i0)
\vphantom{\tan^{-1}
\frac\Lambda{\sqrt{q^2/4-mq_0-i0}}}
\right.\nn\\ && \left. \hspace{8mm}
-c_2\frac{\sqrt
    m}{\Lambda}(q^2/4m-q_0-i0)^{3/2}\tan^{-1}
\frac\Lambda{\sqrt{q^2/4-mq_0-i0}}\right\}.
\label{eq:pwavebubble}
\end{eqnarray}
The integral scales as $q^3$ for large momenta. Thus, two cut-off
dependent parameters have been defined,
\begin{equation}
c_1=\frac{mg^2\Lambda^3}{9\pi^2}, \hspace{10mm}
c_2=\frac{m^2g^2\Lambda}{3\pi^2}.
\label{eq:c1andc2}
\end{equation}
$c_1$ has dimensions of energy while $c_2$ is a very important
dimensionless parameter as will become clear below.

The propagator of bosons becomes
\begin{eqnarray}
D_{\mu\nu}(q,q_0) & = & \nn \\ && \hspace{-12mm}
\frac{\delta_{\mu\nu}}{(1+c_2)(q_0-q^2/4m-\omega_0+i0)+c_2\frac{\sqrt m}
\Lambda(q^2/4m-q_0-i0)^{3/2}\tan^{-1}
\frac\Lambda{\sqrt{q^2/4-mq_0}}}, \nn \\ &&
\label{eq:molpropp}
\end{eqnarray}
which is diagonal in spin indices. It is seen that $c_1$ disappears
from the theory, it is absorbed in the {\it physical
  detuning}\footnote{In the two-channel $s$-wave model, a coefficient
  proportional to the cut-off appeared to shift the bare detuning to
  the physical detuning, see Eq. (\ref{eq:physdetuning}). No analogue
  of $c_2$ is present in the $s$-wave problem.} as
\begin{equation}
\omega_0 = \frac{\epsilon_0-c_1}{1+c_2} \leq0.
\end{equation}
$\omega_0$ is an important parameter in the theory; a negative
$\omega_0$ implies the presence of a physical bound state in the
two-body system. On going through $\omega_0=0$ the bound state
disappears and is replaced by a resonance.

\subsection{$p$-wave scattering}
At low energies, the scattering amplitude between two identical
fermionic atoms may be approximated by [see Eq. (\ref{eq:partialwav})]
\begin{equation}
f_p(k) = \frac{k^2}{-v^{-1}+\frac12k_0k^2-ik^3},
\end{equation}
with $k$ being the relative momentum of the two scattering
atoms. Whereas the $s$-wave scattering amplitude $f_s(k)\to -a$ as
$k\to0$, $f_p(k)\to0$ in the low-energy limit, demonstrating how
$s$-wave scattering, if allowed, would dominate the scattering
problem. The scattering volume $v$ is the analogue of the $s$-wave
scattering length, diverging and changing sign across the Feshbach
resonance. The parameter $k_0$ is a characteristic momentum and is
analogous to the effective range $r_0$.

In the model above, with the choice of cut-off function
Eq. (\ref{eq:hardcut}), the scattering amplitude is
\cite{Gurarie2005,Gurarie2007a}
\begin{equation}
f_p(k) = \frac1{\frac{6\pi}{mg^2k^2}(1+c_2)\left(\omega_0-
\frac{k^2}m\right)-ik},
\label{eq:pwavescatamp}
\end{equation}
from which it is possible to identify
\begin{equation}
  v = - \frac{m g^2}{6 \pi \left(1+c_2 \right) \omega_0},
  \hspace{10mm}
  k_0 = - \frac{ 4 \Lambda \left(1 + c_2 \right)}{\pi c_2}.
\label{eq:par}
\end{equation}
Again it is seen how $\omega_0=0$ is the physical position of the
resonance, the point at which the scattering volume changes sign. For
$\omega_0<0$ the system has a low-lying bound state and as $\omega_0$
changes sign, the bound state turns into a resonance.

\section{Properties of $p$-wave Feshbach resonances}
In the finite density atomic gas there will be three length scales,
and thus it is possible to construct two dimensionless parameters.
The interparticle separation is of order $1/\sqrt{m\epsilon_F}$,
a second length scale is provided by $m^2g^2$, while a third
is the range of the forces $R_e\sim1/\Lambda$. The first
dimensionless parameter is $c_2$ defined in Eq. (\ref{eq:c1andc2}). The
second is
\begin{equation}
  \gamma_{\rm p} = m^{3/2}g^2\sqrt{\epsilon_F}.
\end{equation}
In order for the results to be independent of the precise nature of
short-distance physics, the interparticle separation must be greater
than $R_e$. This translates into the requirement
\begin{equation}
  \gamma_{\rm p} \ll c_2.
\end{equation}

The value of $\gamma_{\rm p}$ determines the width of the Feshbach
resonance in the same manner as $\gamma_{\rm s}$ determines the width
of $s$-wave resonances. The main difference between these is that
while $\gamma_{\rm s}$ increases with decreasing density, the opposite
is true for $\gamma_{\rm p}$. Thus, $p$-wave resonances may be made
arbitrarily narrow by reducing the particle density and it is possible
to construct a pertubative theory, accurate across the Feshbach
resonance, for densities achievable in current experiments.

That the $p$-wave resonances are intrinsically narrow comes from the
presence of the centrifugal barrier, which adds a long ranged $1/r^2$
tail to the interatomic potential, as in $U_{\rm eff}(r) =
U(r)+1/m_rr^2$. The width of a low lying resonance may then be
estimated by computing the decay rate through the effective potential
and is found to be dominated by the presence of the long-ranged tail
\cite{Gurarie2007a}.

Note how, even after the shift to a physical detuning, the scattering
amplitude (\ref{eq:pwavescatamp}) still depends on the regularization
scheme through the parameter $c_2$. If $c_2\ll1$ then the ultraviolet
cut-off indeed drops out of the problem. However, if $c_2\gg1$ then
$k_0\sim-\Lambda$ and the bare molecule may be integrated out,
resulting in a $p$-wave single channel model. Feshbach resonances with
$c_2\ll1$ are called {\it weak} while those with $c_2\gg1$ are termed
the {\it strong} resonances. Unlike $\gamma_{\rm p}$, $c_2$ is an
intrinsic property of the Feshbach resonance in question and depends
only on the physics of the resonance, it cannot be changed simply by
changing the density.

Both of the parameters $c_2$ and $\gamma_{\rm p}$ control the
perturbative expansion in powers of the coupling constant $g$. The
usual approach is to apply mean field theory to the system described
by the Hamiltonian (\ref{eq:pwavehamiltonian}). This approach has been
used in Refs. \cite{Gurarie2005,Cheng2005,Gurarie2007a}. The mean
field approach relies on the smallness of the coupling $g$ which means
that both $\gamma_{\rm p}$ and $c_2$ must be small. This is not
necessarily the case since $c_2$ is an uncontrollable parameter.

The breakdown of mean field theory for strong $p$-wave resonances does
not imply that no other techniques exist with which to investigate the
system. Fluctuational corrections arise from two distinct regimes of
momenta, of the order of the inverse interparticle spacing and of the
order of the ultraviolet cut-off. Corrections to the many-body problem
are governed by the former and are small as long as $\gamma_{\rm p}$
is small. At momenta of the order of the ultraviolet cut-off only
virtual particles propagate and thus corrections to mean field theory
from large momenta must be few-body, resulting in a renormalization of
the few-body coupling constants.

The situation is quite similar to the wide resonance $s$-wave
superfluid studied in chapter \ref{chap:bcsbec}. Here, the
interactions between fermions were strong (and the scattering problems
studied dominated by momenta of order $a^{-1}$) and led to non-trivial
renormalizations of the few-body coupling constants, in particular it
was seen how $a_{\rm b}\approx0.60$. However, the many-body physics
was dominated by physics at small momenta ($p\ll a^{-1}$) in a
perturbative expansion in the gas parameter. As long as the gas
parameter was small, the mean field theory remained essentially valid,
with the few-body couplings non-trivially renormalized.

Finally, it should be mentioned that the $p$-wave Feshbach resonance
investigated in the experiment \cite{Gaebler2007} is narrow and
strong. This estimate was performed in Ref. \cite{Gurarie2007a}. The
existing mean field theory is valid for the narrow and weak Feshbach
resonances, and is in principle not valid for this particular Feshbach
resonance. The few-body physics leads to a non-trivial renormalization
of coupling constants, and indeed it also leads to bound states of
three identical fermionic atoms interacting close to a strong Feshbach
resonance \cite{Levinsen2007,Lasinio2007}. A reliable method with
which to investigate the strongly resonant $p$-wave condensates is
thus needed and this is the subject of the next chapter.

\chapter{Strongly resonant $p$-wave condensates
  \label{chap:pwave}}

While the weak and narrow Feshbach resonances are amenable to a mean
field treatment \cite{Gurarie2005,Cheng2005,Gurarie2007a}, it is the
strong and narrow $p$-wave Feshbach resonances which are most relevant
to current experiments and properties of these are calculated in this
chapter. The main results of this chapter were presented in
Ref. \cite{Levinsen2007}.

An interesting effect in the vicinity of a strong resonance is the
appearance of bound molecule-atom (trimer) states with angular
momentum 1. This effect was first noticed by Y. Castin and
collaborators \cite{CastinPrivate}. These trimer states are quite
unusual. They are very strongly bound, with a size of the order of the
short-disance physics, $R_e$, the same as the closed-channel
molecule. The binding energy is very large (of the order of
$\Lambda^2/m$) and only weakly dependent on detuning from the
resonance. Below, the binding energy of the trimer as a function of
the strength of the resonance is found. In particular, the critical
value of the strength of the resonance at which the trimer state
appears is calculated.

The two-channel Hamiltonian (\ref{eq:pwavehamiltonian}) is valid both
for weak and strong resonances. In a manner similar to the
$s$-wave case (see chapter \ref{chap:bcsbec}) it reduces to a
one-channel model in the limit of infinitely strong resonces, i.e. as
$c_2\to\infty$. However, it is interesting to investigate the manner
in which the physics evolves under the strengthening of the Feshbach
resonance, in particular the appearance of three-body bound states for
strong resonances. Thus the starting point will be the two-channel
Hamiltonian.

As discussed in the previous chapter, in order to study the
two-channel Hamiltonian (\ref{eq:pwavehamiltonian}), the coupling
constant between the open and closed channels needs to reflect the
suppression of the molecule wave-function outside the range of
short-distance physics. Different regularization schemes are
possible. The two choices of cut-off functions used below give results
which are in qualitative agreement, although they do not completely
agree quantitatively. Arguments will be given as to why the results
are still qualitatively trustworthy.

A recent experiment \cite{Gaebler2007} studied a gas of $p$-wave
Feshbach molecules in $^{40}K$. Unfortunately, the gas was found to be
quite short-lived, with a lifetime of about 2
ms. Ref. \cite{Gaebler2007} studied whether the short lifetime might
be due to dipolar relaxation but found that the lifetime was shorter
than predicted from losses due to dipolar relaxation alone. The
presence of the trimer state opens up for the possibility of inelastic
collisions in which the scattering of two molecules results in an atom
and a trimer with large kinetic energies. The lifetime of the $p$-wave
superfluid due to these decay processes is estimated below.

One question which is not addressed below is the nature of the true
ground state of the system. In order to properly answer this question,
the molecule-molecule scattering problem needs to be studied and the
few-body coupling constants computed, as in Ref. \cite{Gurarie2007a}
for weak resonances. A diagrammatic technique similar to the one
studied in section \ref{sec:4body} may be employed, but the solution
of the resulting set of integral equations will be technically very
difficult.

In this chapter, first the diagrammatical expansion used to study
few-body problems will be discussed. Next, the three-body problem will
be solved for different choices of the ultraviolet cut-off and it will
be studied how the system evolves under the strengthening of the
Feshbach resonance. Finally, the stability of the gas will be
considered.

\section{The three-body problem  \label{sec:3bodyp}}
In this section, the three-body problem consisting of three identical
fermionic atoms interacting close to a Feshbach resonance is
studied. It will be assumed that the system is tuned to the BEC side
of the resonance, that is
\begin{equation}
\omega_0=\frac{\epsilon_0-c_1}{1+c_2}\leq 0,
\end{equation}
and thus a physical bound two-body state exists.

\begin{figure}[bt]
\begin{center}
\includegraphics[height=1.4 in]{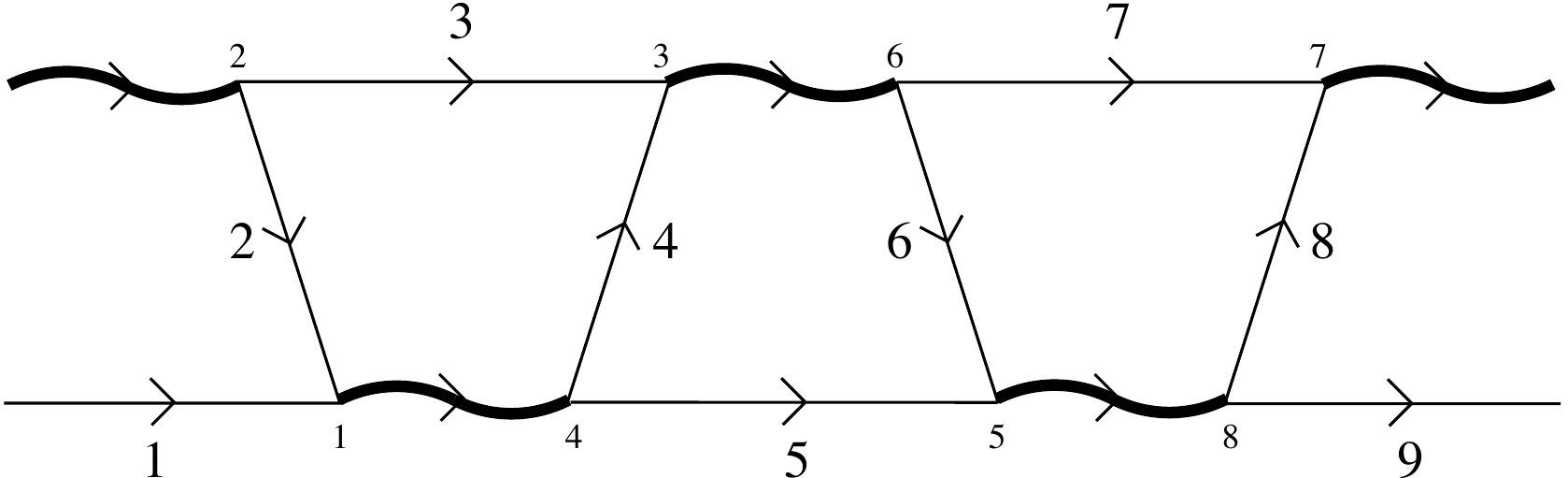}
\caption{The numbering of fermion propagators and vertices as
  explained in the text.}
\label{fig:numbering}
\end{center}
\end{figure}

It is convenient to explicitly write down the Feynman rules governing
the diagrammatic expansion. These may be derived from the Hamiltonian,
Eq. (\ref{eq:pwavehamiltonian}). In the Hamiltonian, each vertex
converting two atoms into a molecule (or vice versa) contains two
fermionic operators. Thus the vertices commute and they may be
arranged in any desired order. The atom-molecule scattering problem
contains precisely one line of fermion propagators.\footnote{This is
  also true for the two-boson irreducible vertex $\Gamma$ appearing in
  the molecule-molecule scattering problem of section
  \ref{sec:4body}. The Feynman rules derived here apply equally well
  to molecule-molecule scattering.} 
Associate a number with each fermion line, from incoming to outgoing,
and also a number with each vertex, as illustrated in
Fig. \ref{fig:numbering}. Then the Feynman rules are read off from the
Hamiltonian:
\begin{enumerate}[(i)]
\item At vertex $i$ where a molecule of spin $\mu$ is created
  associate a factor of $(p_{i+1}-p_i)_\mu/2$ where $p_i$ is the
  numbered fermion momentum. At vertex $i$ where a molecule is
  annihilated, associate a factor $-(p_{i+1}-p_i)_\mu/2$.
\item Fermion contractions give the same signs as in the $s$-wave
  problem. In the three-body problem this is a factor $(-1)^{n+1}$
  where $n$ is the number of loops.
\item Each vertex gives a combinatorial factor of 2 arising from
  indistinguishability of the fermions.
\item Each fermion line gives a factor $iG$ and each molecule a factor
  of $iD_{\mu\nu}$.
\item Finally, each vertex joining fermions of momenta $p_i$ and
  $p_{i+1}$ gives rise to a factor $ig(|\vec p_i-p_{i+1}|/2)\equiv
  ig\,\xi(|\vec p_i-\vec p_{i+1}|/2\Lambda)$.
\end{enumerate}

The atom-molecule scattering problem is studied in a similar manner to
the corresponding $s$-wave problem of section \ref{sec:3body}. The
kinematics is chosen as follows: An incoming molecule of spin $\mu$
has four-momentum $\left(\vec 0,E\right)$ and the incoming fermion
$(\vec 0,0)$. The outgoing molecule has spin $\nu$ and four-momentum
$\left(\vec p,p_0+E\right)$ and the outgoing atom $(-\vec p,-p_0)$. To
calculate the $s$-wave scattering length, the total energy $E$ reduces
to a function $\kappa_0(\omega_0)\leq0$ (this ensures that the
molecular propagator is on-shell and will be defined precisely below)
and to look for bound states the total energy must be taken less than
$\kappa_0$. This is similar to the $s$-wave problem where $E\leq
E_{\rm b}=-\frac1{ma^2}$.

\begin{figure}[bt]
\begin{center}
\includegraphics[height=1.4 in]{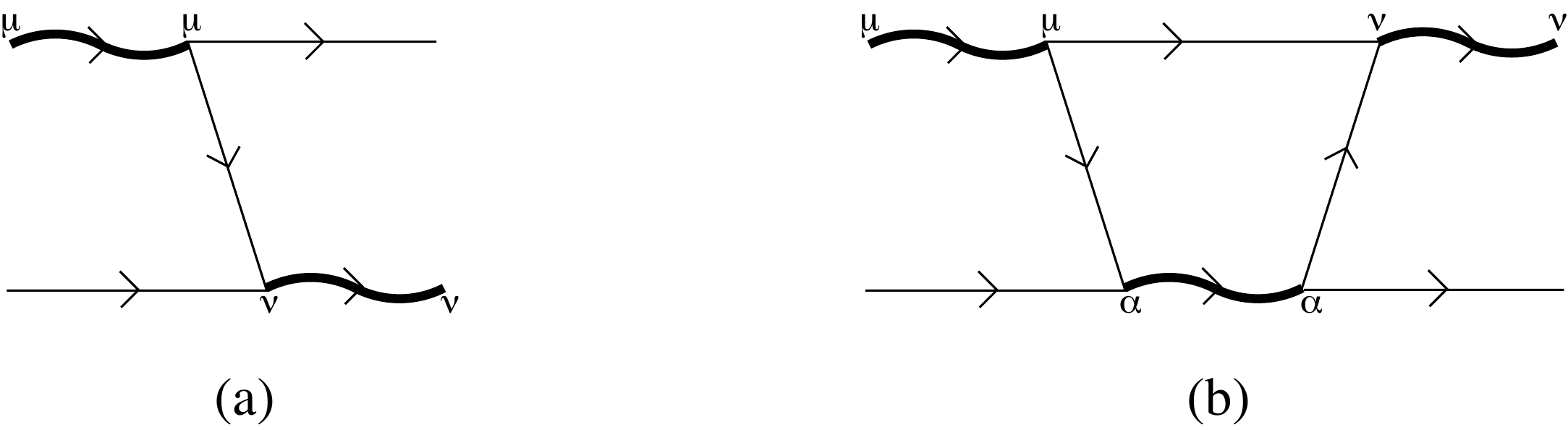}
\caption{The Born diagram (a) and the first loop diagram (b)
  contributing to atom-molecule scattering. The molecular spin has
  been made explicit while the kinematics is explained in the text.}
\label{fig:3bodybornp}
\end{center}
\end{figure}

According to the Feynman rules above, the contribution to the
scattering amplitude from the Born diagram of
Fig. \ref{fig:3bodybornp}a has the value
\begin{equation}
-iT^{(1)}_{\mu\nu}(\vec p,p_0) =
2iZ\,g^2
p_\mu p_\nu
G(\vec p,p_0+E)
\xi\left(|\vec p|/\Lambda\right)\xi\left(|\vec p|/2\Lambda\right).
\label{eq:pwaveborn}
\end{equation}
For proper normalization of the external propagators in the scattering
amplitude, the residue $Z$ of the molecular propagator at its pole is
included (as in the $s$-wave three-body problem, see section
\ref{sec:3body}). The Born diagram is proportional to $p_\mu p_\nu$
and goes to zero as $\vec p,p_0\to0$ (except in the case of scattering
at zero energy with vanishing detuning).

The first loop diagram, shown in Fig. \ref{fig:3bodybornp}b, takes the
value
\begin{eqnarray}
-iT^{(2)}_{\mu\nu}(\vec p,p_0) & = &
8Z\,g^4\int\frac{d^4q}{(2\pi)^4}
q_\mu q_\alpha
G(\vec q,q_0+E)G(-\vec q,-q_0)
\xi\left(|\vec q|/\Lambda\right)\xi\left(|\vec q|/2\Lambda\right)
\nn \\
&& \hspace{0mm}\times
D(\vec q,q_0+E)
G(\vec p+\vec q,p_0+q_0+E)
\nn \\ && \times(p+q/2)_\alpha(p/2+q)_\nu
\xi\left(|\vec p+\vec q/2|/\Lambda\right)
\xi\left(|\vec p/2+\vec q|/\Lambda\right).
\end{eqnarray}
Repeated spin indices indicate a summation over the corresponding
spin. The propagator of molecules has been defined as
$D_{\mu\nu}(q,q_0)\equiv \delta_{\mu\nu}D(q,q_0)$ where it has been
used that the molecular propagator is diagonal in spin indices
(independent of the particular choice of regularization).

Using the Feynman rules above, it is seen that the generalization to an
integral equation is
\begin{eqnarray}
\hspace{-6mm}T_{\mu\nu}(\vec p,p_0) & = & 
-2Z\,g^2G(\vec p,p_0+E)p_\mu p_\nu
\xi\left(|\vec p|/\Lambda\right)\xi\left(|\vec p|/2\Lambda\right)
 \nn \\
&& 
\hspace{-23mm}
-4ig^2\int\frac{d^4q}{(2\pi)^4}T_{\mu\alpha}(\vec q,q_0)D(\vec
q,q_0+E)G(-\vec q,-q_0)
\nn \\ && \hspace{-18mm}\times
G(\vec p+\vec
q,p_0+q_0+E)(p+q/2)_\alpha(p/2+q)_\nu
\xi\left(|\vec p+\vec q/2|/\Lambda\right)
\xi\left(|\vec p/2+\vec q|/\Lambda\right).
\end{eqnarray}
Exactly as in the $s$-wave problem, it is possible to integrate over
the loop frequency with the result $q_0=-q^2/2m$. Let $p_0=-p^2/2m$ to
achieve the same frequency dependence in both $T$-matrices of the
equation. Define the ``on-shell'' scattering amplitude
$T_{\mu\nu}(\vec p)\equiv T_{\mu\nu}(\vec p,-p^2/2m)$. In order to
proceed, it is advantageous to scale out the cut-off by measuring
momenta in units of the cut-off, $\Lambda$. Additionally, measure the
total energy in units of $\Lambda^2/m$. Then the integral equation is
\begin{eqnarray}
  T_{\mu\nu}(\vec p) & = & -2Z\,g^2m\,G(\vec
  p,-p^2/2+E)p_\mu p_\nu \nn \\ &&
\hspace{-10mm}
  -\frac{2m^2g^2\Lambda}{\pi^2}
  \int_0^\infty q^2dq\int\frac{d\Omega_{\vec q}}{4\pi}
  T_{\mu\alpha}(\vec q)D(\vec q,-q^2/2+E)
  G(\vec p+\vec q,-p^2/2-q^2/2+E) \nn \\ && \hspace{-5mm}\times
  (p+q/2)_\alpha(p/2+q)_\nu
  \xi\left(|\vec p+\vec q/2|\right)
  \xi\left(|\vec p/2+\vec q|\right).
\label{eq:pwaveinteq}
\end{eqnarray}
In writing this equation, for simplicity the dimensions of the
propagators $G$ and $D$ have been extracted and the remaining $G$ and
$D$ are dimensionless. The integration $\int d\Omega_{\vec q}$
denotes an integration over directions of $\vec q$.

The tensor $T_{\mu\nu}(\vec p)$ must have the general form
\begin{equation}
  T_{\mu\nu}(\vec p) = T^1(p)\delta_{\mu\nu}+T^2(p)\frac{p_\mu
    p_\nu}{p^2}\equiv \sum_{i=1,2}T^i(p)u^i_{\mu\nu}(\vec p),
\end{equation}
with a set of basis tensors defined as $u^1_{\mu\nu}(\vec
p)=\delta_{\mu\nu}$ and $u^2_{\mu\nu}(\vec p)=p_\mu p_\nu/p^2$. It is
then possible to derive two coupled integral equations for the
coefficients $T^1$ and $T^2$. Multiply Eq. (\ref{eq:pwaveinteq}) by
$u^j_{\mu\nu}(\vec p)$ and sum over spin indices. On the left hand
side a product of basis tensors appears:
\begin{equation}
u^j_{\mu\nu}(\vec p)u^i_{\mu\nu}(\vec p) = \left( \begin{array}{cc}
3 & 1 \\ 1 & 1 \end{array}\right)_{ji} \equiv U_{ji}.
\end{equation}
The matrix $U$ is clearly invertible. Upon further multiplying the
integral equation by the inverse matrix it becomes
\begin{eqnarray}
  T^j(p) & = & -2Z\,g^2m\,G(\vec
  p,-p^2/2+E)p^2\delta_{2,j} \nn \\ &&
  -\frac{2m^2g^2\Lambda}{\pi^2}
  \int_0^\infty q^2dq
D(\vec q,-q^2/2+E)a_{ji}(p,q,E)T^i(q),
\label{eq:pwaveinteq2}
\end{eqnarray}
which is a set of two coupled integral equations. The matrix $a$ is
dimensionless and is given by
\begin{eqnarray}
  a_{ji}(p,q,E) & = & U^{-1}_{jk}\int
  \frac{d\Omega_{\vec q}}{4\pi}
  G(\vec p+\vec q,-p^2/2-q^2/2+E)
  u^k_{\mu\nu}(\vec p)u^i_{\mu\alpha}(\vec q)
  \nn \\ && \times
  (p+q/2)_\alpha(p/2+q)_\nu
  \xi\left(|\vec p+\vec q/2|\right)
  \xi\left(|\vec p/2+\vec q|\right).
\label{eq:matrixa}
\end{eqnarray}
In order to proceed, it is necessary to specify the precise form of
the cut-off function $\xi$, as this affects both the integration
kernel, through the matrix $a$, and the molecular propagator.

\subsection{The hard cut-off \label{sec:hard}}
The coupling constant is now assumed to be constant up to the
ultraviolet cut-off after which it drops to zero. That is,
\begin{equation}
g(|\vec p|) \equiv g\,\Theta(\Lambda-|\vec p|),
\end{equation}
where $\Theta$ is the usual step function, and properties of the gas
calculated with this assumption.

With the hard cut-off, the molecular propagator was found in
Eq. (\ref{eq:molpropp}) to be
\begin{equation}
D(q,q_0) \hspace{-1mm}=\hspace{-1mm}
\frac{\delta_{\mu\nu}}{(1+c_2)(q_0-q^2/4m-\omega_0+i0)
\hspace{-1mm}+\hspace{-1mm}
\frac{c_2\sqrt m}
\Lambda(q^2/4m-q_0-i0)^{3/2}\tan^{-1}
\frac\Lambda{\sqrt{q^2/4-mq_0}}}.
\end{equation}
The physical on-shell molecule in the presence of the medium must have
a pole as its four-momentum vanishes \cite{Hugenholtz1959}. That is,
the molecule is affected by the detuning $\omega_0$ and should be
evaluated at a four-momentum $\left(\vec
  q,q_0+\kappa_0(\omega_0)\right)$ such that $D^{-1}\left(\vec
  0,\kappa_0(\omega_0)\right)=0$. From Eq. (\ref{eq:molpropp}) it is
seen that $\kappa_0(\omega_0)$ satisfies the implicit relation
\begin{equation}
\omega_0 = \kappa_0(\omega_0)+\frac{c_2}{1+c_2}\frac{\sqrt m}{\Lambda}
\left(-\kappa_0(\omega_0)\right)^{3/2}\tan^{-1}
\frac\Lambda{\sqrt{-m\kappa_0(\omega_0)}}.
\end{equation}
For $\omega_0\ll \frac{\Lambda^2}m$ the function $\kappa_0$ reduces to
the physical detuning.

The residue of the molecular propagator at its pole is also found from
the molecular propagator. It takes the value
\begin{equation}
Z = \left[1+c_2-\frac12\frac{c_2\kappa_0}{\Lambda^2/m-\kappa_0}
-\frac{3c_2\sqrt{-m\kappa_0}}{2\Lambda}\tan^{-1}
\frac{\Lambda}{\sqrt{-m\kappa_0}}\right]^{-1},
\end{equation}
and reduces to $Z\approx 1/(1+c_2)$ as $\omega_0\to0$.

\subsubsection{The simplest hard cut-off}
The solution of the integral equation (\ref{eq:pwaveinteq2}) requires
the determination of the matrix $a$, Eq. (\ref{eq:matrixa}). Note that
this matrix {\it does not} depend on the strength of the resonance,
$c_2$. It is instructive to first consider the approximation in which
the momenta appearing in the integral equation are cut off at the
value $\Lambda$. Strictly speaking this approach is not correct, as
the momenta going through the vertices are not treated in a manner
invariant under Galilean transformations. Instead it corresponds to
the approximation $\xi\left(|\vec p+\vec q/2|\right) \xi\left(|\vec
  p/2+\vec q|\right)\to \xi(|\vec p|)\xi(|\vec q|)$ in
Eq. (\ref{eq:matrixa}).

Using this approximation, the matrix $a$ may be evaluated to give
\begin{eqnarray}
a_{11}(p,q,E) & = &
\frac{E-p^2-q^2}{4p^2} \nn \\ &&
-\frac{(p^2-pq+q^2-E)(p^2+pq+q^2-E)}
{8p^3q}\log\frac{p^2-pq+q^2-E}{p^2+pq+q^2-E}, \nn \\
a_{12}(p,q,E) & = & \frac{6p^4+p^2(5q^2-12E)+3(q^2-E)(q^2-2E)}
{12p^2q^2} \nn \\ &&\hspace{-15mm}+\frac{(2p^2+q^2-2E)(p^2-pq+q^2-E)
(p^2+pq+q^2-E)}{8p^3q^3}
\log\frac{p^2-pq+q^2-E}{p^2+pq+q^2-E}, \nn \\
a_{21}(p,q,E) & = &
\frac{-3E-2p^2+3q^2}{4p^2}+\frac{3(q^2-E)^2-p^2E}{8p^3q}
\log\frac{p^2-pq+q^2-E}{p^2+pq+q^2-E}, \nn \\
a_{22}(p,q,E) & = &
\frac{2E-2p^2-q^2}{8p^3q^3}
\left[2pq(2p^2+3q^2-3E)
\vphantom{+\left(p^2(2p^2+4q^2-5E)+3(q^2-E)^2\right)
\log\frac{p^2-pq+q^2-E}{p^2+pq+q^2-E}}\right.
\nn \\ && \hspace{4mm}
\left.\vphantom{2pq(2p^2+3q^2-3E)}
+\left(p^2(2p^2+4q^2-5E)+3(q^2-E)^2\right)
\log\frac{p^2-pq+q^2-E}{p^2+pq+q^2-E}
\right].
\label{eq:simplesta}
\end{eqnarray}
Note that since the total energy $E\leq0$ there are no poles in the
angular integration.

\begin{figure}[bt]
\begin{center}
\includegraphics[height=2.2 in]{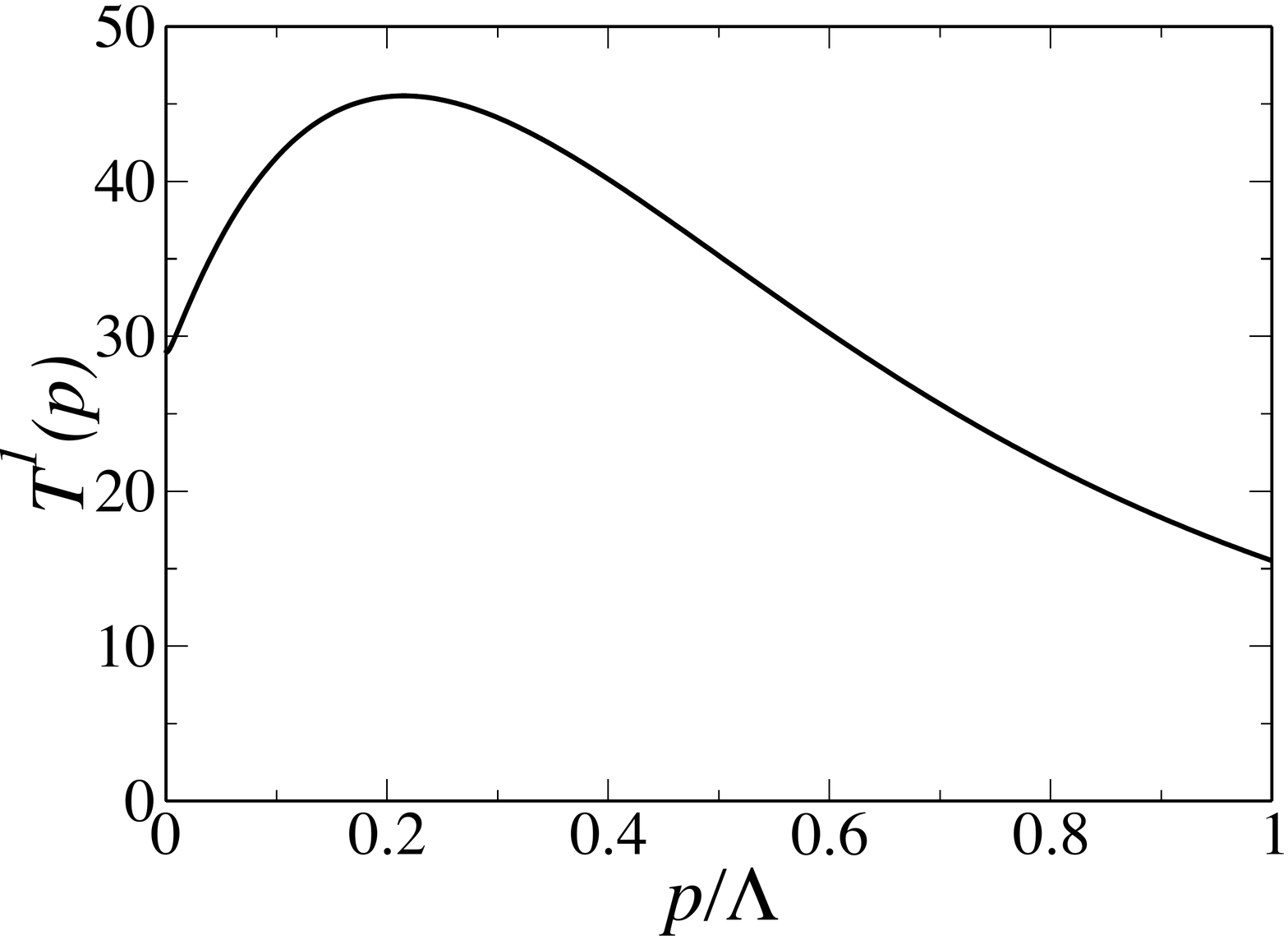}
\includegraphics[height=2.2 in]{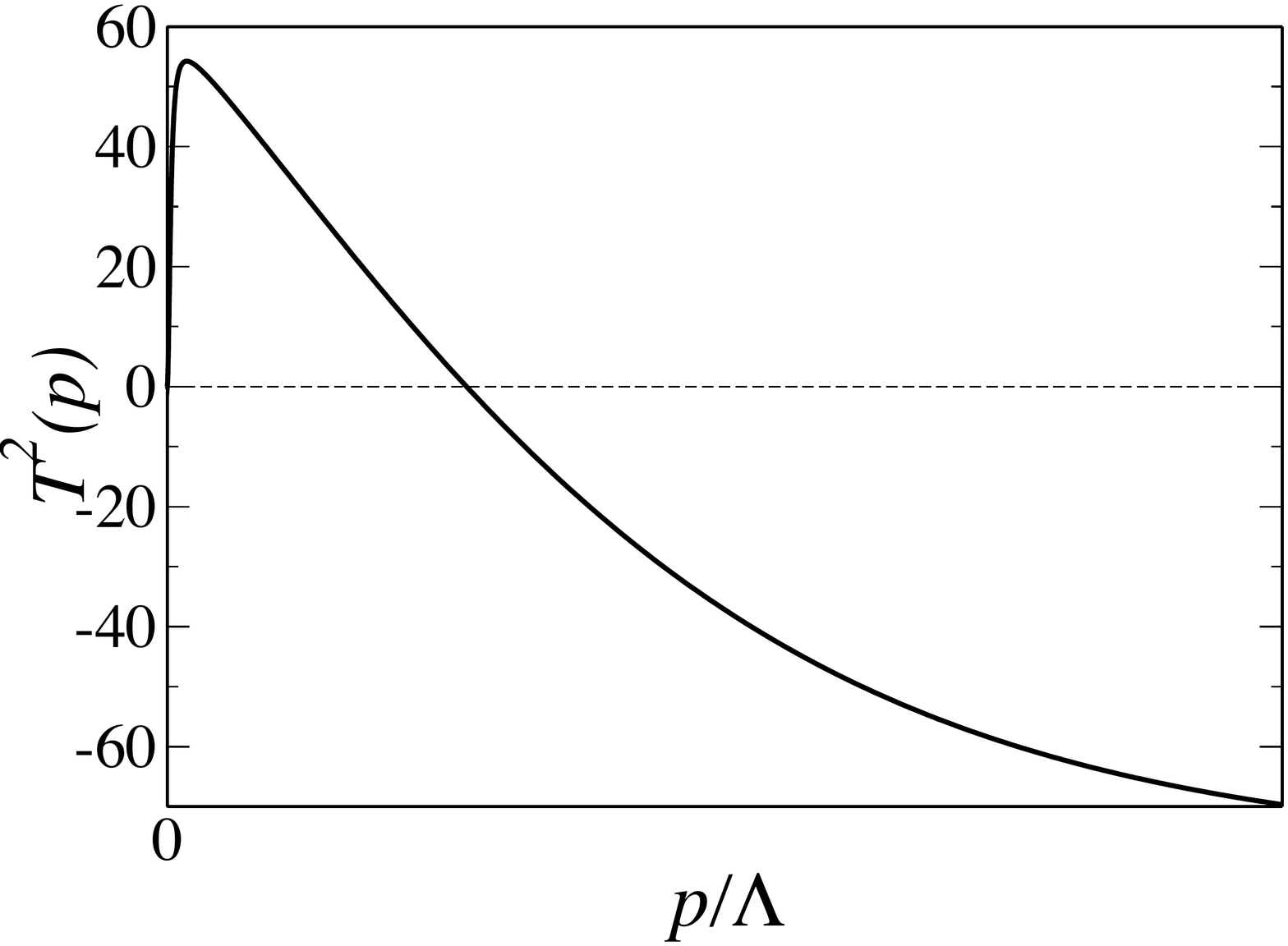}
\caption{The functions $T^1(p)$ and $T^2(p)$ in units of
  $\frac1{m\Lambda}$ evaluated using the simplest hard cut-off.}
\label{fig:thard0}
\end{center}
\end{figure}

The integral equation (\ref{eq:pwaveinteq2}) may then be solved at any
given value of the strength of the Feshbach resonance, $c_2$, the
total energy, $E$, and the detuning $\omega_0$. Fig. \ref{fig:thard0}
shows the functions $T^1(p)$ and $T^2(p)$ in the limit of very strong
resonances with a small detuning. The $s$-wave scattering length is
determined from
\begin{equation}
a_{\rm bf} = \frac{m}{3\pi}T^1(0)
\end{equation}
evaluated at $E=\kappa_0(\omega_0)$. For the specific choice of
parameters
\begin{equation}
a_{\rm bf} \approx 3\Lambda^{-1}, \hspace{10mm}\mbox{at }c_2\to\infty,\,
\omega_0=-10^{-5}\Lambda^2/m.
\end{equation}
The scattering length only weakly depends on the detuning as will be
discussed below.

\begin{figure}[bt]
\begin{center}
\includegraphics[height=2.3 In]{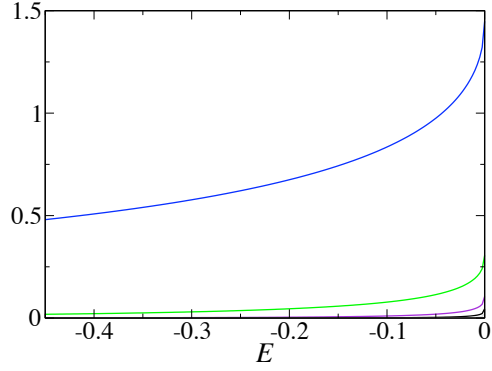}
\caption{The largest eigenvalues of the integration kernel in
  Eq. (\ref{eq:pwaveinteq2}) as a function of total energy $E$ in
  units of $\Lambda^2/m$. The eigenvalues are evaluated at
  $c_2\to\infty$ and $\omega_0=-10^{-5}\Lambda^2/m$. The largest
  eigenvalue is seen to go through 1 at $E\approx-0.04\Lambda^2/m$,
  corresponding to the appearance of the bound trimer.}
\label{fig:eigenvalues}
\end{center}
\end{figure}

Bound trimer states are found as solutions to the homogenous integral
equation at $E<\kappa_0$. One bound state is found in the problem, at
\begin{equation}
E_3\approx -0.04\Lambda^2/m, \hspace{10mm}\mbox{at }c_2\to\infty,\,
\omega_0=-10^{-5}\Lambda^2/m.
\end{equation}
The binding energy is seen to be proportional to $\Lambda^2$, thus
this three-body bound state is of size of the order of $R_e$. The
method used to find the binding energy is the following (also
discussed in chapter \ref{chap:bcsbec}): A bound state in the problem
shows up as a pole of the scattering amplitude at a negative total
energy, the binding energy $E_3$. This in turn corresponds to a
solution of the homogenous integral equation at that energy, thus to
an eigenvalue equal to 1 of the integration
kernel. Fig. \ref{fig:eigenvalues} shows the behavior of the largest
eigenvalues as functions of the total energy $E$. It is seen that an
eigenvalue equal to 1 is obtained for a total energy of $E_{\rm
  b}\approx-0.04\Lambda^2/m$.

\subsubsection{The invariant hard cut-off}
It is now interesting to see if the simplified approach above indeed
captured the correct physics. Therefore, the matrix $a$ of
Eq. (\ref{eq:matrixa}) is now evaluated in an invariant way, using
\begin{equation}
\xi\left(|\vec p+\vec q/2|\right) \xi\left(|\vec p/2+\vec
  q|\right)=\Theta\left(\Lambda-p^2-q^2/4-\vec p\cdot\vec q\right)
\Theta\left(\Lambda-q^2-p^2/4-\vec p\cdot\vec q\right).
\end{equation}
It is seen that the maximum possible value of the momenta is now
$2\Lambda$ and the integral equation is solved on the interval
$p\in[0,2]$. The matrix $a$ may again be obtained exactly. For
$p,q\leq\Lambda$ it reduces to the result (\ref{eq:simplesta})
above. For larger $p,q$ the limits of angular integration depend on
the cut-off and, although the integration is simple to perform, the
resulting expressions are not very enlightening and will not be
written here.

\begin{figure}[bt]
\begin{center}
\includegraphics[height=2.2 In]{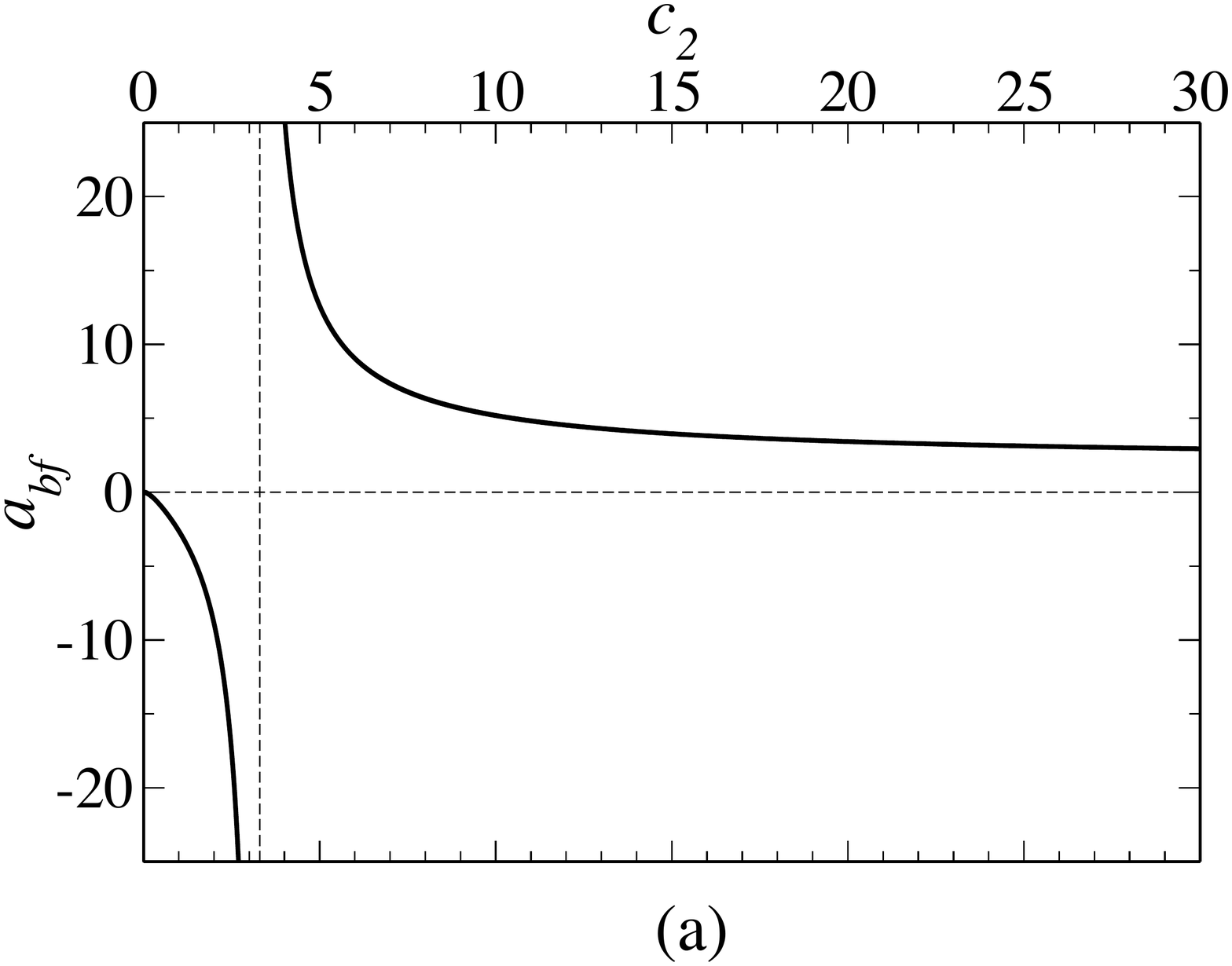}
\includegraphics[height=2.2 In]{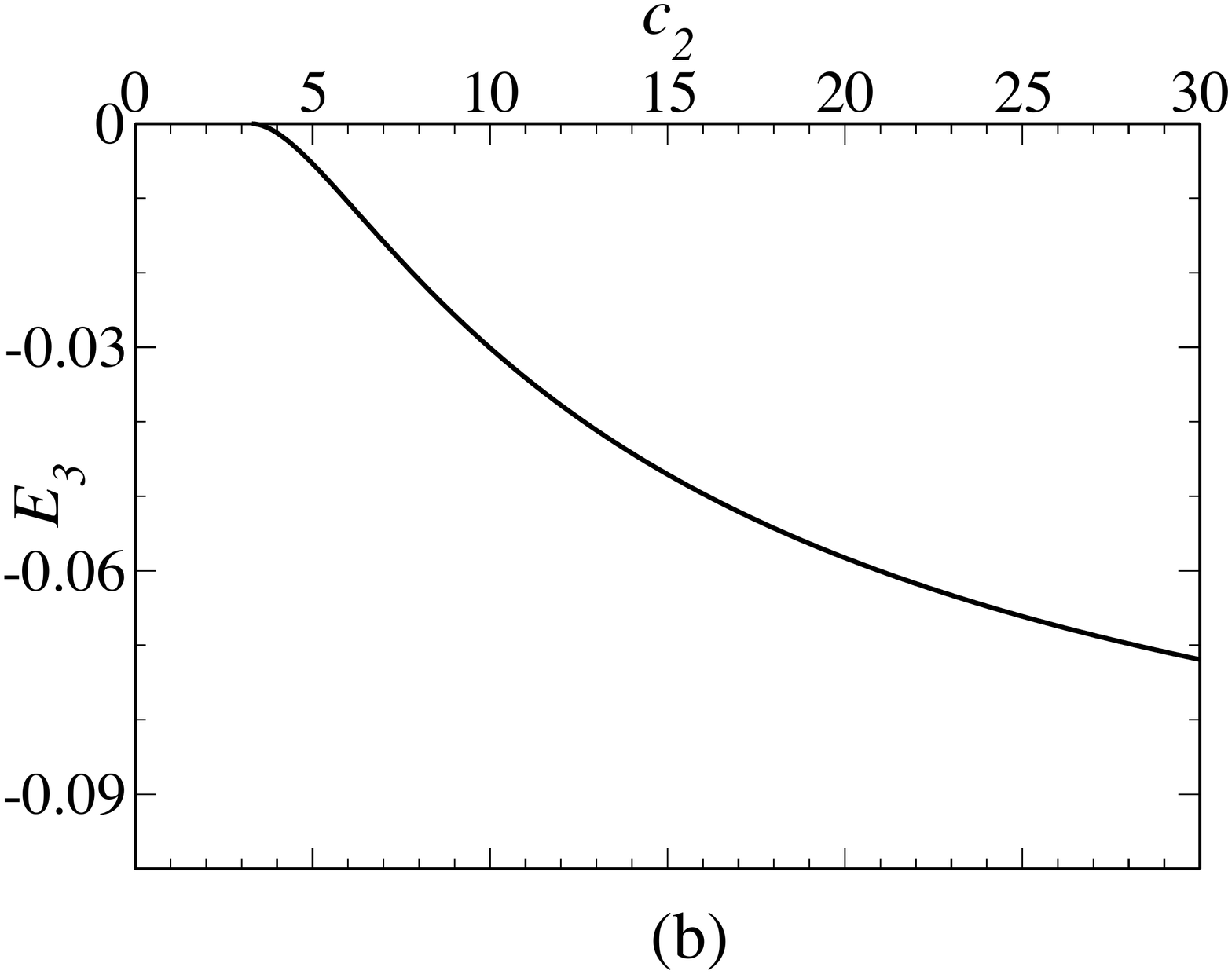}
\caption{(a) The atom-molecule scattering length in units of
  $\Lambda^{-1}$ and (b) the binding energy in units of $\Lambda^2/m$,
  both as functions of $c_2$. The position of the appearance of the
  bound state is seen to be at $c_{2,\rm crit}\approx 3.3$.}
\label{fig:abfvsc2}
\end{center}
\end{figure}

In the limit of very strong resonances ($c_2\to\infty$) the integral
equation (\ref{eq:pwaveinteq2}) reduces to
\begin{eqnarray}
\hspace{-11mm}  T^j(p) & = & -2Z\,g^2m\,G(\vec
  p,-p^2/2+E)p^2\delta_{2,j} \nn \\ &&
  \hspace{-18mm}
  -\frac{6}{\pi^2}
  \int_0^\infty q^2dq
  \frac1{E-3q^2/4-\omega_0+
    (3q^2/4-E)^{3/2}\tan^{-1}
    \frac1{\sqrt{3q^2/4-E}}}
  a_{ji}(p,q,E)T^i(q).
\end{eqnarray}
It is of interest to study the behavior of the physics as the
strength of the resonance is increased. Had $c_2$ been zero, the only
diagram which would have contributed is the Born diagram whose value
was computed in Eq. (\ref{eq:pwaveborn}) above. The Born contribution
does not have a pole at a negative energy and thus the bound state is
absent for weak resonances. It follows that the bound state found for
strong resonances must develop at a certain critical value of
$c_2$. This is indeed the case as depicted in
Fig. \ref{fig:abfvsc2}. The scattering length $a_{\rm bf}$ vanishes
for $c_2=0$ since the Born term is purely proportional to $p_\mu
p_\nu$, see Eq. (\ref{eq:pwaveborn}). $a_{\rm bf}$ becomes negative
for small values of $c_2$ and diverges to $-\infty$ at the critical
value
\begin{equation}
c_{2,\rm crit} \approx 3.3.
\end{equation}
The scattering length then reemerges from $+\infty$ to go
asymptotically towards the value of
\begin{equation}
a_{\rm bf}\approx 1.9\Lambda^{-1},
\end{equation}
as $c_2$ approaches $\infty$. The critical value of $c_2\approx 3.3$
is exactly the value at which the bound state appears. For large
values of $c_2$ the binding energy saturates at
\begin{equation}
E_3\approx -0.11\Lambda^2/m.
\end{equation}

The values obtained here in the large $c_2$ limit are somewhat
different from the ones obtained above using the simplified hard
cut-off, but the results are still qualitatively consistent.

\subsection{Softening the cut-off \label{sec:soft}}
The above results may be tested by applying a different cut-off
function. In this section, the coupling constant $g(|\vec p|)\equiv
g\,\xi(|\vec p|)$ is chosen to be
\begin{equation}
g(|\vec p|) = g\,\exp\left(-p^2/\Lambda^2\right),
\end{equation}
which is more smooth than the hard cut-off used above.

The first order of the day is to compute the propagator of molecules
using the new cut-off. To this end, the fermionic loop $\Sigma$ is
needed and according to Eq. (\ref{eq:pwavebubble}) it becomes
\newcommand{\erfc}{{\rm Erfc}}
\begin{eqnarray}
  \Sigma_{\mu\nu}(p,p_0) & = & \delta_{\mu\nu}\frac{g^2}{3\pi^2}
  \int dq\frac{q^4\exp\left(-2q^2/\Lambda^2\right)}
  {p_0-q^2/m-p^2/4+i0} \nn \\
  & \hspace{-27mm} = & %
\hspace{-17mm}
  -\frac{mg^2}{3\pi^2}\delta_{\mu\nu}\left\{\frac{\sqrt{2\pi}}{16}
    \Lambda^3 
    -\frac{\sqrt{2\pi}}4m\Lambda(q^2/4m-q_0-i0)
\right.\nn \\ && \left.
    +\frac\pi2m^{3/2}
    (q^2/4m-q^0-i0)^{3/2}
    \exp\left(2\frac{q^2/4m-q_0}{\Lambda^2/m}\right)\erfc\sqrt{2
      \frac{q^2/4m-q_0}{\Lambda^2/m}}\right\} \nn \\
  & \hspace{-27mm}
\equiv & 
\hspace{-17mm}
\delta_{\mu\nu} \left\{-c_1'+c_2'(q^2/4m-q_0-i0)-c_2'\frac{\sqrt
      m}{\Lambda}(q^2/4m-q_0-i0)^{3/2}F\left(\frac{
        q^2/4m-q_0}{\Lambda^2/m}\right)\right\}. \nn \\ &&
\end{eqnarray}
The coefficients $c_1'$ and $c_2'$ are the analogues of $c_1$ and
$c_2$ calculated in chapter \ref{chap:pwavesetup} for the hard
cut-off. They are slightly changed and take the values
\begin{equation}
c_1'= \frac{mg^2\Lambda^3}{12(2\pi)^{3/2}}, \hspace{1cm}
c_2' = \frac{m^2g^2\Lambda}{3(2\pi)^{3/2}}.
\end{equation}
The function $F$ replaces the $\arctan$ present in the hard cut-off
propagator and is given by
\begin{equation}
F(x) = \sqrt{2\pi} \exp(2x)\erfc\sqrt{2x}.
\end{equation}
$\erfc$ is the complementary error function
\begin{equation}
\erfc(x)\equiv1-\frac{2}{\sqrt\pi}\int_0^x e^{-t^2}dt.
\end{equation}
Note that $F(x)_{x\to0}\approx\sqrt{2\pi}$ while
$F(x)_{x\to\infty}\approx x^{-1/2}$.

Using Eq. (\ref{eq:dressedp}) the molecular propagator becomes
\begin{equation}
D_{\mu\nu}(\vec q,q_0) = \frac{\delta_{\mu\nu}}{(1+c_2')
(q_0-q^2/4m-\omega_0)+c_2'\frac{\sqrt m}{\Lambda}(q^2/4m-q_0-i0)^{3/2}
F\left(\frac{q^2/4m-q_0}{\Lambda^2/m}\right)}.
\end{equation}
As above, $\omega_0=\frac{\epsilon_0-c_1'}{1+c_2'}$. The on-shell
molecular propagators must be evaluated at $D(\vec
q,q_0+\kappa_0\left(\omega_0)\right)$ with $\kappa_0$ satisfying the
implicit relation
\begin{equation}
  \omega_0 = \kappa_0(\omega_0)+\frac{c_2'}{1+c_2'}\frac{\sqrt m}\Lambda
  (-\kappa_0(\omega_0))^{3/2}
  F\left(-\frac{m\kappa_0(\omega_0)}{\Lambda^2}\right).
\end{equation}
The residue of the molecular propagator at its pole takes the value
\begin{equation}
Z = \left[1+c_2'-\frac{2c_2'}{\sqrt\pi}\frac{m\kappa_0(\omega_0)}{\Lambda^2}
-c_2'\left(\frac32-\frac{2m\kappa_0(\omega_0)}{\Lambda^2}\right)
\sqrt{\frac{-m\kappa_0(\omega_0)}{\Lambda^2}}
F\left(-\frac{m\kappa_0(\omega_0)}{\Lambda^2}\right)\right]^{-1}
\end{equation}
and reduces to $Z\approx1/(1+c_2')$ as the detuning vanishes.

\begin{figure}[bt]
\begin{center}
\includegraphics[height=2.2 in]{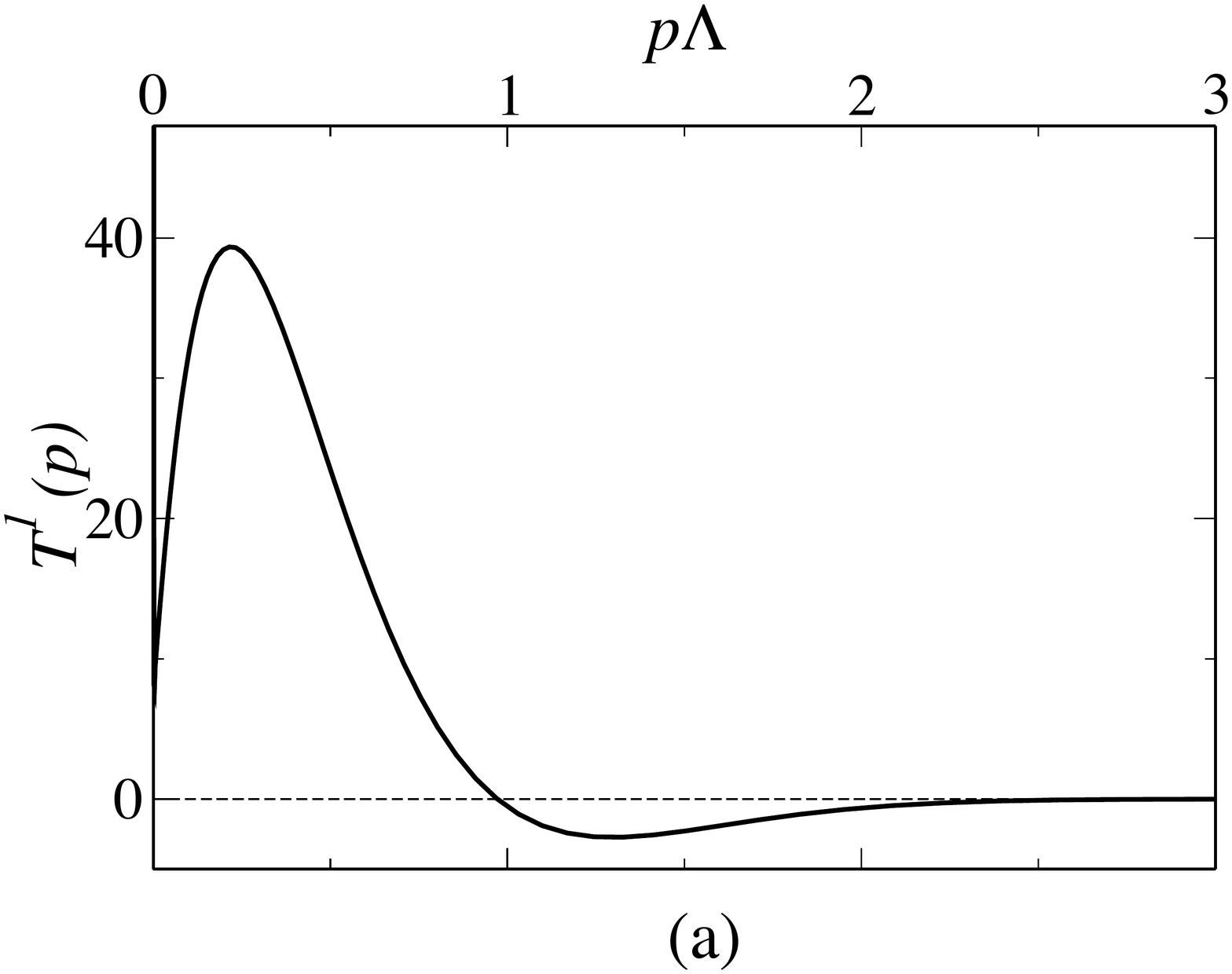}
\includegraphics[height=2.2 in]{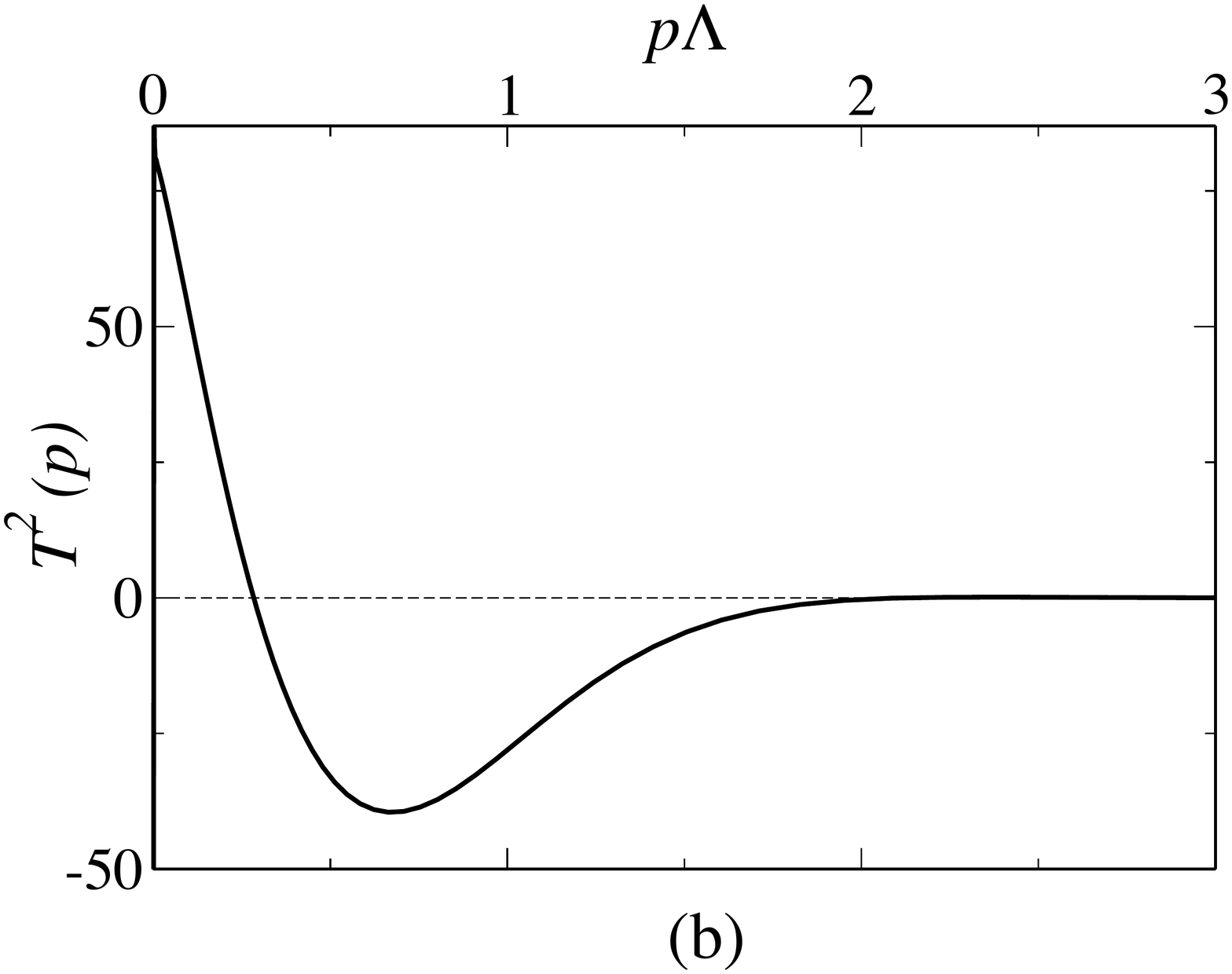}
\caption{The functions $T^1(p)$ and $T^2(p)$ in units of
  $\frac1{m\Lambda}$ evaluated using the ``soft'' cut-off.}
\label{fig:t1t2soft}
\end{center}
\end{figure}

In the limit $c_2'\to\infty$ and $\omega_0\to0$ the scattering
amplitudes $T^1(p)$ and $T^2(p)$ are shown in
Fig. \ref{fig:t1t2soft}. These are obtained by solving the integral
equation (\ref{eq:pwaveinteq2}) using the exponential cut-off. The
functions are seen to approximately match those obtained in the
simplified hard cut-off problem, see Fig. \ref{fig:thard0}.

 

\begin{figure}[bt]
\begin{center}
\includegraphics[height=2.2 In]{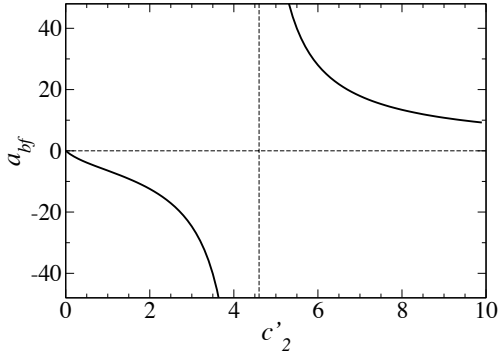}
\caption{Behavior of the scattering length in units of $\Lambda^{-1}$
  under a change of strength of the resonance.}
\label{fig:pwavesoft2}
\end{center}
\end{figure}

Fig. \ref{fig:pwavesoft2} depicts the scattering length
as a function of the strength of the resonance. The detuning has here
been taken to vanish. The critical strength of the resonance, above
which the bound trimer state is present, is found to be
\begin{equation}
c_{2,\rm crit}' \approx 4.6.
\end{equation}

The results presented above for the hard and exponential cut-offs do
not match completely, yet the qualitative behavior is found to be the
same. It should be emphasized that the binding energy for strong
resonances goes as $\Lambda^2/m$ and is very large. The binding energy
is thus sensitive to the physics at large momenta, i.e. to the
specific choice of the regulator $g(|\vec p|)$. The question is then,
does the existence of the trimer depend on the choices of $g(|\vec
p|)$ presented above, that is, is it an artifact of the model and not
the actual physical phenomena? To answer this question, consider the
system close to the formation of the trimer, i.e. $c_2$ close to
$c_{2,\rm crit}$. Here, the binding energy of the trimer may be
arbitrarily small and is a low energy phenomenon, insensitive to the
choice of regulator (on the other hand, the precise value of $c_{2,\rm
  crit}$, being proportional to $\Lambda$, does depend on the choice
of $g(|\vec p|)$). As $c_2$ is increased from $c_{2,\rm crit}$, the
binding energy $E_3$ quickly becomes large. However, once the trimer
has formed, changing the regularization scheme to make the model more
realistic at short distances does not make the trimer
disappear. Indeed, in order for the trimer to disappear at some
$c_2>c_{2,\rm crit}$, the binding energy would have to first become
large and then again drop to zero. The disappearance of the trimer
would then again be a low energy phenomenon and would be predicted by
the theory, contrary to what was found above. In this sense, even
though the binding energy of the trimer is large, the existence of the
trimer is a result of low energy physics. In conclusion, although the
precise value of the binding energy of the trimer depends on the
choice of regulator, the presence of the trimer state for strong
resonances does not.

\section{Stability of the $p$-wave superfluid \label{sec:pwavestab}}
It is important to estimate the lifetime of the $p$-wave
superfluid. In the strongly resonant BEC regime studied here, one of
the main decay channels will be inelastic collisions in which two
molecules turn into an atom and a trimer. In the process, the large
binding energy of the trimer will be released as kinetic energy and
the particles are lost from the system.

\begin{figure}[bt]
\begin{center}
\includegraphics[height=1 in]{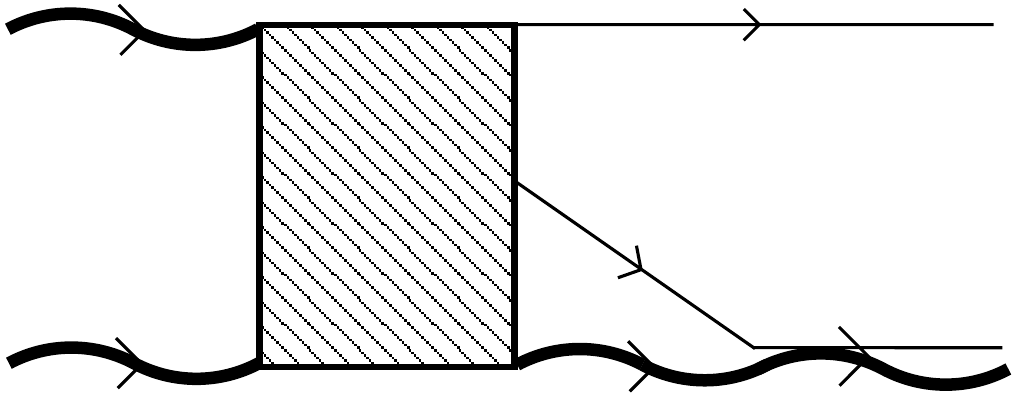}
\caption{The processes which lead to the inelastic losses,
  Eq. (\ref{eq:inel}). The trimer is illustrated as the binding of a
  fermion and a molecule.}
\label{fig:inel}
\end{center}
\end{figure}

As in the case of $s$-wave collisonal losses studied in section
\ref{sec:relaxation} the loss rate will be written as
\begin{equation}
\dot n_{\rm b} = -\alpha_{\rm in}n_{\rm b}^2.
\end{equation}
$n_{\rm b}$ is again the density of bosonic molecules. The inelastic
relaxation rate is given by
\begin{equation}
\alpha_{\rm in} = \frac2 m\left<k_i \sigma_{\rm in}(k_i)\right>.
\label{eq:inel}
\end{equation}
Here, $k_i$ is defined as the relative momentum of the incident
molecules and the average is taken over $k_i$. The inelastic
scattering cross section is given by \cite{LL}
\begin{equation}
\sigma_{\rm in} = \int |f_{\rm in}|^2d\Omega\frac{k_f}{k_i}.
\end{equation}
$f_{\rm in}$ is the inelastic scattering amplitude into the final
relative momentum $k_f$ and is of order $R_e$, just like the
three-body scattering length above was found to be proportional to
$1/\Lambda\sim R_e$. This may be seen from estimating the behavior of
diagrams of the type shown in Fig. \ref{fig:inel}. These diagrams are
similar to the form studied in the $s$-wave four-fermion problem but
in the $p$-wave formalism. The final relative momentum arises from
the kinetic energy released in the binding of three fermionic atoms
into a trimer, thus $k_f\sim \Lambda\sim 1/R_e$ and
\begin{equation}
\alpha_{\rm in} \sim \frac{R_e}m.
\end{equation}

Additionally, there are inelastic decay channels similar to those
studied in section \ref{sec:relaxation} in which a molecule decays
into a deeply bound molecular state in the presence of an additional
atom or molecule. Estimates similar to the above show that the decay
rate due to these processes are of the same order as $\alpha_{\rm
  in}$. Note that these are present both for weak and strong
resonances, unlike the decay to the trimer state.

Consider the experiment described in Ref. \cite{Gaebler2007} in which
it was found that the lifetime of diatomic molecules was quite short,
of the order of 2 ms. It is interesting to estimate whether this could
be explained by the losses due to the inelastic processes. The decay
rate is $\Gamma_{\rm in}\sim\alpha_{\rm in}n$ and is density
dependent. The atomic density in the experiment was $n\approx 7\times
10^{-12}\mbox{cm}^{-3}$ and $R_e$ estimated by the size of the molecules to
be .02 $a_0$. Thus the decay rate is of the order of 10 Hz,
corresponding to a lifetime of about 100 ms. However, the prefactor to
the estimate is uncertain and it is also unknown whether the losses
observed in Ref. \cite{Gaebler2007} would depend on the density. In
conclusion, it is uncertain whether the observed lifetime of diatomic
molecules in Ref. \cite{Gaebler2007} is limited by inelastic
collisions.

Shortly after Ref. \cite{Levinsen2007}, containing the work described
in this chapter, became available on-line, the paper
\cite{Lasinio2007} by Jona-Lasinio {\it et al} appeared in which
similar conclusions were obtained for the $p$-wave superfluid.

\chapter{Conclusions and outlook
  \label{chap:conclusion}}

In this thesis, the $s$- and $p$-wave paired superfluids have been
investigated. In the wide-resonance $s$-wave crossover, properties of
the BEC regime were computed in the gas parameter expansion. Higher
order corrections were found, and the mass imbalanced BEC was
considered. It was studied how $p$-wave resonances naturally fall into
categories of weak and strong depending on the precise details of the
Feshbach resonance, and it was demonstrated how the strongly resonant
superfluid is unstable towards the formation of strongly bound trimers
with angular momentum 1. The rate of losses arising from collisional
relaxation was estimated and it was seen how the predicted lifetime of
the gas was slightly overestimated compared with the relevant
experiment.

A prominent element throughout the thesis has been the constant
interplay between few- and many-body physics. This was observed in the
$s$-wave BEC regime, where few-body physics determined the
molecule-molecule scattering length $a_{\rm b}$ in terms of the
atom-atom scattering length $a$. Once $a_b$ was known, the many-body
physics was derived in terms of this few-body parameter. The decay
rate of the superfluid due to collisional relaxation in both the $s$-
and the $p$-wave superfluids was also demonstrated to depend on
few-body physics.

Possible future directions of study in the $s$-wave case include
finite temperature extensions of the work. A study of the excitation
spectrum at momenta of the order of the inverse scattering length, for
which the underlying structure of the composite bosons is probed, might
prove fruitful. Finally, it would be interesting to compute higher
order corrections to the chemical potential and ground state energy
which explicitly demonstrate the underlying fermionic structure of the
BEC of composite bosons.

An exciting prospect is to restrict the $p$-wave superfluid to two
spatial dimensions. Here, the BCS phase has been shown to be
topological and will support vortices with non-Abelian excitations
\cite{Read2000,Gurarie2005}, both necessary ingredients for a fault
tolerant quantum computer \cite{kitaev2003}. It remains to be seen
whether a stable, two-dimensional $p$-wave superfluid can be achieved.

\bibliographystyle{plain}
\bibliography{thesis}

\end{document}